\documentclass[review]{elsarticle}
\usepackage[margin=1.25in]{geometry}
\usepackage{lineno}
\usepackage[caption = false]{subfig}
\usepackage{caption}
\usepackage{subfig}
\usepackage{color}
\usepackage{amssymb,amsmath, graphicx}
\usepackage{tikz}
 \usepackage{amssymb}
 \usepackage{graphics,graphicx}
\usepackage{graphicx}
\usepackage{epstopdf}
\usepackage{amsmath}
\usepackage{caption}
\usepackage{multirow}
\usepackage{array}
\usepackage{nomencl}
\makenomenclature
\usepackage{setspace}
\usepackage{caption}
\usepackage{float}
\usepackage{comment}
\usepackage{mathtools}
\usepackage{soul}

\journal{Journal of \LaTeX\ Templates}

\def\St{\textrm{St}}
\newcommand{\bc}{\begin{center}}
\newcommand{\ec}{\end{center}}

\newcommand{\be}{\begin{equation}}
\newcommand{\ee}{\end{equation}}

\newcommand{\bea}{\begin{eqnarray}}
\newcommand{\eea}{\end{eqnarray}}
\newcommand{\beas}{\begin{eqnarray*}}
\newcommand{\eeas}{\end{eqnarray*}}

\def\avg#1{{\left<{#1}\right>}}

\begin{document}

\begin{frontmatter}

\title{Numerical Simulation of Vortex-Induced Vibration With Bistable Springs : Consistency with the Equilibrium Constraint}
%%\tnotetext[mytitlenote]{Fully documented templates are available in the %%elsarticle package on \href{http://www.ctan.org/tex-archive/macros/latex/contrib/elsarticle}{CTAN}.}


\author[mymainaddress]{Rameez Badhurshah}
\author[mymainaddress]{Rajneesh Bhardwaj}
\author[mysecondaryaddress]{Amitabh Bhattacharya \corref{mycorrespondingauthor}}


\cortext[mycorrespondingauthor]{Corresponding author}
\ead{bhattach@gmail.com}

\address[mymainaddress]{Department of Mechanical Engineering, Indian Institute of Technology Bombay, Mumbai, 400076, India}
\address[mysecondaryaddress]{Department of Applied Mechanics, Indian Institute of Technology Delhi, Hauz Khas, New Delhi, 110016, India}

\begin{abstract}
We present results from two-dimensional numerical simulations based on Immersed Boundary Method (IBM) of a cylinder in uniform fluid flow attached to bistable springs undergoing Vortex-Induced Vibrations (VIV). The elastic spring potential for the bistable springs, consisting of two potential wells, is completely defined by the spacing between the potential minima and the depth of the potential wells. We perform simulations of VIV with linear spring, as well as bistable springs with two different inter-well separations, over a wide range of reduced velocity. As expected, large oscillation amplitudes correspond to lock-in of the lift force with the natural frequency of the spring-mass system. The range of reduced velocity over which lock-in occurs is significantly higher for VIV with bistable springs compared to VIV with linear springs, although the maximum possible amplitude appears to be independent of the spring type. For VIV with bistable springs, the cylinder undergoes double-well oscillations in the lock-in regime. Range of reduced velocity over which lock-in occurs increases when the inter-well distance is reduced. The vortex shedding patterns and amplitude trends look similar at the same equivalent reduced velocity for the different springs. The results here are consistent with our prior theory, in which we propose a new ``Equilibrium-Constraint (EC)" based on average kinetic energy budget of the structure. For a given spring potential, the intersection of natural frequency curves with the EC curve yields the possible range of reduced velocities over which lock-in should occur. Our numerical simulations show a collapse of the amplitude-versus-structure frequency data for all the simulations onto roughly the same curve, thus supporting the existence of the EC, and providing an explanation for the trends in the VIV oscillations. The present study provides fundamental insights into VIV characteristics of bistable springs, which may be useful for designing broadband energy harvesters. 
\end{abstract}

\begin{keyword}
\texttt{Vortex-Induced Vibration}\sep Bistable spring \sep Lock-in \sep Fluid structure interaction
\end{keyword}

\end{frontmatter}

%\linenumbers

\section{Introduction}
Vortex-Induced Vibrations (VIVs) resulting from fluid-structure interaction is a well-studied phenomenon. A large body of literature has discussed the suppression of VIVs to protect the structure from failure due to fatigue. On the other hand, several authors have attempted to harness useful energy from structural vibrations of bluff bodies placed in uniform flow \cite{barrero2012extracting, zhang2017design, bernitsas2008vivace, bhattacharya2016power, garg2019vortex}. The response of such structures depends largely upon 
mass ratio, damping ratio of structure, the Reynolds number, and reduced velocity of the flow. The ``lock-in" regime, where the vortex shedding frequency synchronizes with the structure frequency, plays a significant role in determining the range over which significant oscillation amplitudes of the structure can be observed \cite{williamson2004vortex}. \par

VIV of cylinders attached to linear Hookean spring has been extensively studied. However, the effect of spring nonlinearity on VIV characteristics is relatively less-studied. Experimental results by Mackowski and Williamson on VIV of cylinder attached to nonlinear hardening/softening spring demonstrated that the cylinder can sustain large amplitude vibrations over a larger range of flow velocities, compared to VIV with linear springs \cite{Mackowski}. Also, by defining a generalized value of reduced velocity, namely equivalent reduced velocity, the performance of cylinder with nonlinear spring could be predicted using prior knowledge of VIV response of cylinder attached with linear spring. \par

A nonlinear spring system may enable VIV responses at more than one meta-stable state for the given flow velocity. In this work, we focus on bistable springs, for which the force potential consists of two mimimas. Thus, a cylinder attached to bistable spring may oscillate within single-well (small-amplitude oscillation) or transit between the two wells (large-oscillation amplitude) \cite{harne2013review}. It has been known that spring-mass systems consisting of bistable springs can show a significant response over a large range of excitation frequencies \cite{huynh2017numerical}. Huynh and Tjahjowidodo performed numerical and experimental studies on VIV of a cylinder with bistable spring to develop a bifurcation map showing the regions corresponding to chaotic vibrations \cite{huynh2017experimental}. Zhang \emph{et al.} strategically placed permanent magnets at the free end of a cantilevered cylinder to enhance the power output of piezoelectric harvester from VIV of the cylinder  \cite{zhang2017design}. The experimental results report a 138 $\%$ and 29 $\%$ increment in the synchronization region and harvested power, respectively. Bistability can also arise naturally during 
angular oscillations induced by VIV \cite{bhattacharya2016power, lugt1983autorotation, greenwell2014autorotation} \par

In prior work \citep{badhurshah2019lock}, we have used a Wake Oscillator Model (WOM) \cite{facchinetti2002viv} to study VIV of cylinders attached to a bistable spring. Compared to VIV with linear springs, we found a significant increase in the range of reduced velocities over which lock-in occurs for VIV with bistable springs, especially when the distance between the potential wells is reduced. To explain this result, we proposed a theory in which we balanced the production and dissipation of mechanical energy, assuming harmonic motion of the cylinder during lock-in. We then obtained a constraint between the displacement amplitude and the structure frequency during lock-in, which we termed as the ``Equilibrium Constraint" (EC). The EC can be represented by a curve in the amplitude-versus-frequency plane, and is \emph{independent} of the type of spring, provided the cylinder motion is approximately harmonic. We carefully characterized the natural frequency of bistable springs as a function of the inter-well separation and the amplitude of oscillation. During lock-in, the intersection of the EC curve and the natural frequency curve gives the possible amplitude(s) of oscillation and structure frequency for a given reduced velocity. This theory is able to successfully predict that, for VIV with bistable springs having small inter-well separation, the range of reduced velocity over which lock-in occurs can increase dramatically. We also performed numerical simulations of the WOM equations for VIV with both linear and bistable springs, as reprted in Ref. \cite{badhurshah2019lock}. We found an excellent qualitative agreement with the predictions from our theory with results from the WOM simulations. The WOM simulations also showed the possible existence of the EC curve, since the data points from VIV with different spring types appeared to collapse on to the same curve in the amplitude-versus-frequency plane. 
\par Recently, Ellingsen and Amandolese \cite{ellingsen2020amplitude}, have also utilized the kinetic energy budget of the cylinder, along with the WOM equations, to demarcate the synchronization regimes for VIV of cylinder attached to linear spring. As part of the theoretical analysis, a constraint between the amplitude and frequency of the structure was obtained, which looks very similar to the EC curve derived in Badhurshah \emph{et al.} \citep{badhurshah2019lock}. However, the analysis presented in Ref. \cite{ellingsen2020amplitude} is specific to VIV with linear springs, and does not explore spring nonlinearity.  
\par While WOM has been used often to model VIV \cite{violette2010linear, farshidianfar2010modified, xu2010new}, it does not model the details of the flow field accurately, and is not calibrated for VIV with bistable springs. In this work, we therefore perform fully-resolved two-dimensional Computational Fluid Dynamics (CFD) simulations to obtain a more accurate characterization for VIV of cylinder attached to bistable springs at a high mass ratio ($m^*=25.46$). Our simulations are performed using a well-validated solver, based on Immersed Boundary method (IBM) \citep{mittal2008versatile}. To further understand the validity of the theory proposed in \cite{badhurshah2019lock}, we will also try to verify the existence of the EC curve using the CFD simulations, which can then in turn explain the results from the CFD simulations at least qualitatively. We will also compare and examine the vortex shedding patterns for VIV with linear and bistable springs respectively.
\par CFD based simulations for cylinders undergoing VIV have been performed extensively by several research groups, and have been validated  against experiments \cite{bao2012two, ahn2006strongly, borazjani2009numerical}. However, prior CFD simulations for VIV of cylinders attached to nonlinear spring available are limited. Recently, Wang \emph{et al.} \cite{wang2019effect} performed 2D simulations to investigate VIV response of cylinder attached with spring having cubic nonlinearity in displacement for low mass ratio and low Reynolds number $(60-220)$. Similiar to Ref. \cite{Mackowski}, the spring parameter $\lambda$ was varied over a range of positive and negative values, which allowed exploration of VIV with softening and hardening springs, respectively. The VIV response was further discussed in detail by categorizing them in either initial excitation, lower branch, upper branch, or desynchronization regimes. Also, it was reported that the VIV responses of cylinders with linear springs and cubic nonlinearity springs overlapped when data was plotted against equivalent reduced velocity. However, no theoretical explaination was provided here on how the spring nonlinearity increased the lock-in regime, and the study did not consider the effect of bistability in the spring potential. Furthermore, this study was performed at low mass ratio, for which our prior theory \cite{badhurshah2019lock} is not strictly applicable.  \par

We have organized the rest of the paper as follows. In section \ref{thebac} we  summarize the EC based theory from Badhurshah \emph{et al.} \cite{badhurshah2019lock} and also provide some new insights on the effects of structure damping on the EC. In section \ref{objectives}, we list the objectives of our CFD study. In section \ref{nummet}, we outline the methodology for our CFD study, and also present validation of the IBM based solver. Next, we discuss the main results from our CFD simulations (Section \ref{result}), including lock-in characteristics and vortex shedding patterns. We discuss the consistency of the results from CFD simulations with our prior theory (Section \ref{ampfreq}). The main conclusions from the work are presented in Section \ref{conclu}.

\begin{table}
\begin{center}
\begin{tabular}{|c | l |}
\hline
\textbf{Symbol} & \textbf{Description}  \\
\hline
$A_0$ & Dimensional characteristic displacement amplitude ($=\max(|Y|)-|\avg{Y}|$)
\\
\hline
$a_0$ & Nondimensional characteristic displacement amplitude ($=\max(|y|)-|\avg{y}|$) 
\\
\hline
$C_{D}$ & Drag coefficient ($C_D=\frac{2 F_X}{\rho_f U^2}$)\\
\hline
$C_{L}$ & Lift coefficient ($C_L=\frac{2 F_Y}{\rho_f U^2}$)\\
\hline
$C_{L0}$ & Amplitude of coefficient of lift for stationary cylinder due to vortex shedding
\\
\hline
$C_{L,vs}(T)$ & Instantaneous lift coefficient of cylinder due to vortex shedding\\
\hline
$D$ & Cylinder diameter \\
\hline
$F_Y(T)$ & Net instantaneous dimensional lift force from fluid 
\\
\hline
$F^*_Y(\tau)$ & Net instantaneous nondimensional lift force from fluid 
\\
\hline
$F_n^*$ & Nondimensional function of amplitude ($=\Omega_n/\Omega_s$)
\\
\hline
$K$ & Linearized spring constant around potential minima ($=4k_1$ for bistable spring)
\\
\hline 
$k_1,k_3$ & Spring constants for bistable spring potential 
\\
\hline
$M$ & Mass number  ($ = \frac{C_{L0}}{2}\frac{1}{8 \pi^2 St^2 \mu}$)\\  
\hline
$m$ & Sum of added mass and solid mass ($=m_s+m_f$) \\
 \hline
$m^*$ & Mass ratio based on structure mass only ($=\frac{m_s}{\rho (\pi/4) D^2 }$) 
 \\
\hline
$m_s, m_f$ & Mass of cylinder, added mass of fluid \\
\hline
$q(T)$ & Wake variable ($=2C_{L,vs}(T)/C_{L0}$) 
\\
\hline
$q_0,\psi$ & Amplitude, phase angle of wake variable
\\
\hline
$r_s, r_f$ & Structure, fluid damping coefficient \\
\hline
St & Strouhal number for vortex shedding ($=\Omega_f D/(2\pi U)$)
\\
\hline
$S(T)$ & Instantaneous lift force on cylinder due to vortex shedding ($=\frac{1}{2}\rho U^2 D C_{L,vs}(T)$)\\
\hline
$T$ & Dimensional time \\
\hline
$t,\tau$ & Nondimensional times ($t=T\Omega_f$, $\tau=TU/D$)\\
\hline
$U$ & Dimensional incident fluid velocity
\\
\hline
$U_r$ & Reduced velocity based on $\Omega_s$ ($=\frac{2\pi}{\Omega_s}\frac{U}{D}=\frac{1}{\St\delta}$)\\
\hline
$U_r^{eq}$ & Equivalent reduced velocity, based on $\Omega_n$ ($=\frac{2\pi}{\Omega_s}\frac{U}{D}=U_r \frac{\Omega_s}{\Omega_n}$)  
\\
\hline
$Y_0$ & Location of potential minima in bistable spring ($=\pm \left[ k_1/(2k_3)\right]^{1/2}$)
 \\
\hline
$Y(T)$ & Dimensional cylinder displacement \\
 \hline
 $Y_{cr}$ & Critical amplitude for bistable spring $(=\sqrt{2}|Y_0|)$\\
\hline
$y_{cr}$ & Nondimensional location of critical amplitude for bistable spring ($=Y_{cr}/D$)
\\
\hline
$y(t)$ & Nondimensional cylinder displacement ($=Y(T)/D$)\\
\hline
$\beta$ & Nondimensional location of potential minima for bistable spring ($=Y_0/D$)\\
 \hline
$\gamma$ & Stall parameter ($=\frac{r_f}{\Omega_f \rho D^2}$)  \\
 \hline
 $\delta$ &
 Reduced  angular frequency ratio   ($= \frac{\Omega_s}{\Omega_f}$) \\
 \hline 
$\mu$ & Mass  ratio based on stucture and added mass ($=\frac{m_s+m_f}{\rho D^2}$)\\
\hline
 $\xi$  &  Damping  ratio ($=\frac{r_s}{2 m \Omega_s}$)\\ 
\hline
$\rho, \rho_s$ & Density of fluid, density of solid \\
\hline
$\phi_s(Y)$ & Dimensional spring potential
\\
\hline
$\varphi_s(y)$ & Nondimensional spring potential (in WOM) 
\\
\hline
$\varphi_s^*(y)$ & Nondimensional spring potential (in CFD simulation)
\\
\hline
$\Omega$ & Fundamental frequency of structure oscillation
\\
\hline
$\Omega_f$ & Vortex shedding frequency for stationary cylinder\\
\hline

$\Omega_n$ & Fundamental natural frequency of spring ($=\Omega_s$ for linear spring)
\\
\hline

$\Omega_s$ & Natural frequency based on linearlized spring constant ($=\sqrt{K/m}$)
\\
\hline
$\omega$ & Nondimensional fundamental frequency of structure oscillation ($=\Omega/\Omega_f$)
\\
\hline
$\omega_n$ & Nondimensional fundamental natural frequency of spring ($=\Omega_n/\Omega_f$)
\\
\hline
$\omega_s$ & Nondimensional linearized spring constant ($=\Omega_s D/U$)
\\

\hline 
\end{tabular}
\end{center}
\vspace{-0.2in}
\caption{\label{nomen} Nomenclature for major symbols used in paper.}
\end{table}

\section{\label{thebac}Theoretical background}

The theory presented in Badhushah \emph{et al.}\citep{badhurshah2019lock}, which is valid when the vortex shedding is locked-in with the structure, balances production and dissipation of the kinetic energy of the structure. An ``Equilibrium Constraint" is thus obtained between the oscillation amplitude and oscillation frequency of the structure. It is then possible to explain the increase in the range of reduced velocity over which lock-in occurs for VIV with bistable springs (compared to VIV with linear springs) by examining the range of oscillation frequencies for which the natural frequency of the spring-mass system coincides with the structure frequency. We recall the salient features of the theory, along with new observations on the EC in the presence of damping in section \ref{eckeb}. The reader may refer to Badhushah \emph{et al.} \citep{badhurshah2019lock} for further details. 

\subsection{\label{strdy} Governing equations for structure}
\begin{figure}[] 
\begin{center}
\includegraphics[width=0.6\textwidth, clip]{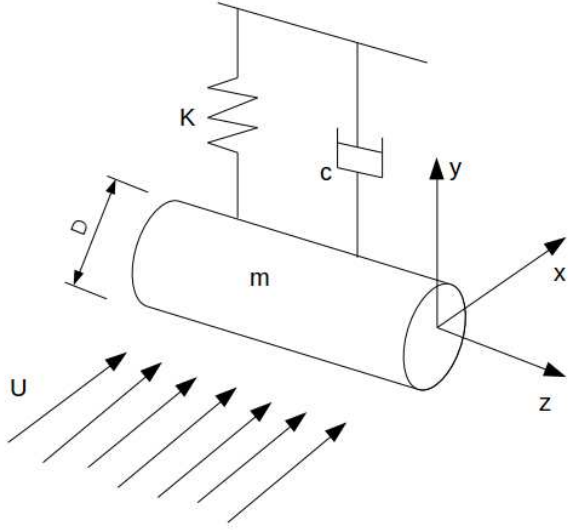}   
\end{center}
\caption{\label{schem}  Schematic sketch showing a cylinder of diameter $D$ attached with spring and damper. The simulations performed in the current study are two-dimensional. The flow is directed along X-direction while the cylinder can vibrate only in Y-direction (cross-flow vibrations).}
\end{figure}

The schematic for fluid-structure interaction involving uniform flow inflow velocity $U$ around a cylinder tethered to a spring and damper, is shown in Fig \ref{schem}. The cylinder is allowed to oscillate only along the transverse ($Y$) direction, and is attached to a spring, as well as a damper.
The acceleration of the cylinder of mass $m_s$ (per unit depth) and diameter $D$ is governed by the following dimensional linear momentum conservation equation:
\begin{equation}
\label{struc}
m_s\ddot{Y} +r_s\dot{Y} + \frac{d\phi_s}{dY} = F_Y
\end{equation}
where $Y(T)$ is the position of the cylinder axis, while $r_s$ and $\phi_s$ represents net spring damping coefficient (per unit depth), and spring force potential (per unit depth), respectively. Here $F_Y(T)$ is the net lift force (per unit depth) acting on the cylinder from the fluid. The form of the spring force potential is given as:
\begin{equation}
\label{us}
\phi_s(Y) = \left\{ \begin{array}{cc}
\frac{1}{2}K Y^2 & \textrm{For Linear Springs} \\
-k_1 Y^2+k_3 Y^4 & \textrm{For Bistable Springs} 
\end{array} \right.
\end{equation}
Where, $K$ is spring constant for the linear spring and $k_1$, $k_2$ are the parameters defining the bistable spring potential. The equilibrium stable points ($Y_0$) of bistable springs, for which $\frac{d\phi_s}{dY}\big|_{Y_0}=0$, $\frac{d^2\phi_s}{dY^2}\big|_{Y=Y_0}>0$ are:
\begin{equation}
Y_0 = \pm \Big(\frac{k_1}{2 k_3}\Big)^\frac{1}{2}
\end{equation}
Furthermore, the equivalent linear spring constant around the stable points, valid for small oscillations around $Y=\pm Y_0$, are:
\begin{equation}
 K= \frac{d^2\phi_s}{d Y^2}\bigg|_{Y=Y_0}=4 k_1
\end{equation}
We denote the linearized natural frequency of the spring as $\Omega_s = \sqrt{\frac{K}{m_s}} $ (for both types of spring) and bistability parameter as $\beta = \frac{Y_0}{D}$. Another important parameter for bistable springs is the magnitude of critical  displacement, $Y_{cr}$, where the spring potential equals zero:
  \begin{equation}
  \label{crdisp}
  Y_{cr}=\sqrt{\frac{k_1}{k_3}} = \sqrt{2} \left|Y_0 \right|
  \end{equation}
For cylinder attached to bistable spring undergoing free oscillations in vaccum (i.e. $r_s=F_Y=0$) with displacement amplitude $A_{0}$, single well oscillations will take place if $A_{0}<Y_{cr}$, while double well oscillations take place if $A_{0}>Y_{cr}$. 

\subsection{\label{eqred}Equivalent reduced velocity}
The reduced velocity in terms of the linearized natural frequency $\Omega_s$ is:
\begin{equation}
\label{redvel}
U_r = \frac{2 \pi}{\Omega_s} \frac{U}{D} 
\end{equation}
For cylinder attached to bistable spring, and cylinder attached to non-linear springs in general, the natural frequency in vacuum is a strong function of maximum oscillation amplitude \cite{Mackowski, badhurshah2019lock}. Hence, we define $U_r^{eq}$, equivalent reduced velocity for VIV with bistable springs, as:
\begin{equation}
\label{eqredvel}
U_r^{eq} = \frac{2 \pi}{\Omega_n} \frac{U}{D} = U_r \frac{\Omega_s}{\Omega_n}
\end{equation}
Here, $U_r^{eq}$ depends on the natural frequency $\Omega_n$ of the bistable spring, which in turn depends on the maximum oscillation amplitude of the cylinder. Thus, $U_r^{eq}$ is not known \textit{a priori}. In Badhurshah \emph{et al.}, the natural frequency $\Omega_n$ was tabulated for a bistable spring-mass system \cite{badhurshah2019lock} for single well and double-well oscillations, as a function of the nondimensional displacement amplitude $a_0=A_0/D$. Using dimensional analysis and a fitting function, the equivalent reduced velocity was related to $U_r$ as follows:
\begin{equation}
\label{eqredvelrel}
U_r^{eq} = \frac{U_r}{F_n^*(a_0/\beta)}  
\end{equation}
where $F_n^*=\Omega_n/\Omega_s$ is a function of $x=a_0/\beta$, as given below:
\begin{equation}
\label{fnwell}
F_n^*(x) =
\left\{
\begin{array}{cc}
F_1(x) = 1-0.5 x^2 -10x ^4&\textrm{for}\quad 0<x<\sqrt{2}-1\\
F_2(x) = 0.6 \sqrt{x^2 -1.6} &\textrm{for}\quad x>\sqrt{2}
\end{array}\right.
\end{equation}
Note that in the limit of small single well oscillations, $\lim_{x\rightarrow 0} F_n^*(x)=1$, while in the limit of large double well oscillations, $\lim_{x\rightarrow \infty} F_n^*(x)=C_\infty x$, where $C_\infty=0.6$. Finally, we observe that, for a given reduced frequency $U_r$, the normalized natural frequency $\omega_n=\Omega_n/\Omega_f$ is given by :
\bea
\label{omneq}
\omega_n&=&\frac{1}{\St U_r} F_n^*\left(\frac{a_0}{\beta}\right)
\eea
For linear springs, $F_n^*=\Omega_n/\Omega_s=1$ may be used in the above expression.

\subsection{\label{eckeb} The Equilibrium Constraint}

We first note that the force from the fluid onto the cylinder $F_Y(T)$ in Eq \ref{struc} may be decomposed into three parts as follows \cite{facchinetti2004coupling}:
\begin{eqnarray}
\label{fdecomp}
F_Y(T)&=&S(T)-r_f \dot{Y}(T) - m_f \ddot{Y}(T) 
\end{eqnarray}
where $S(T)=\frac{1}{2}\rho U^2 D C_{L,vs}(T)$ is the instantaneous lift force due to vortex shedding, $r_f=\gamma \rho \Omega_f D^2$ is the fluid-added damping coefficient, while $m_f=\rho \pi D^2/4$ is the added mass. Here $C_{L,vs}(T)$ is the instantaneous lift coefficient due to vortex shedding, $\gamma=\tilde{C}_D/(4\pi \St)$, $\tilde{C}_D$ is a coefficient of drag, $\Omega_f$ is the angular frequency of vortex shedding in the absence of cylinder motion, while $\St=\Omega_f D/(2\pi U)$ is the characteristic Strouhal number for vortex shedding. The above decomposition has been widely used in the past, especially in the context of the reduced-order Wake Oscillator Models (WOMs) for VIV \cite{blevins1990flow,facchinetti2002viv}. Following the convention of typical WOMs \cite{blevins1990flow,facchinetti2002viv}, we use $\rho$, $D$ and $\Omega_f$ to nondimensionalize Eqn. \ref{struc} as follows:
\begin{equation}\label{yeqnlin}
\frac{d^2 y}{d t^2}+(2\xi\delta+\frac{\gamma}{\mu})\frac{d y}{dt}+ \frac{d\varphi_s}{dy} =M q
\end{equation}
where $t=T\Omega_f$, $y=Y/D$, and $\varphi_s$ is the non-dimensional spring potential, such that:
\bea
\frac{d\varphi_s}{dy}&=&
\left\{
\begin{array}{ccc}
\delta^2 y & \textrm{for} & \textrm{Linear Springs} \\
- \frac{1}{2}\delta^2 y +\frac{1}{2}\frac{\delta^2}{\beta^2} y^3 & \textrm{for} & \textrm{Bistable springs}
\end{array}
\right.
\eea
Here, $\mu=(m_s+m_f)/(\rho D^2)$ is a solid-to-fluid mass ratio which accounts for the added mass, $\xi=\frac{r_s}{2 m \Omega_s}$ is the structure damping ratio, $\delta=\Omega_s/\Omega_f$ is the nondimensional linearized spring constant, $M= \frac{C_{L0}}{2}\frac{1}{8 \pi^2 St^2 \mu}$ is a constant known as ``Mass Number", while $q(t)=2 C_{L,vs}(T)/C_{L0}$ is the wake variable, with $C_{L0}$ being the amplitude of lift coefficient for stationary cylinder. Assuming that the structure oscillations and fluid forces are periodic in time, we can derive the following budget equation for the kinetic energy of the structure from Eqn. \ref{yeqnlin}:
\bea
\label{proddiss}
(2\xi \delta+C_\gamma) \avg{ {\dot y}^2}&=&M \avg{q\cdot \dot{y}}
\eea
where $C_\gamma=\frac{\gamma}{\mu}$, and $\avg{\cdot}$ implies averaging over several oscillation cycles. {
During lock-in, we assume that the oscillations are approximately harmonic, so that $y(t)=a_0 \cos  \omega t$, while $q(t)=q_0 \cos (\omega t+\psi) $. Here $ \omega=\Omega/\Omega_f$ is the non-dimensional structure frequency and $\psi$ is the phase difference between the displacement and lift force.} Substituting the harmonic forms for cylinder displacement and lift force into Eq \ref{proddiss}, we obtain the following approximate constraint between $a_0$ and $\omega$:
\bea
\label{EC}
(2\xi\delta+C_\gamma) a_0 \omega&=&M q_0(a_0,\omega) \sin\psi(a_0,\omega)
\eea 
{We refer the reader to Ref. \cite{badhurshah2019lock} for the expressions $q_0(a_0,\omega)$ and $\psi(a_0,\omega)$, which have been obtained via data generated from one-way coupled simulations of the wake equation in the Wake Oscillator Model (WOM) \cite{facchinetti2002viv}}. We denote the constraint (Eqn. \ref{EC}) between $\omega$ and $a_0$ as the ``Equilibrium Constraint" (EC). Under the assumptions stated above, the EC does not depend on the type of spring. 
\par
\begin{figure}[] 
\begin{center}
\begin{tabular}{cc}
\subfloat[Linear spring]{\label{linecpl}\includegraphics[scale = 0.5]{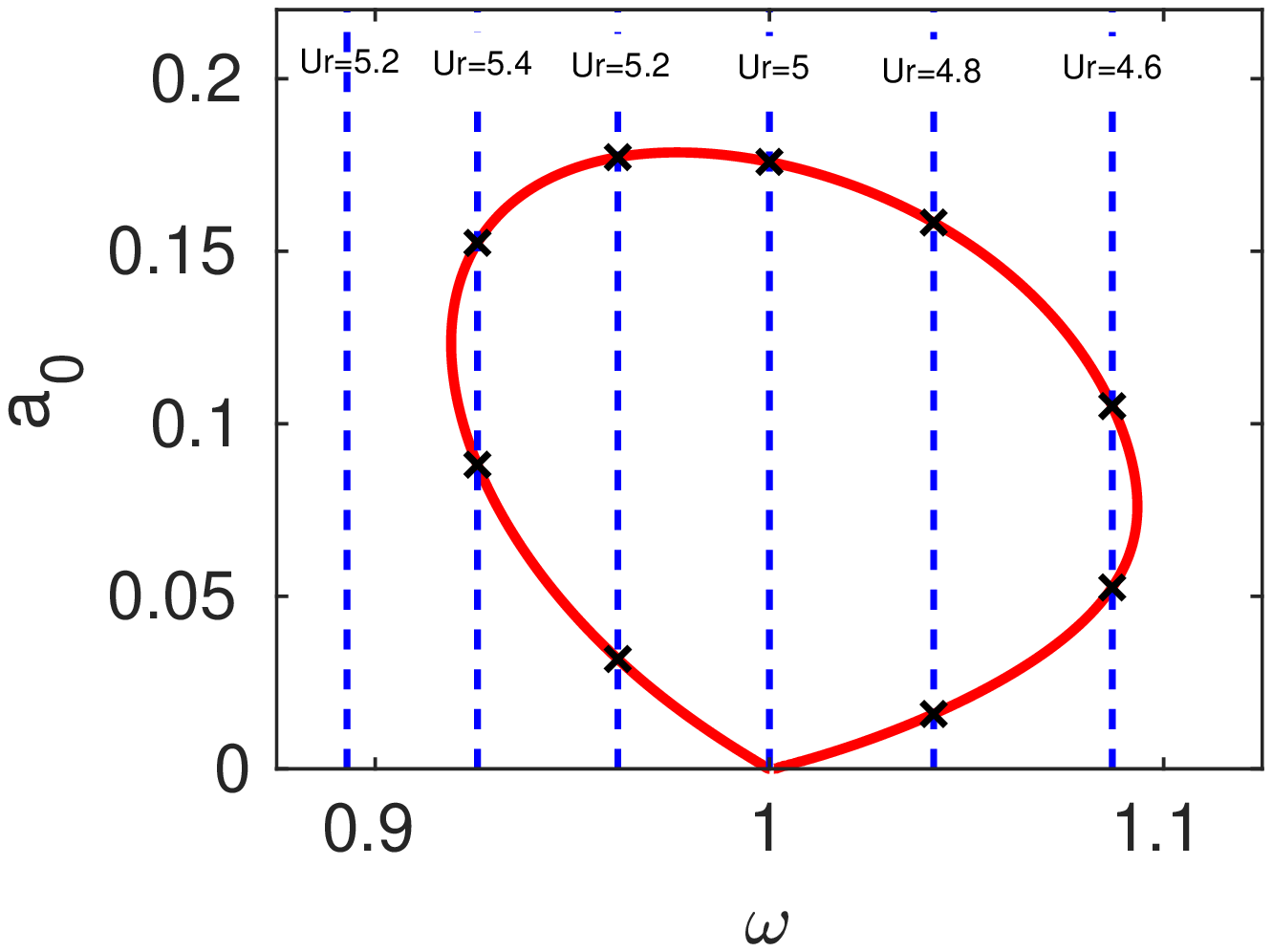}} &  
\subfloat[Bistable spring]{\label{biecpl}\includegraphics[scale = 0.5]{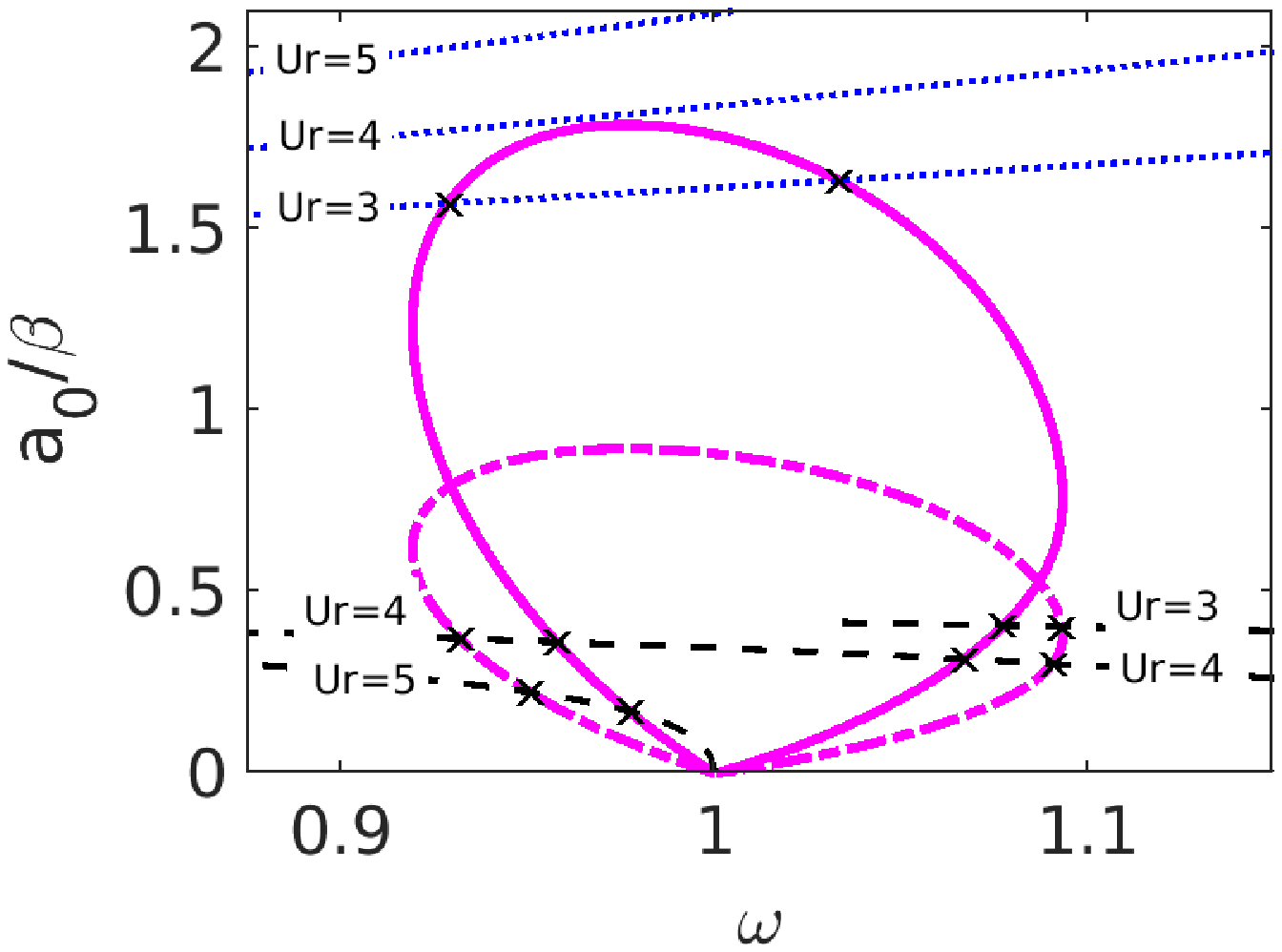}} \\

\end{tabular}
\end{center}
\caption{\label{ecplot} { Schematic plot showing prediction of amplitude and frequency during lockin for VIV with (a) linear spring (b) bistable spring with zero structure damping ($\xi=0$). In (a) the (solid) magenta curve shows zero isocontour of Eq . \ref{EC} which has been termed as Equilibrium constraint (EC) between $a_0$ and $\omega$. The black vertical lines represents the normalized natural frequency $\omega=1/(\St U_r)$ for linear spring for different $U_r$ values. In (b), the EC has been plotted on a $a_0/\beta$-vs-$\omega$ plane for (solid magenta) $\beta=0.1$ and (dashed magenta) $\beta=0.2$. The blue and black lines here denote the natural frequency $\omega=F_n^*(a_0/\beta)/(\St U_r)$ (Eqn \ref{omneq}) for double well oscillations and single well oscillations respectively, for different $U_r$ values. As per Eq. \ref{fnwell}, there is a range $a_0/\beta\in[\sqrt{2}-1,\sqrt{2}]$ over which the natural frequency is not defined. The crosses ($\times$) here represent values of $a_0,\omega$ where the EC curve intersects the natural frequency curves, indicating possibility of lock-in.}}
\end{figure}

 For zero structure damping (i.e. $\xi=0$), the EC reduces to the form $P(a_0,\omega)=0$, where  $P(a_0,\omega)=C_\gamma a_0 \omega - M q_0(a_0,\omega)\sin\psi(a_0,\omega)$. For this limit, it may be noted that the EC is also independent of the mass ratio $\mu$, due to the factor $1/\mu$ canceling out from both $C_\gamma$ and $M$. In the rest of the paper, we will focus on the VIV with zero structural damping (i.e. free oscillations). In Fig \ref{ecplot}(a) the EC (solid magenta, calibrated using WOM with standard parameters \cite{badhurshah2019lock}) has been plotted on the $a_0-\omega$ plane. It is clear that the EC curve spans a small range in $\omega$ ($0.92<\omega<1.12$), and that, for the same value of $\omega$, the EC curve can yield two values of $a_0$.  
\par In the presence of non-zero damping (i.e. $\xi\neq 0$) Eqn. \ref{EC} implies that the shape of the EC curve depends on $2\xi\delta \mu$, which is in fact proportional to $S_G/U_r$, where $S_G=8\pi^2 \St^2 \xi \mu$ is the Skop-Griffin number \cite{skop1975theory,facchinetti2004coupling,khalak1999motions}. Thus, the family of EC curves is in fact parameterized by $S_G/U_r$, and not just $S_G$. However, typically, researchers fix $S_G$ while characterizing VIV with non-zero structure damping. Due to the form of $q_0(a_0,\omega)$ and $\sin \psi(a_0,\omega)$, we have found that the height and width of the EC curve shrinks for positive values of $\St/U_r$ (not shown here). For VIV with linear springs, the range of $U_r$ over which lock-in takes place is limited, and therefore the EC depends primarily on $S_G$. However, when the lock-in range increases (e.g. due to spring non-linearity \cite{Mackowski,badhurshah2019lock,wang2019effect}), the shape of the EC curve can depend significantly on $U_r$ as well. For instance, for a fixed value of $S_G\sim O(1)$, if lock-in occurs at $U_r\gg 1$, then $S_G/U_r$ may become negligible, leading to $(a_0,\omega)$ approximately coinciding with the EC corresponding to zero damping $\xi=0$, which thus also allows for large values of $a_0$. The dependence of the EC on $S_G/U_r$ may explain the experimental results by Mackowski and Williamson \cite{Mackowski} on VIV with nonlinear springs in the presence of damping. Here it was found that, for a fixed mass-damping parameter, large-amplitude oscillation (and therefore high energy extraction efficiency) was observed for VIV with nonlinear springs, especially at larger reduced velocity.

\subsection{\label{locinb} Prediction of possible lock-in range}
For the discussion in this section, as well as the rest of the paper, we will neglect the structure damping (i.e. we will assume $\xi=0$). {The intersection of EC curve with the amplitude-dependent natural frequency curves can give us the range of $U_r$ over which lock-in may occur. In Fig. \ref{linecpl}, the dashed vertical lines denote natural frequency of the linear spring-mass system ($\omega_n = \frac{1}{\St U_r}$), and their intersection with the EC represent the possible lock-in points on the $a_0$-vs-$\omega$ plane. Here, lock-in takes place over a relatively narrow range of $U_r$ ($4.54<U_r<5.43$), since the natural frequency curves are vertically oriented. }
 \par
{For bistable springs (Fig \ref{biecpl}), natural frequency $\omega_n$ depends on $a_0/\beta$ (Eq. \ref{fnwell}). Therefore, we have plotted the EC on an $a_0/\beta$-vs-$\omega$ plane for bistable springs, with representative inter-well separation $\beta=0.1$ and $\beta=0.2$. Clearly, the natural frequency curves (black, blue lines in Fig \ref{biecpl}) are no longer vertical. As a result, the range of $U_r$ over which the EC and natural frequency curves intersect (and therefore enable lock-in) increases for bistable springs. For $\beta=0.2$, the lock-in involves purely small single well oscillations. However, for $\beta=0.1$, the lock-in may involve double-well oscillations as well.} 
In Badhurshah \emph{et al.} \cite{badhurshah2019lock}, we also derived the following approximate expression for lock-in range with bistable springs involving double well oscillations:    
\bea
\label{approxlockin}
\frac{C_\infty \sqrt{2}}{\textrm{St}} \leq U_r \leq \frac{C_\infty}{\textrm{St}}\frac{a_{max}}{\beta} 
\eea 
where $a_{max}$ is the maximum value of $a_0$ in the EC curve. The above expression assumes $a_{max}/\beta\gg 1$, for which the normalized natural frequency curve $\omega_n(a_0)=F^*_n(a_0/\beta)/(\St U_r)\approx C_\infty a_0/(\beta\St U_r)$ is almost a straight line. Since the EC exists over a narrow range in $\omega$, therefore the regime over which double well oscillations take place may be approximated as a line segment joining $(1,y_{cr})$ and $(1,a_{max})$ on the $\omega$--$a_0$ plane. The range of $U_r$ over which this line segment intersects with the approximately linear natural frequency curve then yields the lock-in range in Eqn. \ref{approxlockin}.    

It should be noted that the above theory is strictly valid only for high mass ratio ($m^*\gg 1$), where lock-in of structure with natural frequency ($\Omega=\Omega_n$) is required for high amplitude oscillations to occur \cite{barrero2012extracting}. For low mass ratios, the high amplitude oscillations may be attributed to lock-in of structure with vortex shedding; in this case the oscillation amplitude $a_0$ may not always be determined by the intersection of natural frequency curve with the EC. {We also note that the EC based theory presented here cannot account for lock-in of vortex shedding frequency with harmonics of the natural frequency of spring-mass system.}\par

\section{\label{objectives} Objectives of CFD simulations}  
In Badhurshah \emph{et al.} \cite{badhurshah2019lock}, we performed WOM simulations of VIV with bistable springs, with zero structure damping. We found that, in general, the range of lock-in is larger for VIV with bistable springs, compared to VIV with linear springs. For instance, for $\beta=0.05$, we found that large-amplitude double-well oscillations may occur over $U_r\in[2,15]$. WOM simulations of VIV with linear springs, on the other hand, showed a narrow lock-in range of $U_r\in[4,6.2]$. Regardless of the spring type, the maximum oscillation amplitude due to VIV was the same. Moreover, we found a strong collapse of $a_0$-vs-$\omega$ values onto a single curve, regardless of the spring type, which supported the existence of the EC. Thus, many of the critical features of the theory were at least qualitatively consistent with the results from the WOM simulation. In the present work, we will similarly examine the following in our CFD simulations of VIV with linear and bistable springs, with zero structure damping, for high mass ratio:
\begin{enumerate}
\item Variation of amplitude with reduced velocity for different spring types. 
\item Power Spectral Density of lift force as well as cylinder velocity  
\item Lock-in characteristics, and especially lock-in range
\item Vortex shedding patterns 
\item Consistency of the results with existence of an EC curve which is independent of spring type
\end{enumerate}  

\section{\label{nummet}Numerical methodology}
\subsection{\label{fludy} Immersed Boundary Solver}
The two-dimensional incompressible and unsteady Navier Stokes equations are evolved here for the fluid velocity field $\mathbf{u}(X,Y,Z,t)$, so that we solve for only $u_X,u_Y$ in the $X-Y$ plane, and assign $u_Z=0$. The cylinder itself is constrained to move only in the $Y$ direction. The computational domain is rectangular, with dimensions $L_x=40D$, $L_y=30D$, where $D$ is the dimensional cylinder diameter (Fig \ref{schem}(b)). The center of the cylinder is located at the plane $X=15D$. No-slip boundary condition is applied on the cylinder surface, Neumann (i.e. fully developed) boundary condition is applied at the outflow plane ($X=L_x$), while free slip conditions are applied at the lateral boundaries, $Y=\pm 15D$. The solver allows for fluid-structure interaction, and uses a sharp interface ghost cell based Immersed Boundary Method (IBM) \cite{mittal2008versatile}. The cylinder surface is discretized with equispaced points, while the fluid domain is discretized using a non-uniform Cartesian grid, with the finest grid cells near the cylinder (Figs \ref{mesh}(a),(b)) having same width in both $X$ and $Y$ directions $\Delta X_{min}=\Delta Y_{min}=D \cdot \Delta_{min}$.  
The number of elements in the fluid domain was chosen as $N_x=$385 and $N_y$=257 in $X$ and $Y$ directions respectively. The minimum grid size used here, non-dimensionalized with respect to $D$, is $\Delta_{min}=$0.015. A detailed grid convergence study has been presented in section \ref{gridconv} below to justify this grid resolution.

\begin{figure}[] 
\begin{center}
\includegraphics[width=0.7\textwidth, clip]{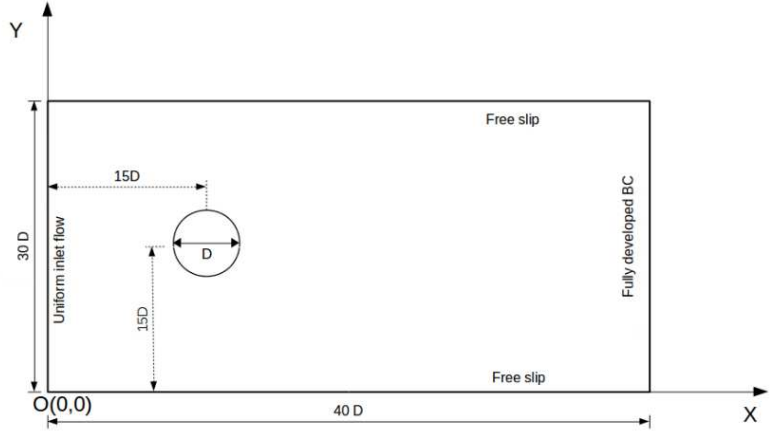} 
\end{center}
\caption{\label{flowdom} Schematic of computational domain along with the boundary conditions and cylinder position chosen for simulating the cylinder undergoing VIV.}
\end{figure}

%Mesh plots
\begin{figure}[H] 
\begin{center}
\begin{tabular}{cc}
\subfloat[]{\label{mesh}}{\includegraphics[scale=0.2]{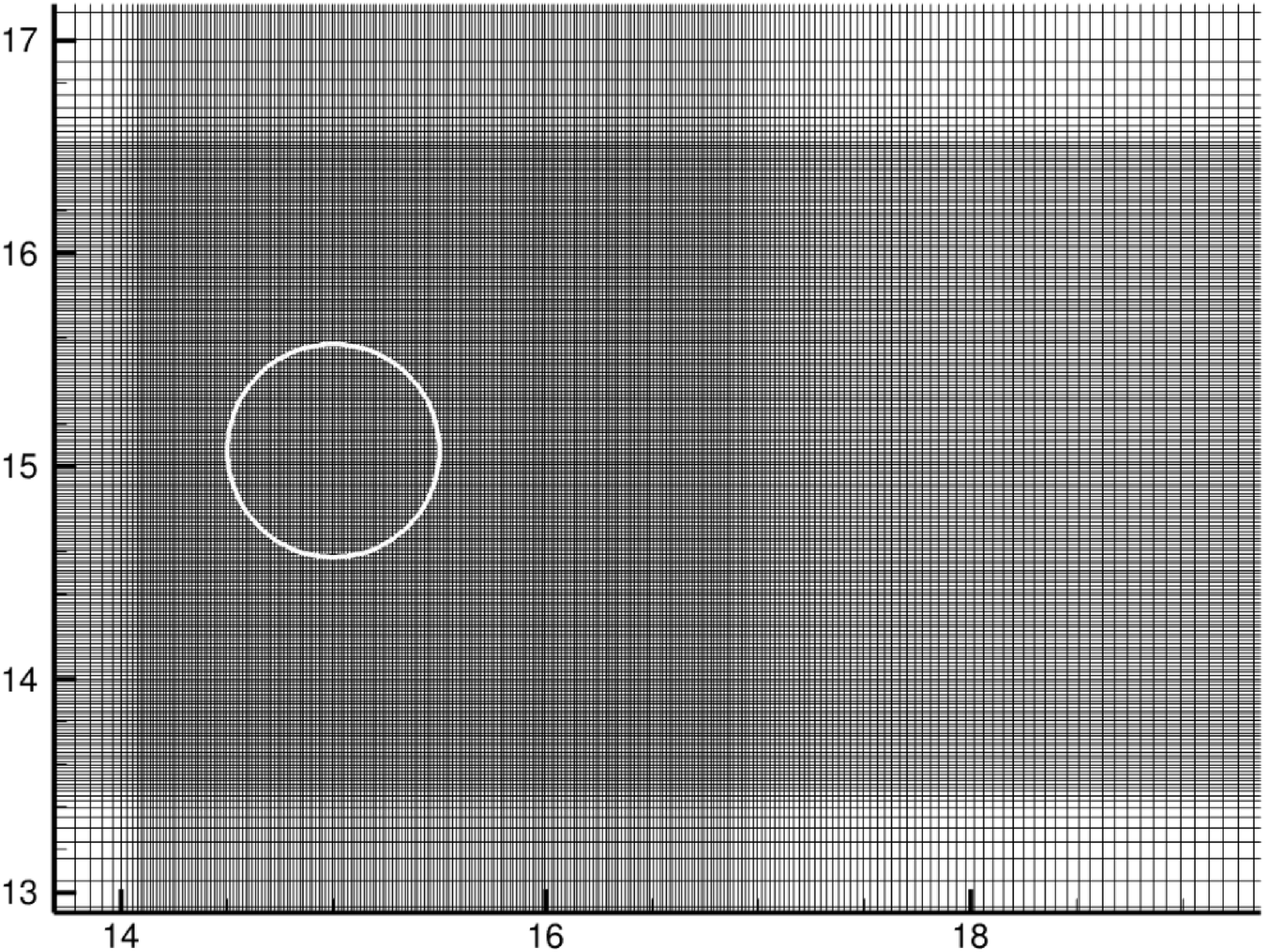}} &  
\subfloat[]{\label{meshzoom}\includegraphics[scale=0.2]{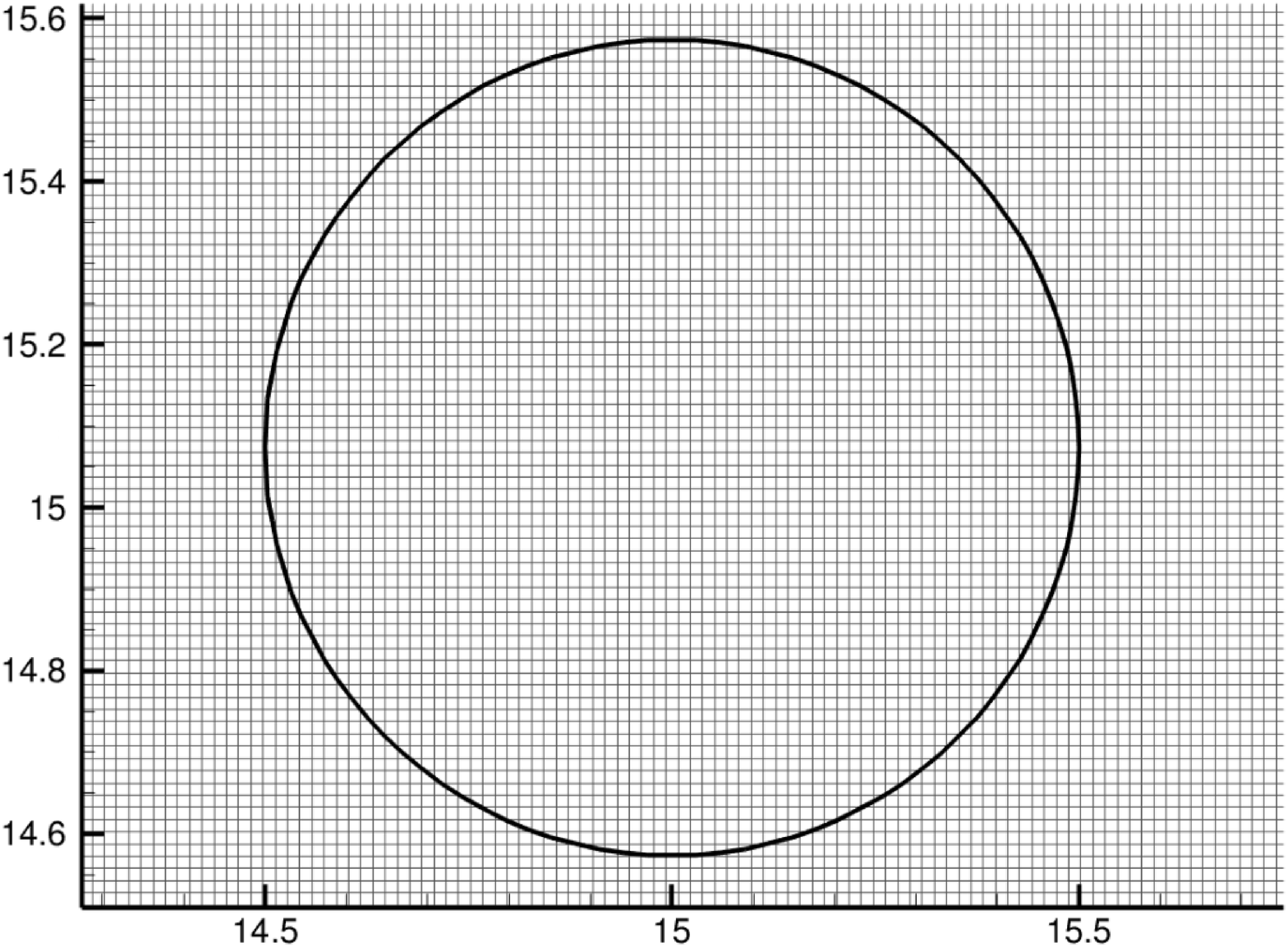}} \\

\end{tabular}
\end{center}
\caption{\label{mesh} Figure showing the mesh near the cylinder for the fluid domain with the details chosen from Mesh 4. Figure \ref{meshzoom} zooms the mesh near the cylinder. {The cylinder surface was discretized with 104 equispaced points.}  }
\end{figure}

In the CFD simulations, we solve for a non-dimensional form of Eqn. \ref{struc} in which the variables are non-dimensionalized with respect to $\rho$, $D$, and $U$:     
  \begin{equation}
  \label{yaccibm}
  \frac{d y^2}{d\tau^2}+ 2  \xi \omega_s \frac{d y}{d \tau} + \frac{d \varphi_s^*}{dy} = \frac {2}{\pi} \frac{F^*_Y}{m^*} 
  \end{equation}
where 
\bea
\frac{d \varphi_s^*}{dy}&=&\left\{
\begin{array}{ccc}
\omega_s^2 y \quad&\textrm{for}\quad &\textrm{Linear Springs}\\
 - \frac{\omega_s^2}{2} y + \frac{\omega_s^2}{2 \beta^2 } y^3\quad &\textrm{for}\quad &\textrm{Bistable Springs}
\end{array}
\right.
\eea
$\tau=TU/D$, $\omega_s=\Omega_s D/U$, $F^*_Y = \frac{F_Y}{\frac{1}{2}\rho U^2 D}$, are the non-dimensional time, linearized natural frequency of spring, and lift force from fluid, while $m^* = \frac{m_s}{\frac{\pi}{4}\rho D^2}$ is the mass ratio based on the cylinder mass only.
We implement Velocity Verlet time stepping scheme to integrate the equation of motion, since it conserves the potential energy of the spring. This method evolves $y$ and $\dot{y}=dy/d\tau$ in tandem with the equation for acceleration (Eqn. \ref{yaccibm}) from time step $\tau_n$ to $\tau_{n+1}$ as follows:
\begin{eqnarray}
y(\tau_{n+1}) &=& y(\tau_n) + \dot{y}(\tau_n) \Delta{\tau} + \frac{1}{2}\ddot{y}(\tau_n) \Delta{\tau}^2\\
\dot{y}(\tau_{n+1}) &=& \dot{y}(\tau_n) + \frac{1}{2}\bigg (\ddot{y}(\tau_n)+\ddot{y}(\tau_{n+1})\bigg) \Delta{\tau}
\end{eqnarray}
where $\Delta \tau=\tau_{n+1}-\tau_{n}$ is a fixed time step, satisfying CFL criterion for the fluid.

\subsection{\label{valid} Validation of IBM based solver}
\begin{figure}[H] 
\begin{center}
\begin{tabular}{c}
{\includegraphics[scale = 0.9]{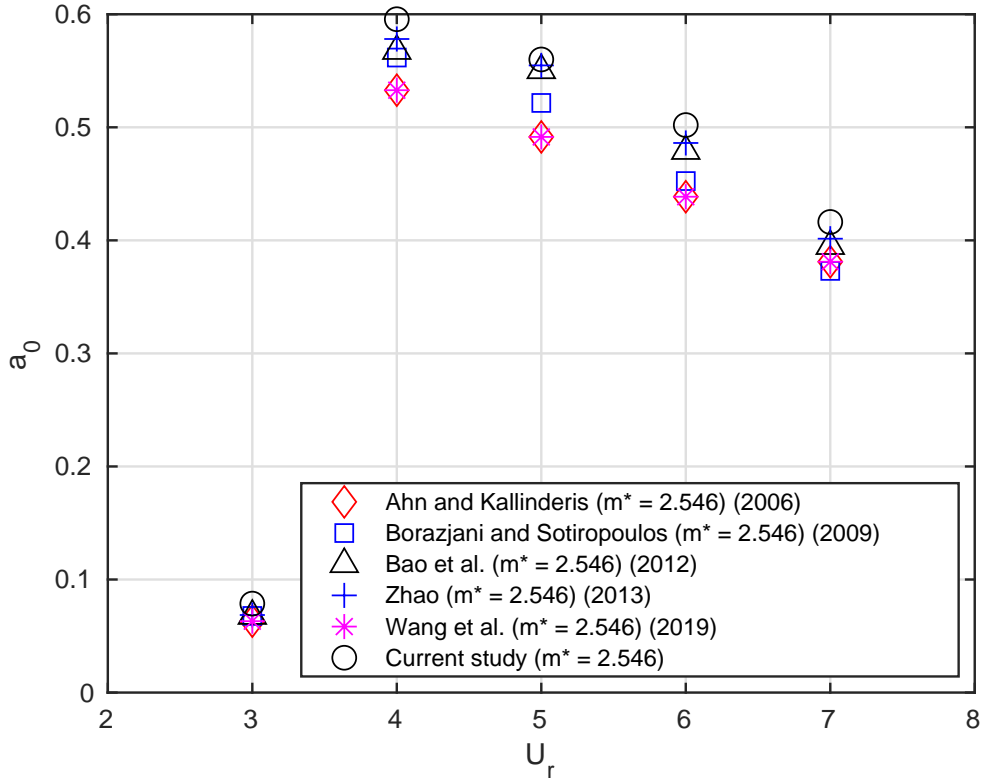}}   
\end{tabular}
\end{center}
\caption{\label{valplot} Validation and verification of maximum amplitude versus reduced velocity for an undamped cylinder attached to linear spring undergoing VIV at $m^* = 2.546$ and $Re = 150$ \cite{ahn2006strongly, borazjani2009numerical, bao2012two,zhao2013flow, wang2019effect}. }
\end{figure}

\begin{figure}[H] 
\begin{center}
\begin{tabular}{c}
{\includegraphics[scale = 0.7]{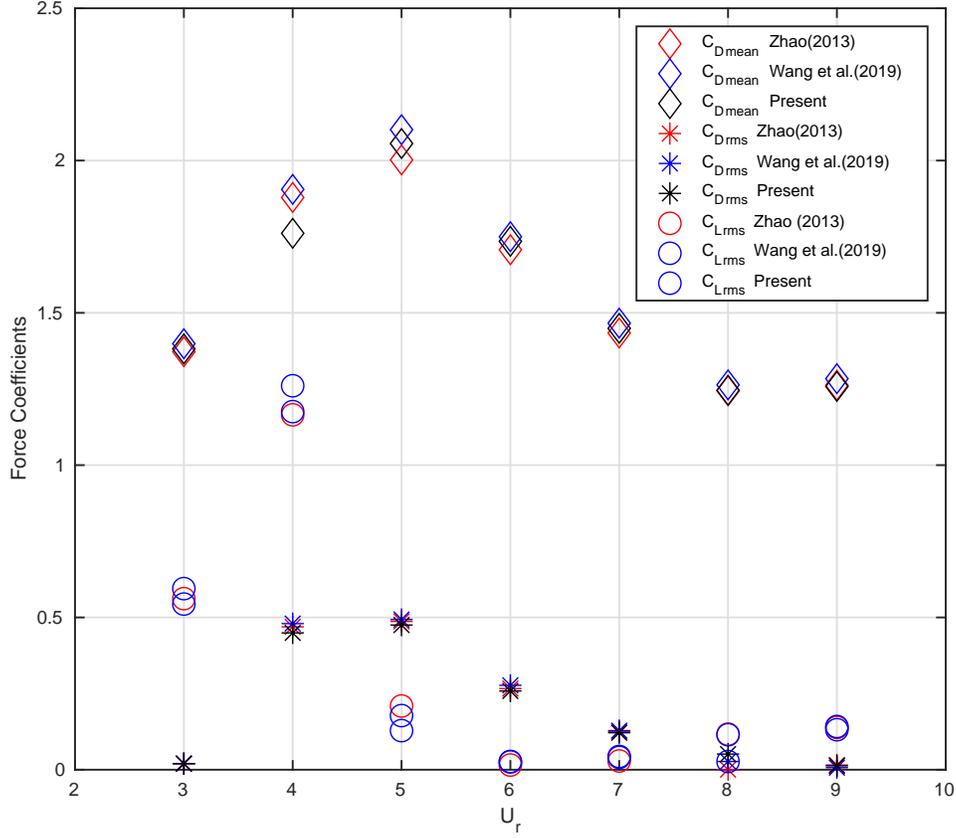}}   
\end{tabular}
\end{center}
\caption{\label{valfcoeff} Validation of mean and RMS force coefficients versus reduced velocity for an undamped cylinder attached to linear spring undergoing VIV at $m^* = 2.546$ and $Re = 150$. The current simulations have been validated with existing literature \cite{zhao2013flow, wang2019effect}. }
\end{figure}

The IBM based solver is validated and verified against benchmark results available for cylinders attached to linear springs undergoing cross-flow VIVs. The mass ratio here is set as $m^* = 2.546$, as chosen from the literature, and the structural damping is set to zero. The Strouhal number for vortex shedding corresponding to the case with fixed circular cylinder with $\textrm{Re}_D=150$ is $\St=0.183$. Fig \ref{valplot} shows a comparison of the cylinder displacement amplitude ($a_{0}=A_{0}/D$) from our simulations against prior literature, in which we have included studies for $m^*=2$ as well \cite{ahn2006strongly, borazjani2009numerical, bao2012two, wang2019effect}, over a range of reduced velocity $U_r=2\pi U/(\Omega_s D)$. We find a reasonable (within 10\%) agreement of $a_{0}$ with prior literature, and the values of $a_{0}$ are closest to that of Bao \emph{et. al} \citep{bao2012two}, who have used a mass ratio of $m^*=2.546$ for their simulations. We also compare the data for mean drag force ($C_{Dmean}$), RMS of lift coefficient ($C_{Lrms}$) and RMS of drag coefficient ($C_{Drms}$) from our simulations with results from Wang \emph{et al.} \cite{wang2019effect} in Fig \ref{valfcoeff}. We again find reasonable agreement between the data sets for force coefficients.

\subsection{\label{parsim} Parameters chosen for CFD simulation}
For CFD simulations in this paper, the mass ratio is fixed as, $m^* = 25.46$ (or $m/(\rho D^2)=20$) with zero structure damping $\xi=0$, as per the discussion in section \ref{locinb}. We perform simulations for VIV with linear springs as well as bistable springs. For VIV with bistable spring, we consider two values of $\beta$, namely, $\beta = 0.1$ and $\beta = 0.2$. The list of reduced velocities for which the simulations have been performed, for each spring type, have been listed in Table \ref{simpar}. For a given spring type, we change $U_r$ by fixing the inlet velocity $U$ and varying the linearized natural frequency $\Omega_s$. Therefore the Reynolds number is fixed at $\textrm{Re}_D=UD/\nu=150$ for all our simulations.

\subsection{\label{initsim} Initial Conditions}
For VIV with linear springs, the oscillation amplitude is known to be insensitive to initial displacement of the cylinder. On the other hand, for VIV with nonlinear springs, it is known that the oscillation amplitude can depend strongly on initial displacement \cite{badhurshah2019lock, Mackowski}. Therefore, we start all the simulations involving VIV with bistable spring from a larger initial displacement $y(0)\approx 0.6$. To provide such an initial condition, we also have to ensure that the initial velocity field is consistent with the high displacement. We found that simply initializing with a large cylinder displacement and uniform velocity field can lead to damping of the cylinder oscillations before the vortex shedding cycle stabilizes. We therefore first carry out a ``spin-up" simulation of VIV with linear spring for $U_r=5$, until periodic oscillations are reached. We then switch the spring potential from linear to bistable function, with the appropriate $\beta$ value and $U_r$, when the cylinder reaches its peak displacement. We will denote $t=0$ as moment when the spring potential is switched. Note that $y_{cr}=\sqrt{2} \beta$ for bistable springs, so that $y(0)>y_{cr}$ is satisfied for all the simulations involving VIV with bistable springs. The initial condition thus allows for double well oscillations for VIV with bistable springs.  

\begin{table}
\begin{center}
\begin{tabular}{|c|c|}
\hline
Spring Type & List of reduced velocities ($U_r$)\\
\hline
Linear & 2, 3, 4, \textcolor{red}{\textbf{4.75, 5, 5.25, 5.75, 6, 7}}, 8, 9, 10, 12, 15, 18, 21, 24\\
\hline
Bistable ($\beta =0.1$) &1, \textcolor{red}{\textbf{2, 3, 4, 5, 6, 7, 8, 9, 10, 12, 15, 18}}, 21, 24 \\
\hline
Bistable ($\beta=0.2$) & 1, \textcolor{red}{\textbf{2, 3, 4, 5, 6, 7, 8, 9}}, 10, 12, 15, 18, 21, 24\\
\hline 
\end{tabular}
\end{center}
\caption{\label{simpar} List of reduced velocities for which simulations have been conducted, for different spring types. Structural damping $\xi=0$ for all the simulations, and mass ratio is $m^*=25.46$. The $U_r$ values with red boldfaced font denote data points corresponding to large displacement amplitudes (i.e. $a_0>0.05$). For VIV bistable springs, large amplitude oscillations correspond to only double well oscillations. }
\end{table}

\subsection{\label{gridconv} Grid convergence test}

\begin{figure}[H] 
\begin{center}
\begin{tabular}{cc}
\subfloat[]{\label{bicomp}} {\includegraphics[scale=0.5]{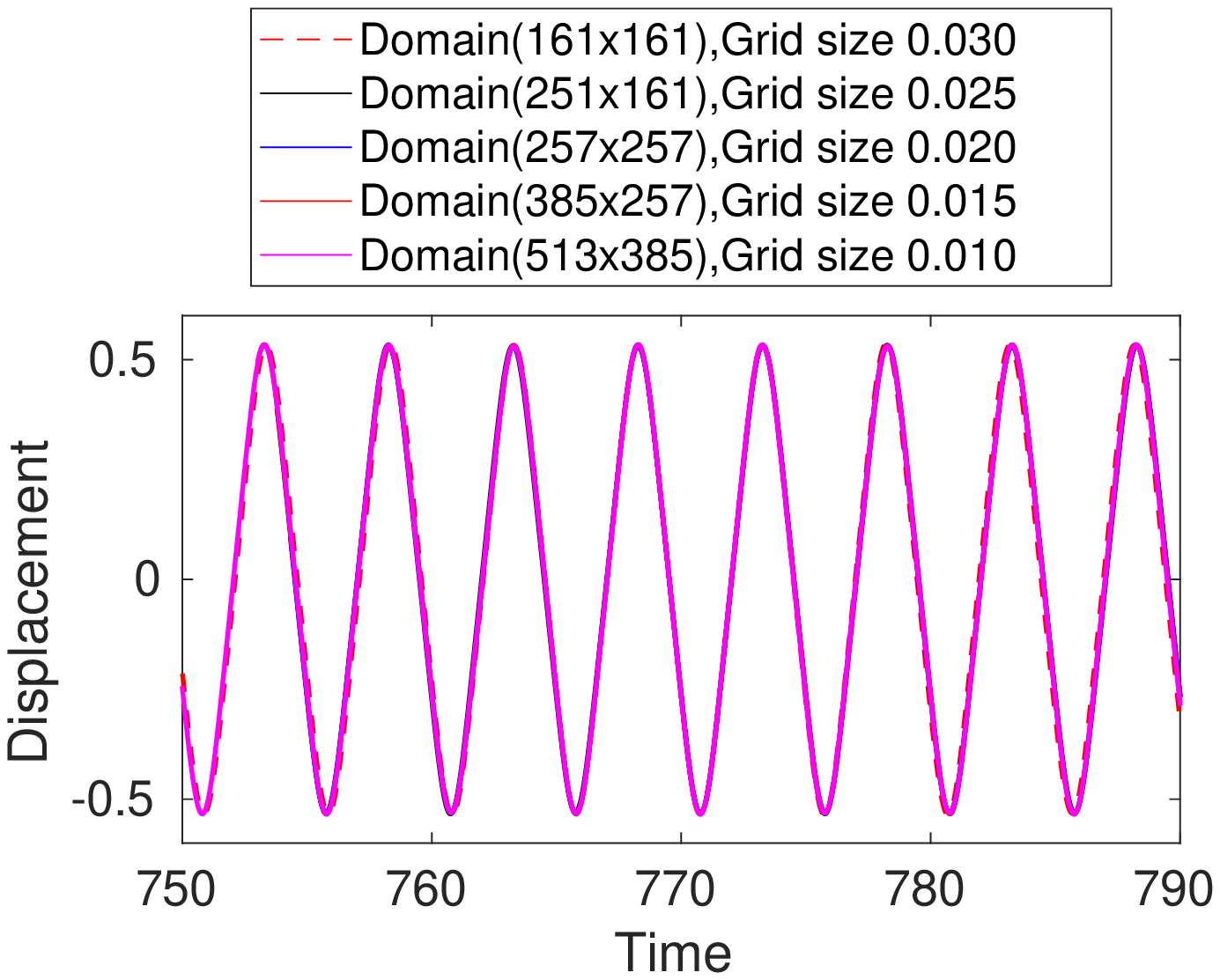}} &  
\subfloat[]{\label{bicompzoom}\includegraphics[scale=0.5]{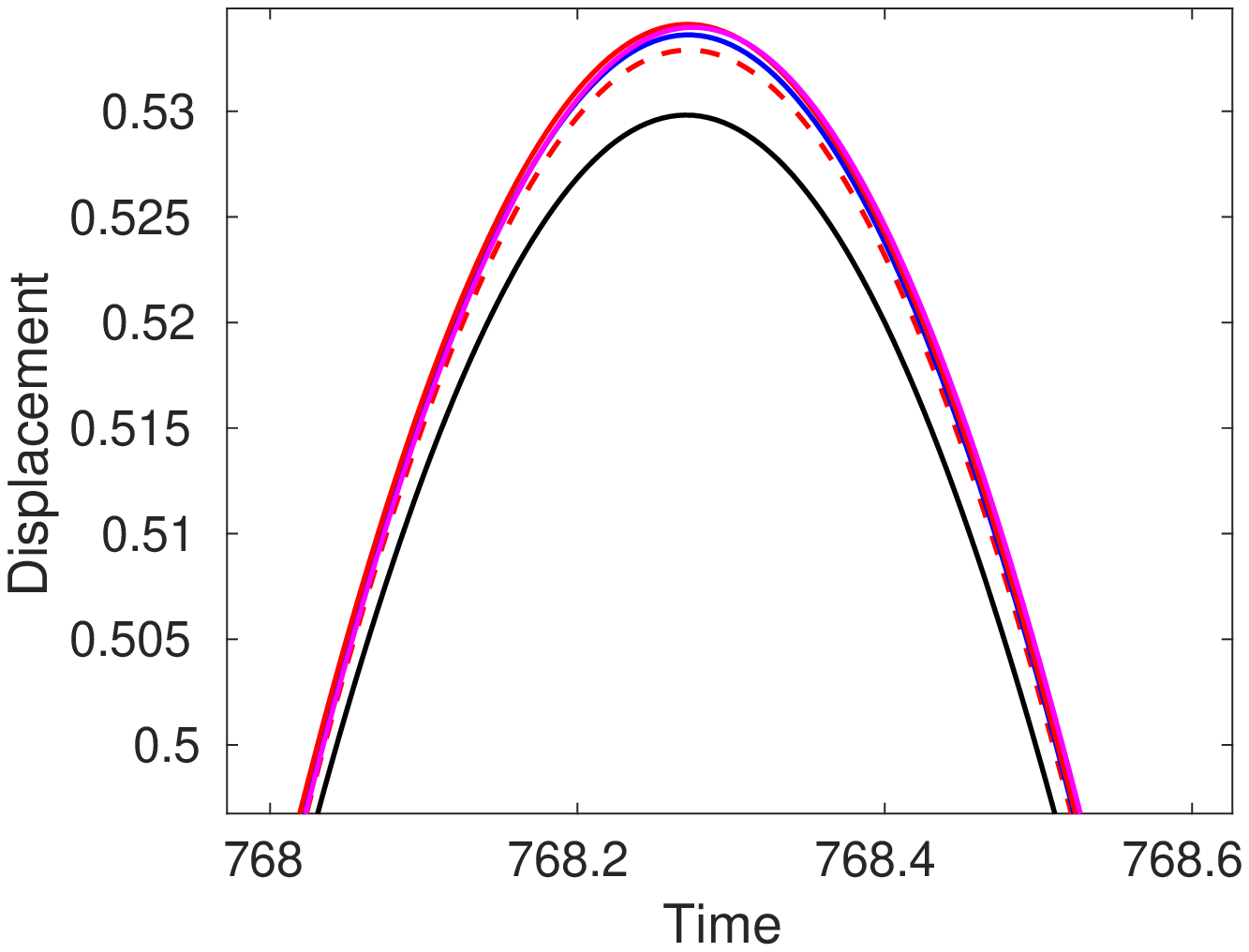}} \\
\end{tabular}
\end{center}
\caption{\label{gdtbi} Displacement versus time plotted for bistable spring ($\beta=0.1$) at $U_r = 15 $ and mass ratio of $m^*=25.64$ for different meshes. The domain size was fixed as $40\times 30$ for all the three cases shown. Fig. \ref{bicompzoom} zooms the graph near one of the peaks.}
\end{figure}

\begin{figure}[H]
\begin{center}
\includegraphics[width=\textwidth]{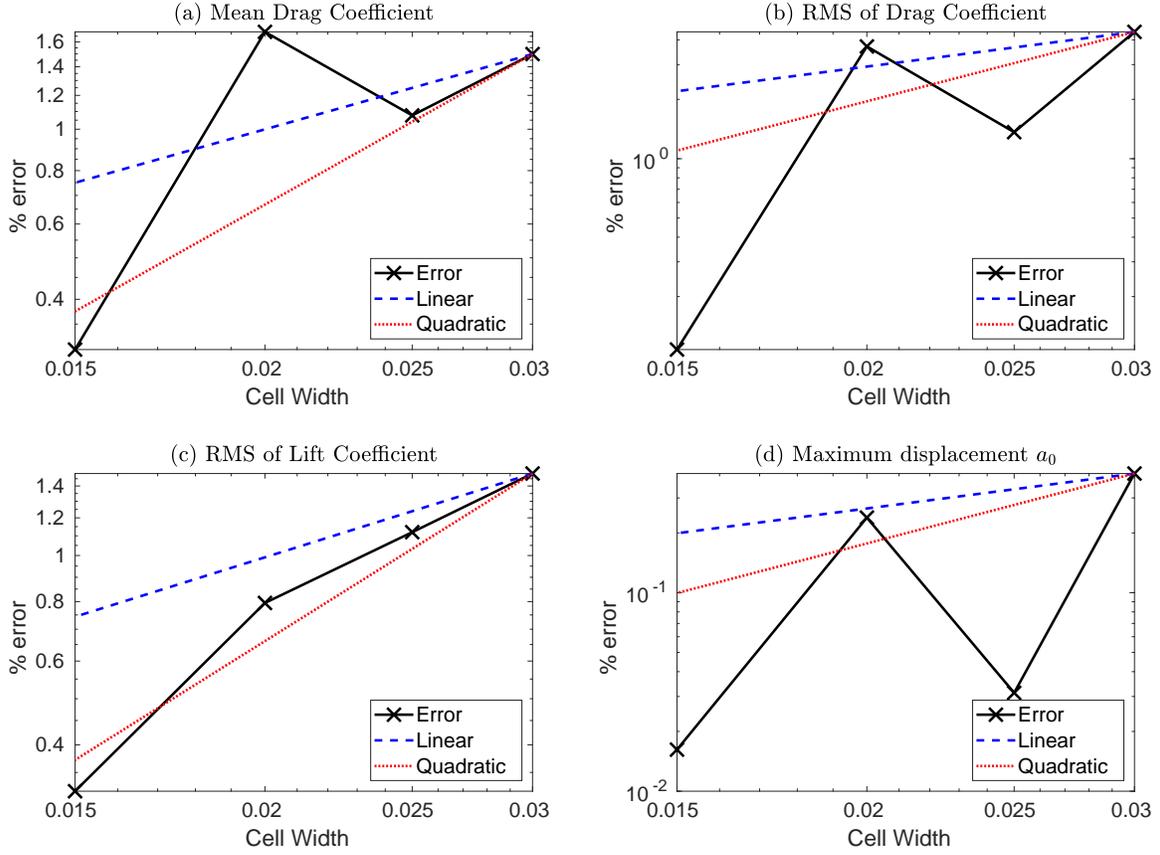}
\caption{\label{biconvfig} Graphs showing grid convergence trends for (a) $C_{Dmean}$, (b) $C_{Drms}$, (c) $C_{Lrms}$ and (d) $a_0$ with respect to cell width $\Delta_{min}$, for VIV with bistable spring ($\beta=0.1$). Here $\%$ error for each quantity is taken with respect to values for Mesh 5 ($\Delta_{min}=0.01$). Reference lines show linear ($--$) and quadratic ($\cdots$) convergence of error. }
\end{center}
\end{figure}

We perform a grid convergence test to select the optimal computational mesh for the numerical simulations. We choose to conduct the test for VIV with bistable spring ($\beta=0.1$) and $U_r=15$, since this relatively severe case involves high-amplitude non-harmonic oscillations of the cylinder. The size of the finest grid cells near the cylinder ($\Delta_{min}$) is varied over a large range ($\Delta_{min}\in[0.01, 0.03]$) for the test. The different mesh sizes used for the test, along with the drag coefficient, lift coefficient, and $a_0$ measured from the respective simulations, have been listed in Table \ref{tabbiforce}. Fig \ref{gdtbi} shows the time series plot of $y(t)$ for the different mesh sizes. Both Table \ref{tabbiforce} and Fig \ref{gdtbi} indicate reasonably good grid convergence for the drag and lift coefficients, $a_0$ as well as displacement time series of the the cylinder. Graphs of $\%$ error with respect to finest grid considered and $\Delta_{min}$ in Fig \ref{biconvfig} however demonstrates that the grid convergence for these quantities is not monotonic. This non-monotonic trend in error could be due to the fact that the grid cells in the wake region behind the cylinder do not preserve aspect ratio $\Delta x/\Delta y$ at any given location for different mesh types, due to the nature of multi-block stretched mesh being used. Nevertheless, the overall order of accuracy is always more than 1 for all the force coefficients and $a_0$. Since the error in all quantities between Mesh 4 and Mesh 5 is smaller than $1\%$, we therefore choose Mesh 4  ($N_x=385,\,N_y=257,\,\Delta_{min}=0.015$) for the rest of the simulations in this paper. The size of the coarsest grid cell for Mesh 4, near the outer boundary, is $\Delta_{max}=0.295$. 

\begin{table} [H]
\begin{center}
\begin{tabular}{|c|c|c|c|c|c|c|c|}
\hline
Mesh & $N_x$ & $N_y$ & $\Delta_{min}$ & ${C_D}_{mean}$& ${C_D}_{rms}$ & ${C_L}_{rms}$ & $a_0$ 
\\
\hline
1 & 161 & 161 & 3.0E-02 & 1.6695 & 0.1804 & 1.6173 & 0.5360 \\
\hline
2 & 257 & 161 & 2.5E-02 & 1.6767 & 0.1861 & 1.6233 & 0.5339 \\
\hline
3 & 257 & 257 &2.0E-02 & 1.6663 & 0.1817 & 1.6286 & 0.5336 \\
\hline
4 & 385 & 257 & 1.5E-02 & 1.6897 & 0.1885 & 1.6469 & 0.5345 \\
\hline
5 & 513 & 385 & 1.0E-02 & 1.6949 & 0.1887 & 1.6417&  0.5344 \\
\hline
\end{tabular}
\end{center}
\caption{\label{tabbiforce}  Drag and lift coefficients and maximum displacement computed for VIV with bistable spring ($\beta=0.1$), for different mesh sizes $N_x\times N_y$ used to discretize the fluid domain. }
\end{table}

\section{\label{result}Results and Discussion}

This section presents numerical results based on the CFD simulations of VIV with linear and bistable springs.  We report the numerical data once the cylinder reaches a periodic oscillatory state. The statistics here is based on data collected over time interval of more than $200$ nondimensional time units in the periodic state.  The nondimensional timestep used here is $\Delta t=0.005$. \par

\subsection{\label{timeseries} Time series of displacement, and phase portraits}

\begin{figure}[] 
\begin{center}
\begin{tabular}{ccc}
\subfloat[Linear  ($U_r=3$)]{\label{lindisp3}\includegraphics[width=0.3\textwidth]{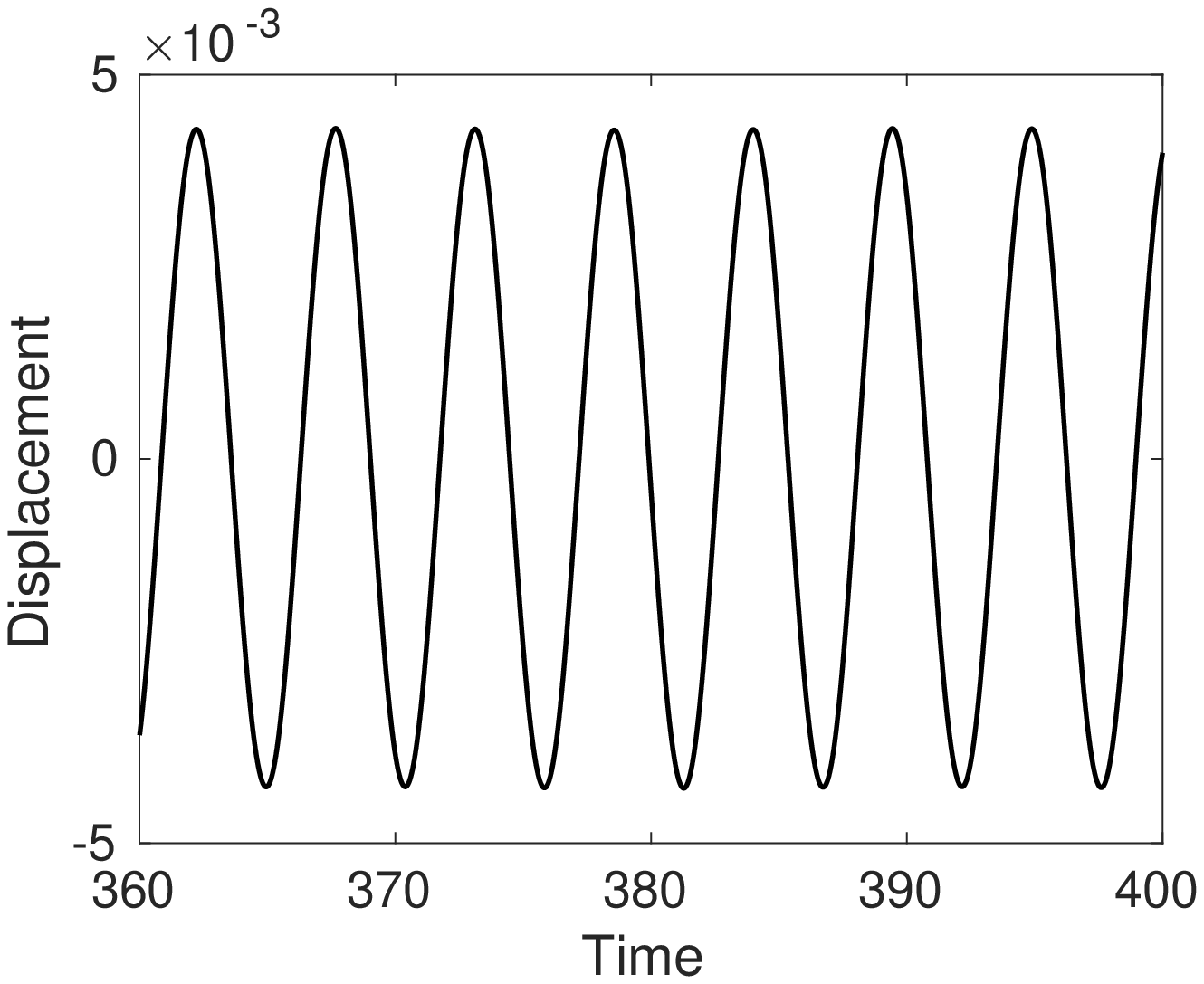}} &
\subfloat[Linear ($U_r=5$)]{\label{lindisp5}\includegraphics[width=0.3\textwidth]{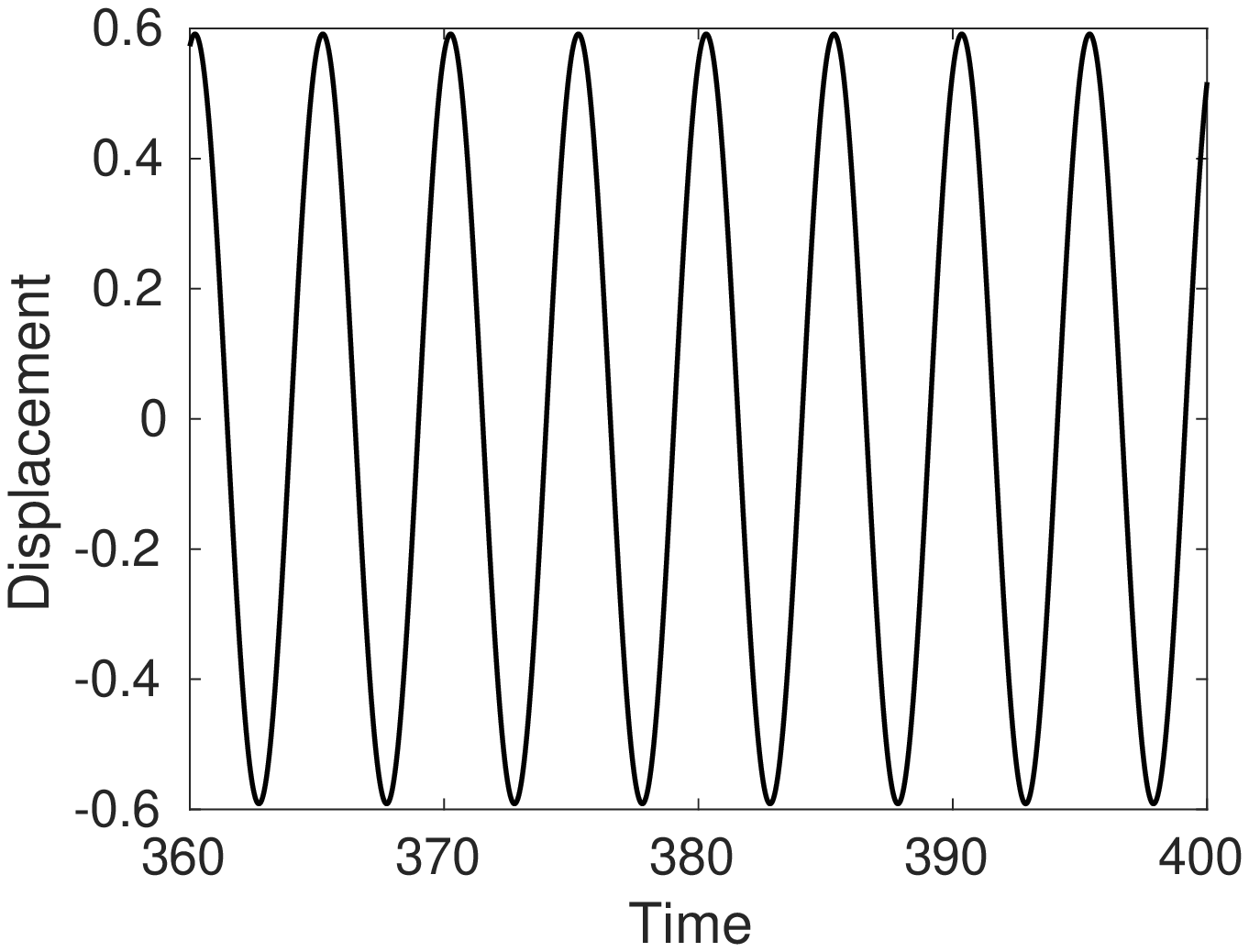}} &
\subfloat[Linear  ($U_r=9$)]{\label{lindisp7}\includegraphics[width=0.3\textwidth]{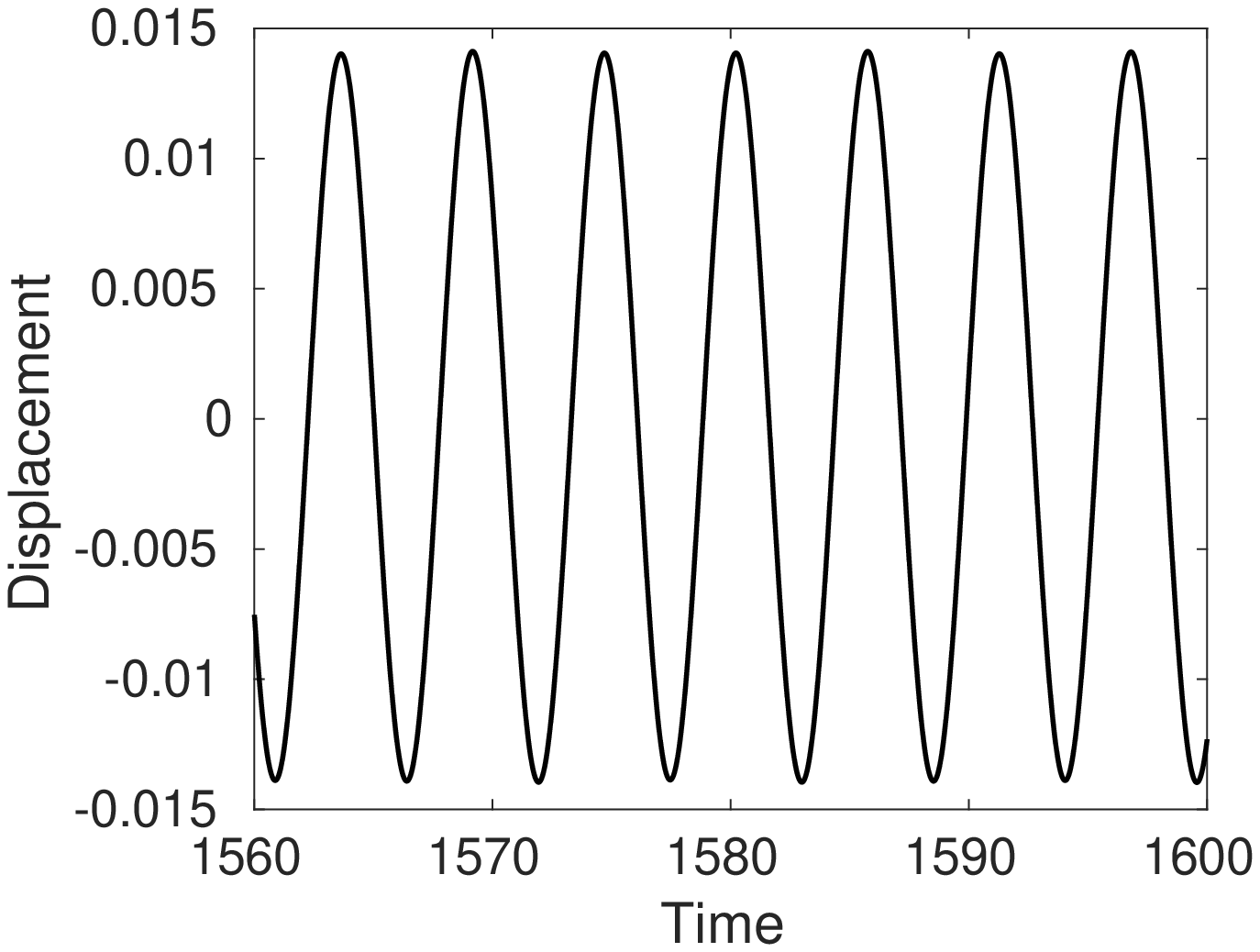}} 
\\
\subfloat[Bistable, $\beta=0.1$ ($U_r=3$)]{\label{b1disp3}\includegraphics[width=0.3\textwidth]{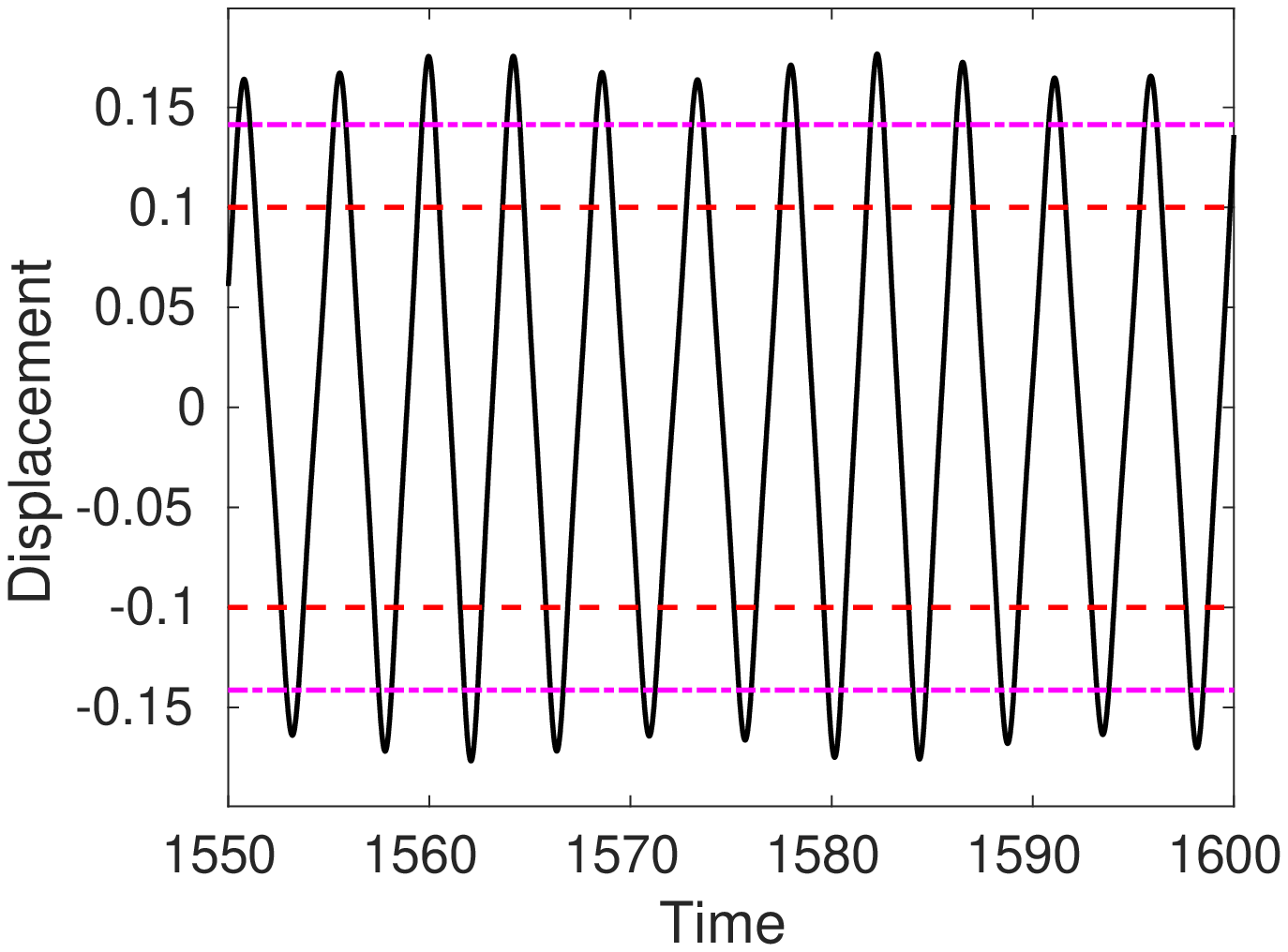}} &
\subfloat[Bistable , $\beta=0.1$ ($U_r=18$)]{\label{b1disp18}\includegraphics[width=0.3\textwidth]{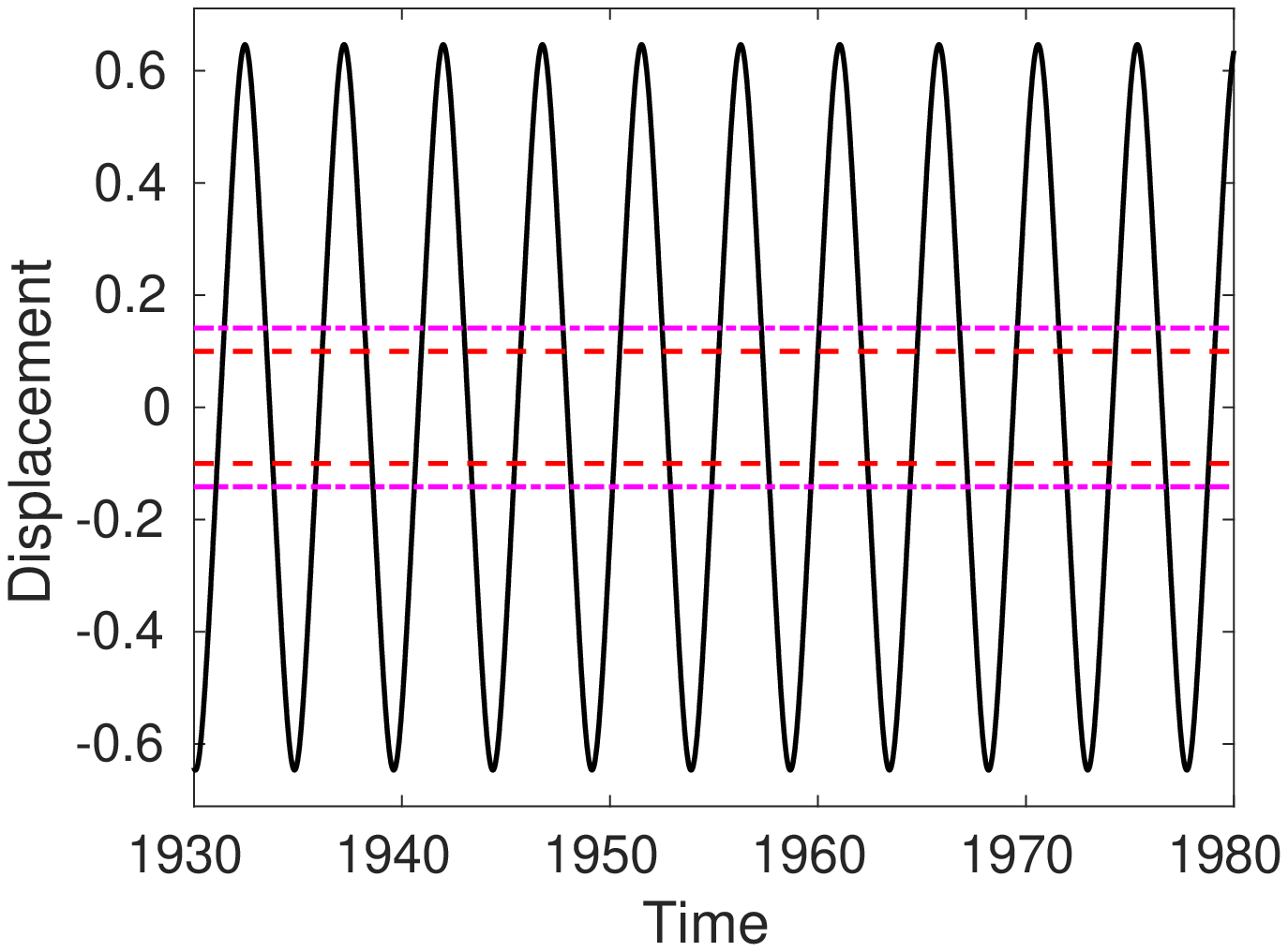}} &
\subfloat[Bistable , $\beta=0.1$ ($U_r=21$)]{\label{b1disp21}\includegraphics[width=0.3\textwidth]{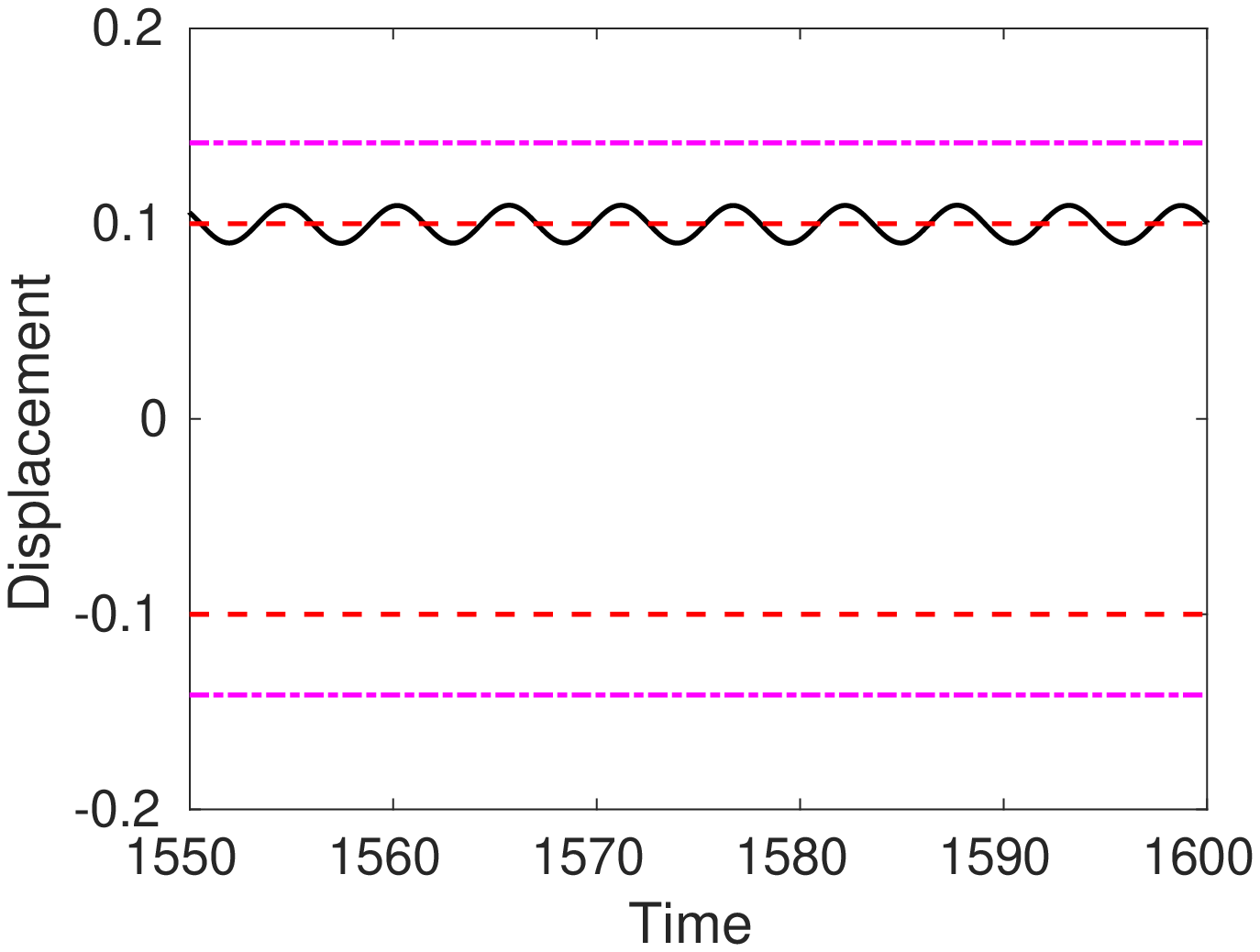}} 
\\
\subfloat[Bistable, $\beta=0.2$ ($U_r=3$)]{\label{b2disp3}\includegraphics[width=0.3\textwidth]{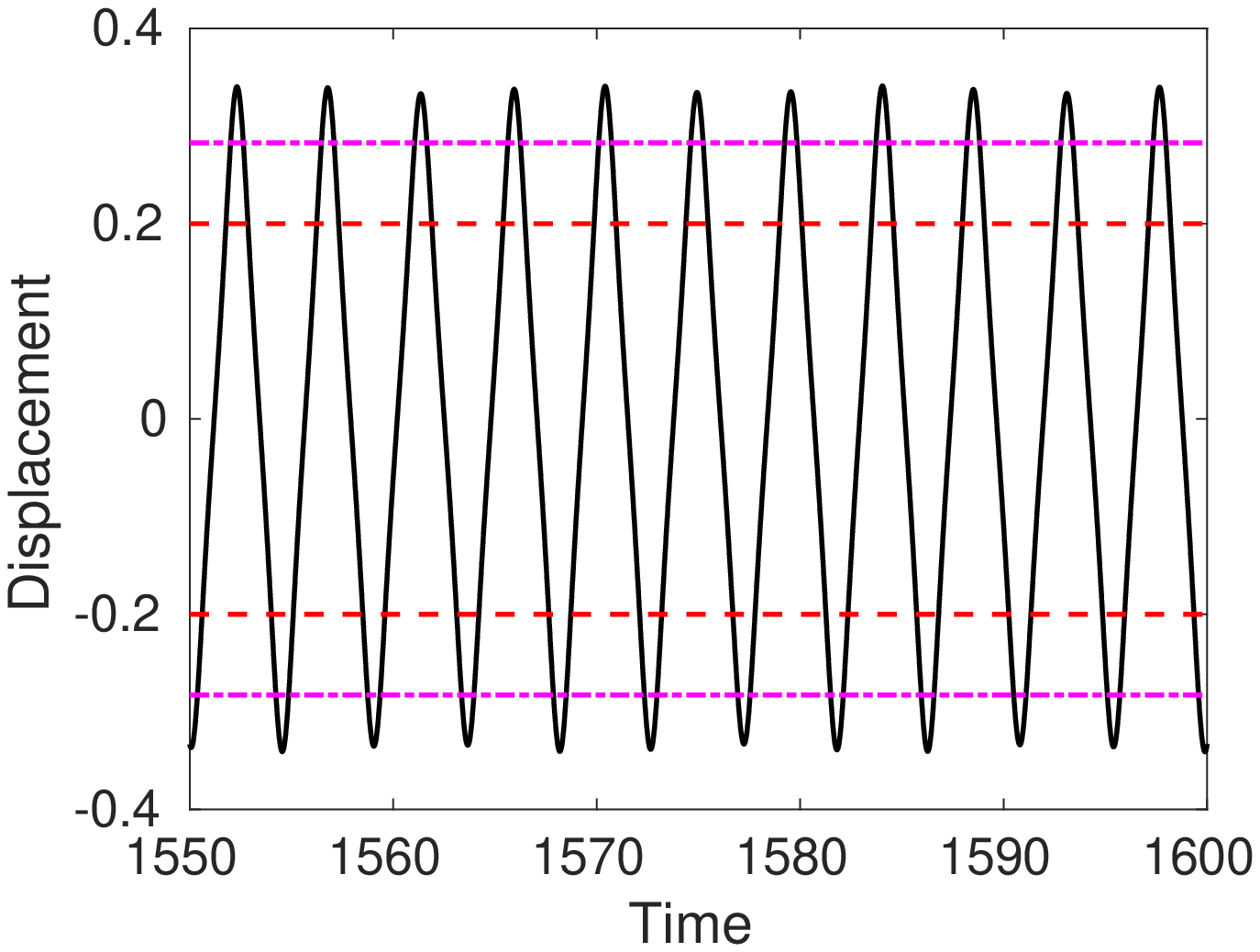}} &
\subfloat[Bistable , $\beta=0.2$ ($U_r=9$)]{\label{b2disp9}\includegraphics[width=0.3\textwidth]{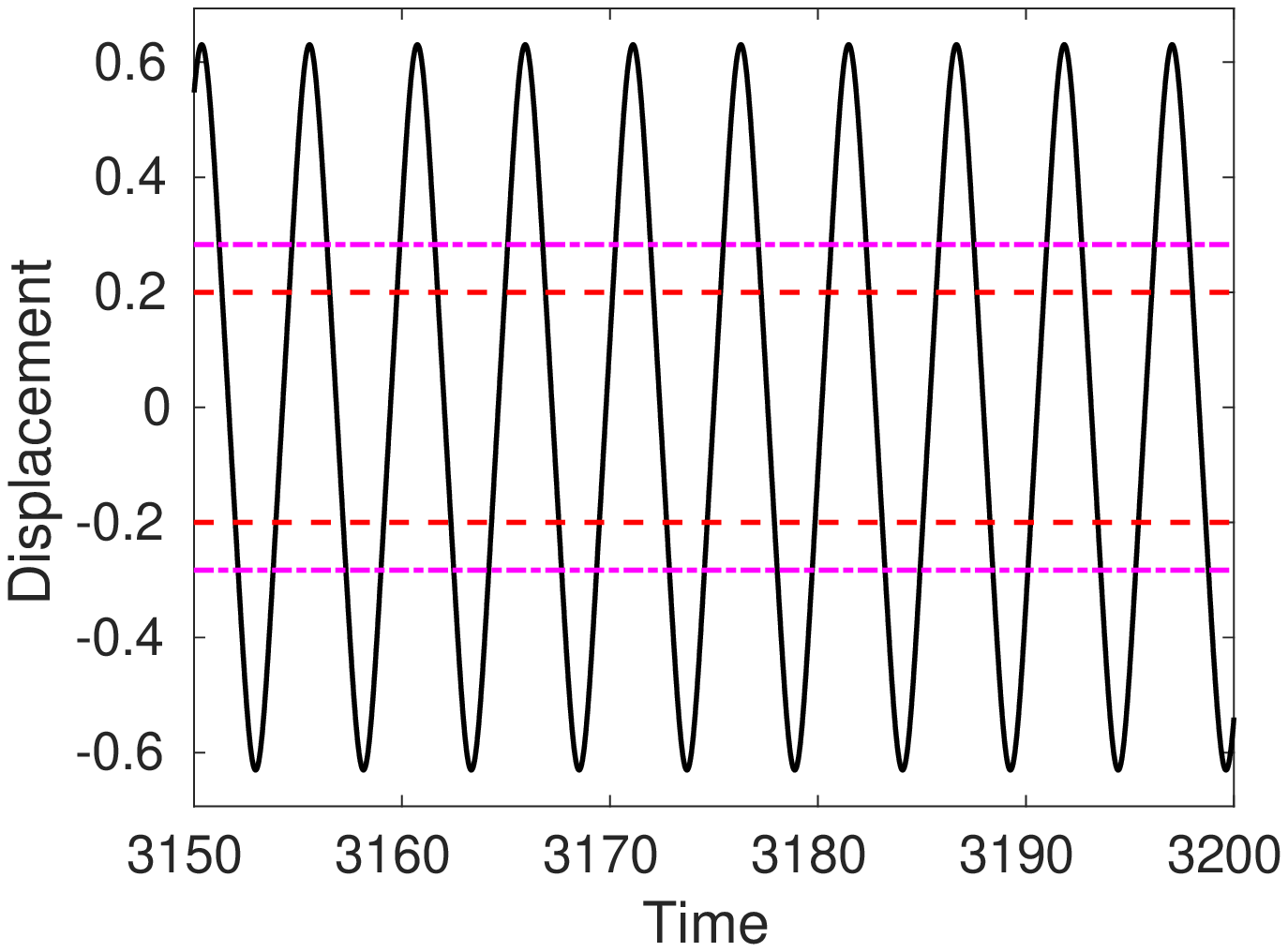}} &
\subfloat[Bistable , $\beta=0.2$ ($U_r=12$)]{\label{b2disp12}\includegraphics[width=0.3\textwidth]{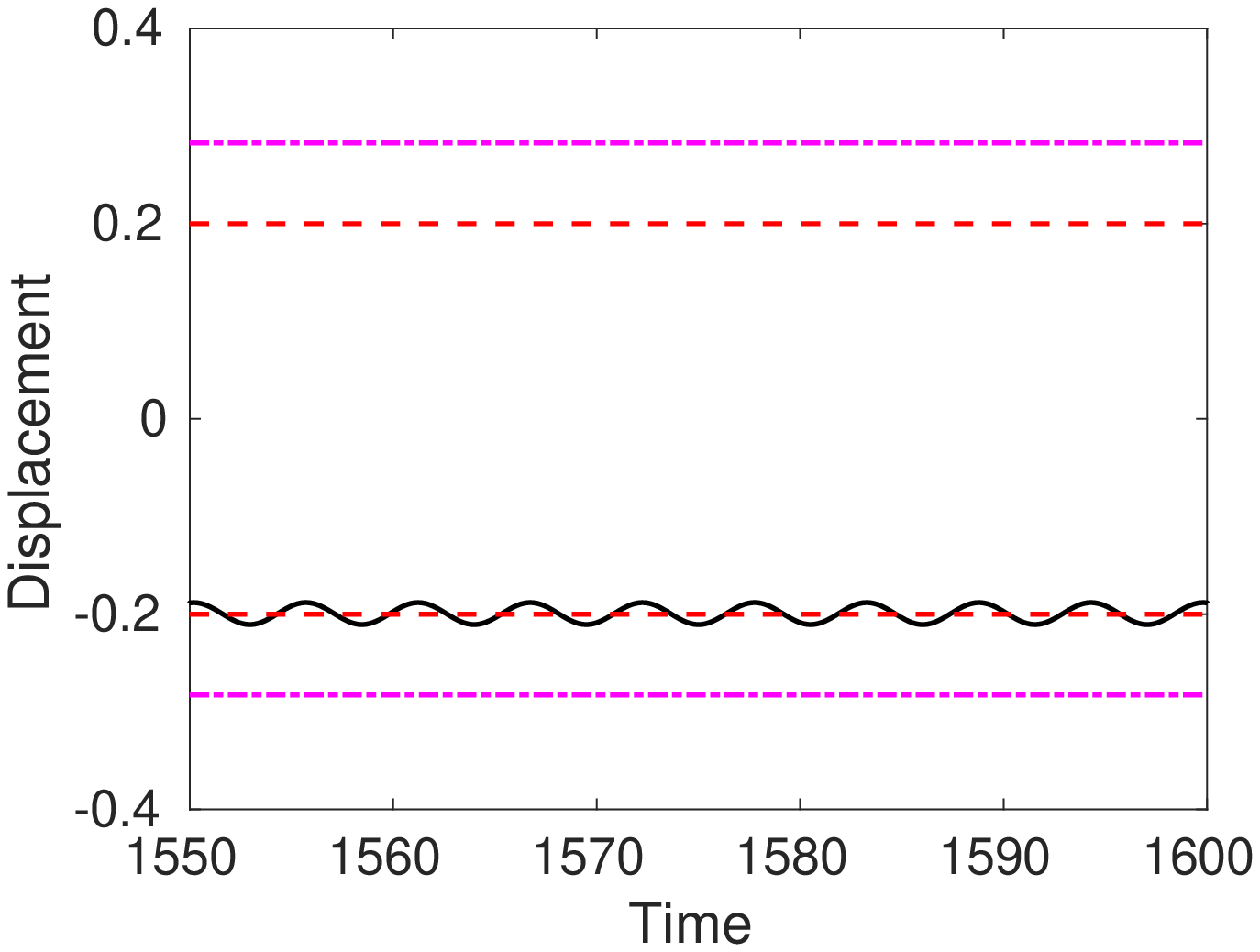}} 
\\
\end{tabular}
\end{center}
\caption{\label{tseries} Representative displacement time series $y(\tau)$ for linear and bistable springs. {The dashed red lines represent the position of the stable point ($y=\beta$) while the magenta lines represent non-dimensional critical displacement $y=y_{cr}$, given by the non-dimensional form of Eq. \ref{crdisp}.}} 
\end{figure}

In Fig. \ref{tseries}, we have plotted the time series of cylinder displacement, $y(t)$, after a periodic state is reached, for VIV with different springs and reduced velocities. As expected, VIV with linear springs exhibits the smallest range of $U_r$ over which large amplitude oscillations occur (Figs \ref{lindisp3},\ref{lindisp5},\ref{lindisp7}). For VIV with bistable spring, for lower inter-well separations ($\beta=0.1$), we can see large amplitude oscillations for both $U_r=3$ and $U_r=18$ (Figs. \ref{b1disp3}, \ref{b1disp18}), indicating a large range of $U_r$ over which high amplitude oscillations can occur. We also observe low frequency beating for the $U_r=3$ case (Fig. \ref{b1disp3}), perhaps due to the fact that the amplitude of oscillation is quite close to $y_{cr}=\sqrt{2}\beta=0.14$ here. For $U_r=21$, we observe low amplitude single-well oscillation (Fig \ref{b1disp21}). Thus, there is a somewhat abrupt reduction in displacement amplitude between $U_r=18$ and $U_r=21$ . For higher inter-well separations ($\beta=0.2$), we notice trends which are similar to the cases with low inter-well separation ($\beta=0.1$), except that the jump from high to low amplitude oscillations appears to occur at lower reduced velocity, between $U_r=9$ and $U_r=12$.  

\begin{figure}[] 
\begin{center}
\begin{tabular}{ccc}
\subfloat[Linear  ($U_r=3$)]{\label{linpp5}\includegraphics[width=0.3\textwidth]{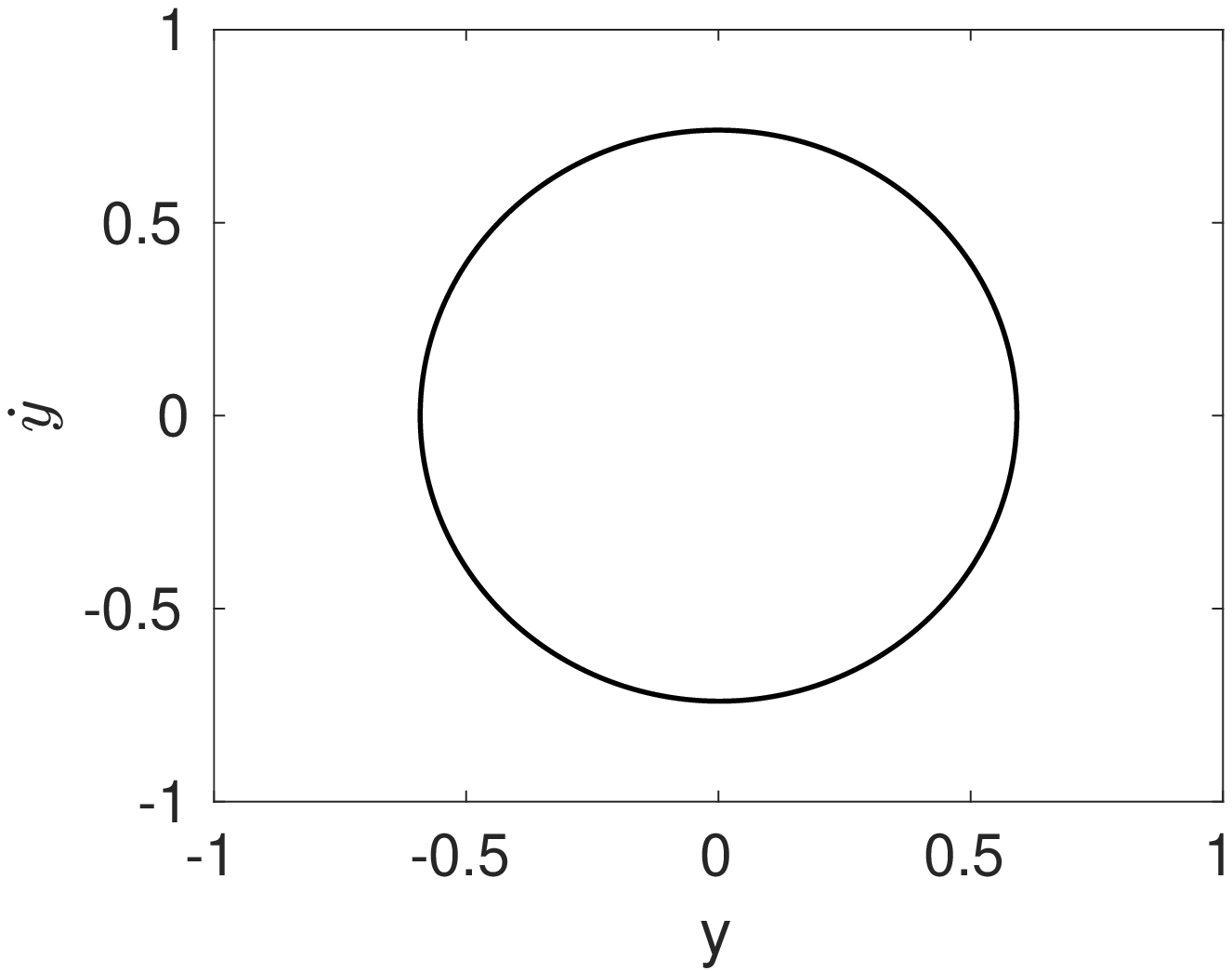}} &
\subfloat[Bistable, $\beta=0.1$ ($U_r=3$)]{\label{b1pp3}\includegraphics[width=0.3\textwidth]{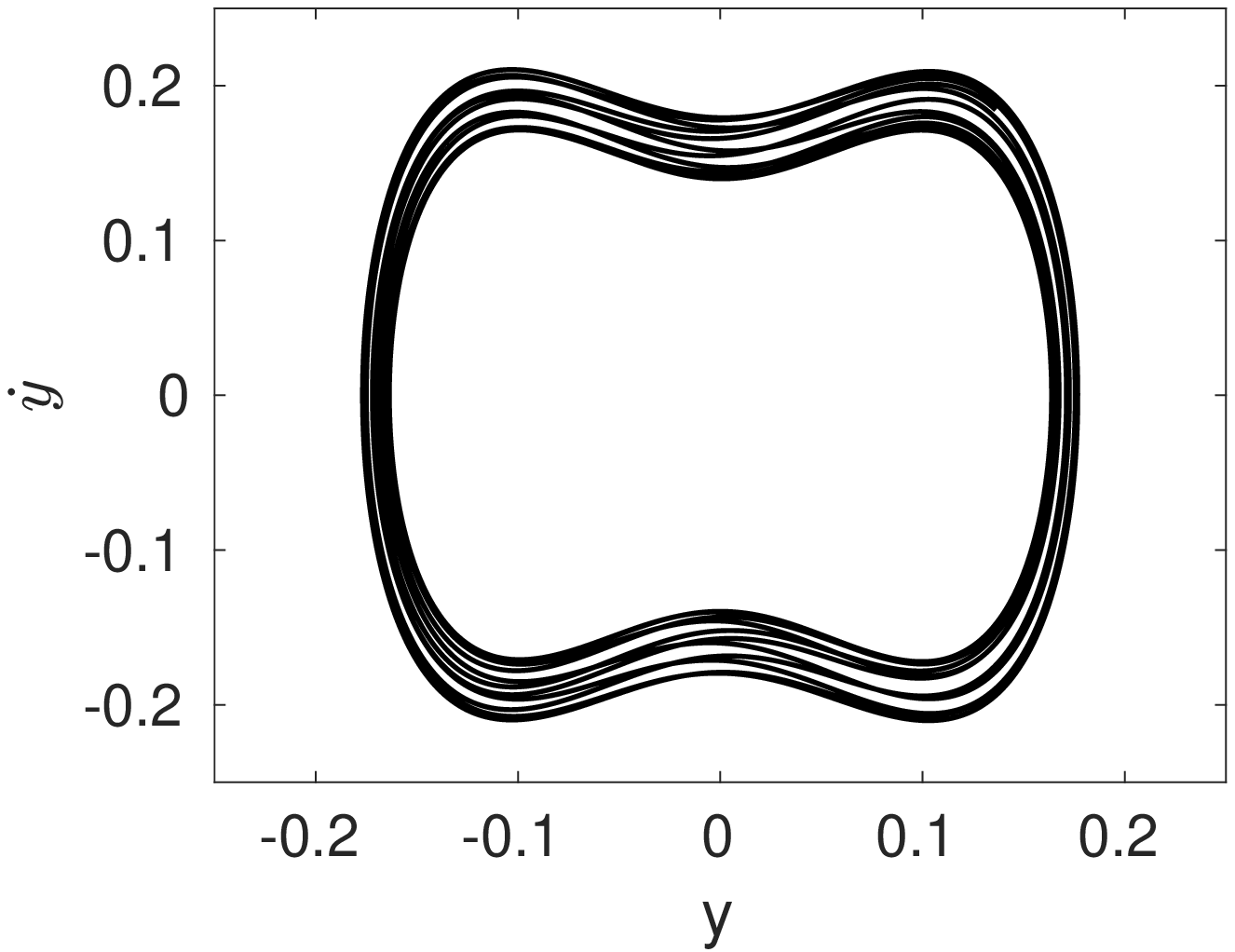}} &
\subfloat[Bistable, $\beta=0.1$ ($U_r=18$)]{\label{b1pp18}\includegraphics[width=0.3\textwidth]{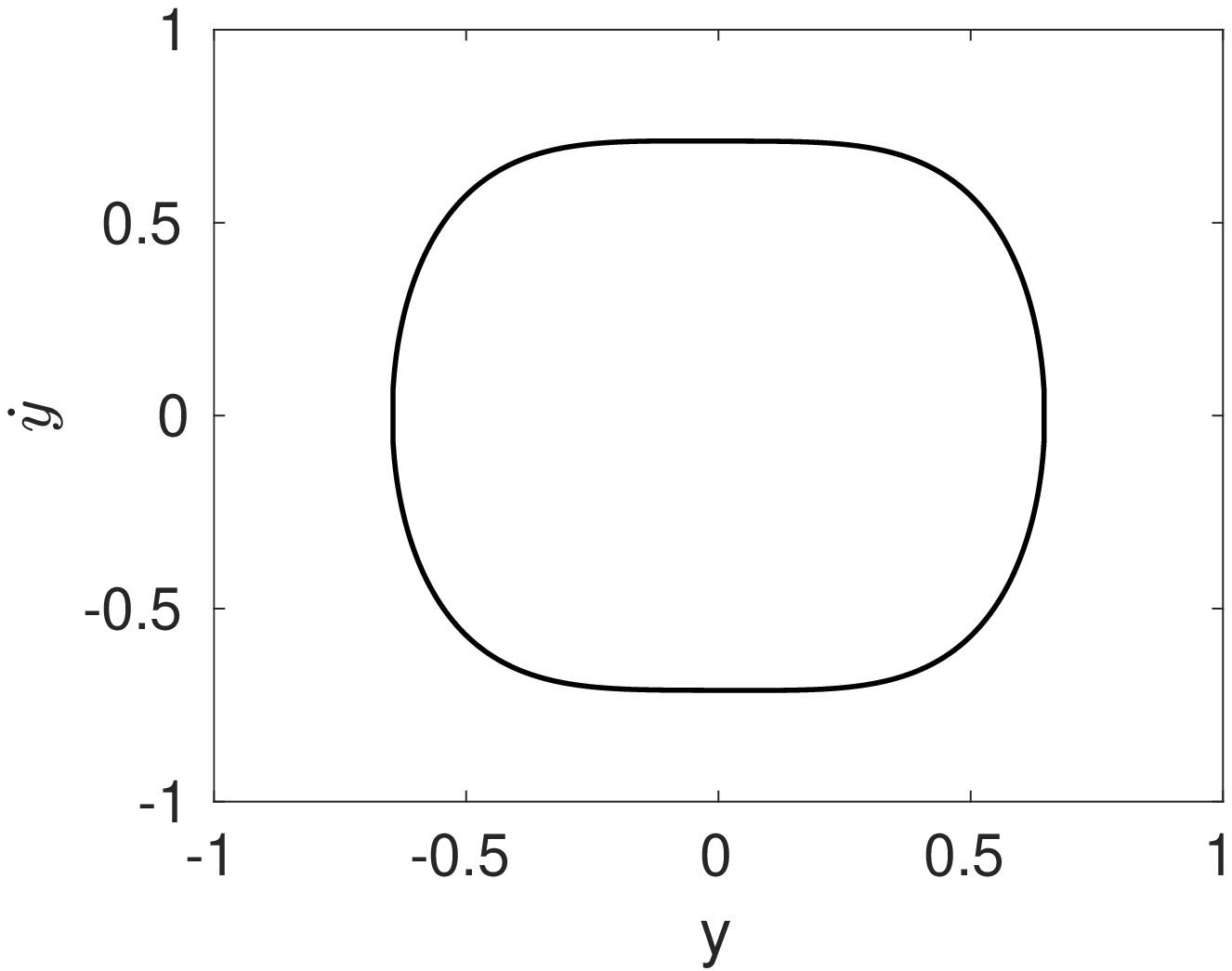}} \\
\end{tabular}
\end{center}
\caption{\label{ppfigs} Representative phase $\dot{y}$ vs $y$ phase portraits for linear and bistable springs.} 
\end{figure}

Fig \ref{ppfigs} show $\dot{y}$-vs-$y$ phase portraits for VIV with linear springs and bistable springs ($\beta=0.1$) for some of the representative high-amplitude cases. Not surprisingly, the phase portrait for VIV with linear spring has an elliptical shape, suggesting an approximately harmonic oscillation (Fig. \ref{linpp5}). On the other hand, the phase portrait for bistable spring departs significantly from an elliptical shape (Fig. \ref{b1pp3},\ref{b1pp18}). Here, for $U_r=3$, the phase portrait is again consistent with a low frequency beating phenomenon, and also has a two lobed structure, suggesting that the limit cycle is in a state which is very close to single-well oscillations. 

\subsection{\label{ampchar} Displacement Amplitude}

\begin{figure}[] 
\begin{center}
\begin{tabular}{c}
\subfloat[\label{ampur} ]{\includegraphics[width=4in,clip]{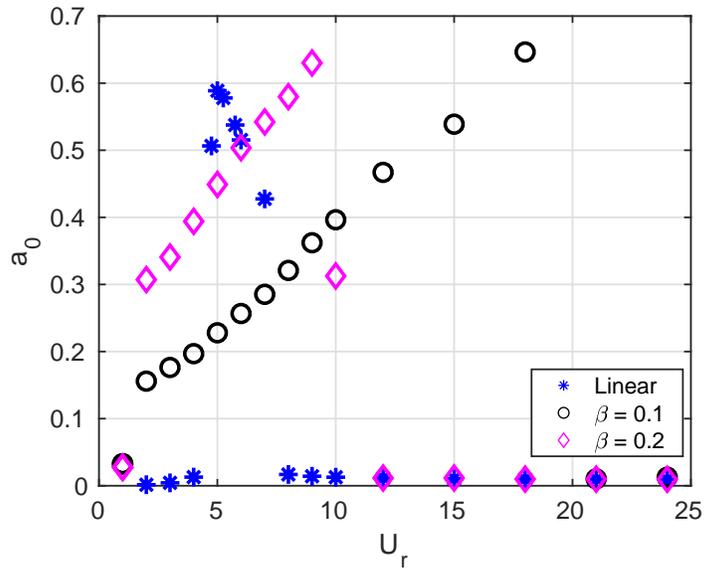}} \\
\subfloat[\label{ureqvsur} ]{\includegraphics[width=3.5in,clip]{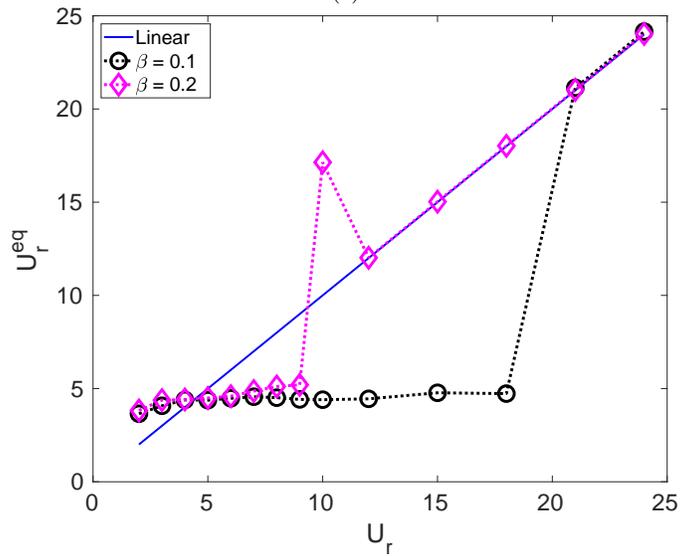}} 
\end{tabular}
\end{center}
\caption{\label{ampversusred} (a) Plot of displacement amplitude $a_0$ versus reduced $U_r$ velocity for different types of springs (linear spring, bistable spring with $\beta=0.1$ and bistable spring with $\beta=0.2$). (b) Equivalent reduced velocity $U_r^{eq}$ versus reduced velocity $U_r$ for different types of springs. Symbols at data points have been connected with lines for visual guidance.  }
\end{figure}

\begin{figure}[H] 
\centering
\begin{tabular}{cc}
\subfloat[\label{ampurstar}Amplitude versus $U_r^{eq}$]{\includegraphics[width=7cm, clip]{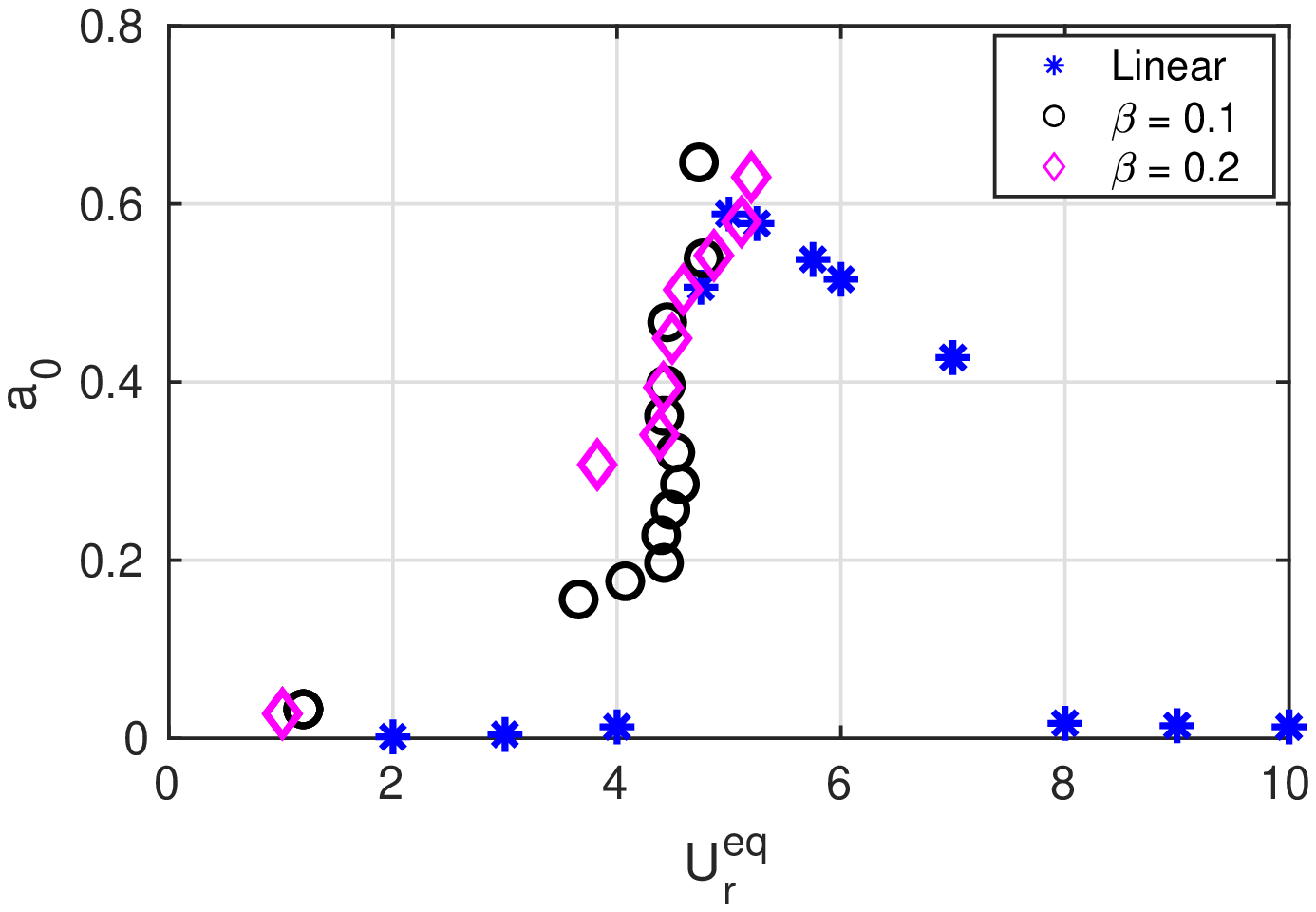}} &
\subfloat[\label{cdmeanureq}Mean drag coefficient versus $U_r^{eq}$]{\includegraphics[width=7cm, clip]{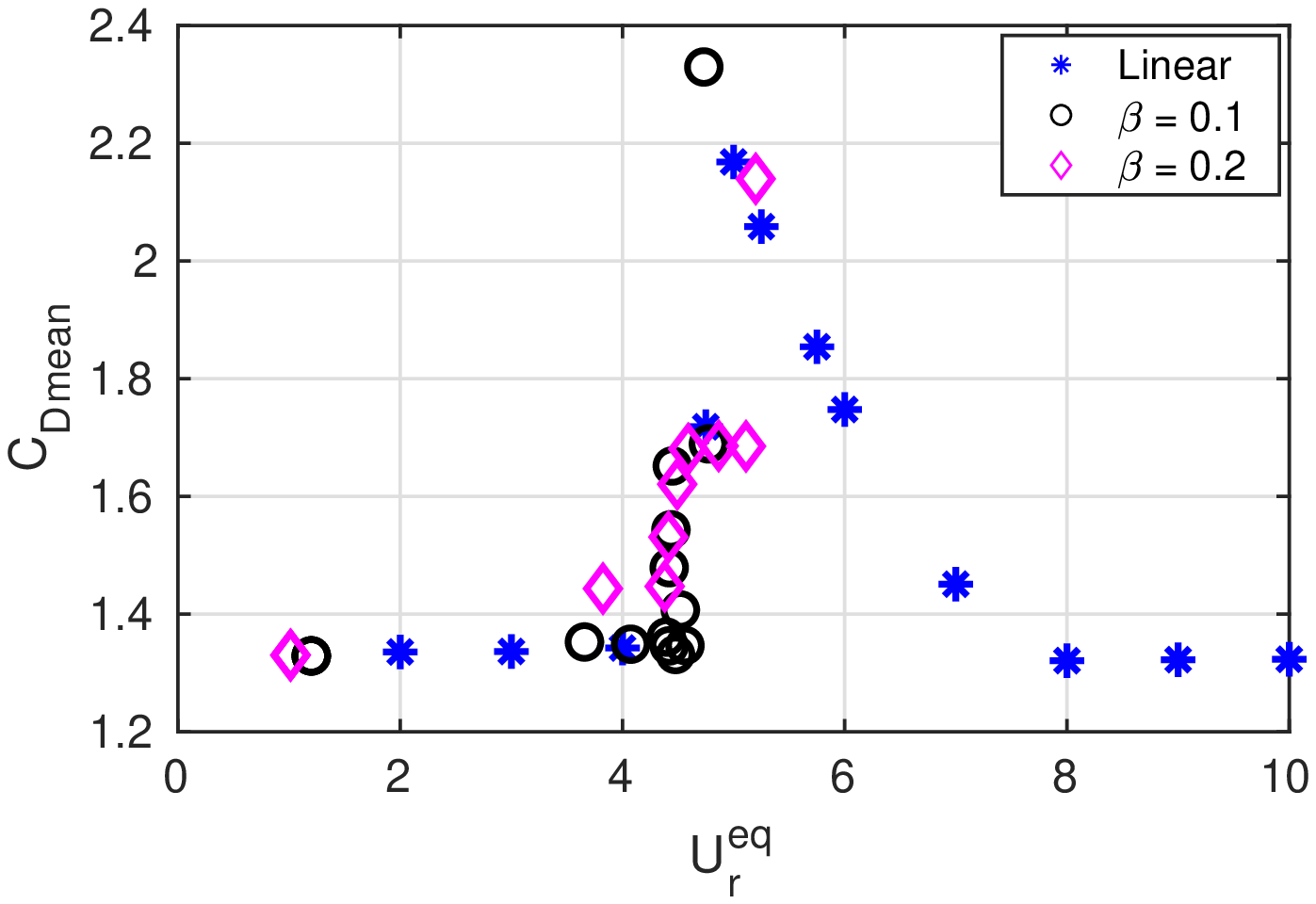}} \\
\subfloat[\label{cdrmsureq}RMS drag coefficient versus $U_r^{eq}$]{\includegraphics[width=7cm, clip]{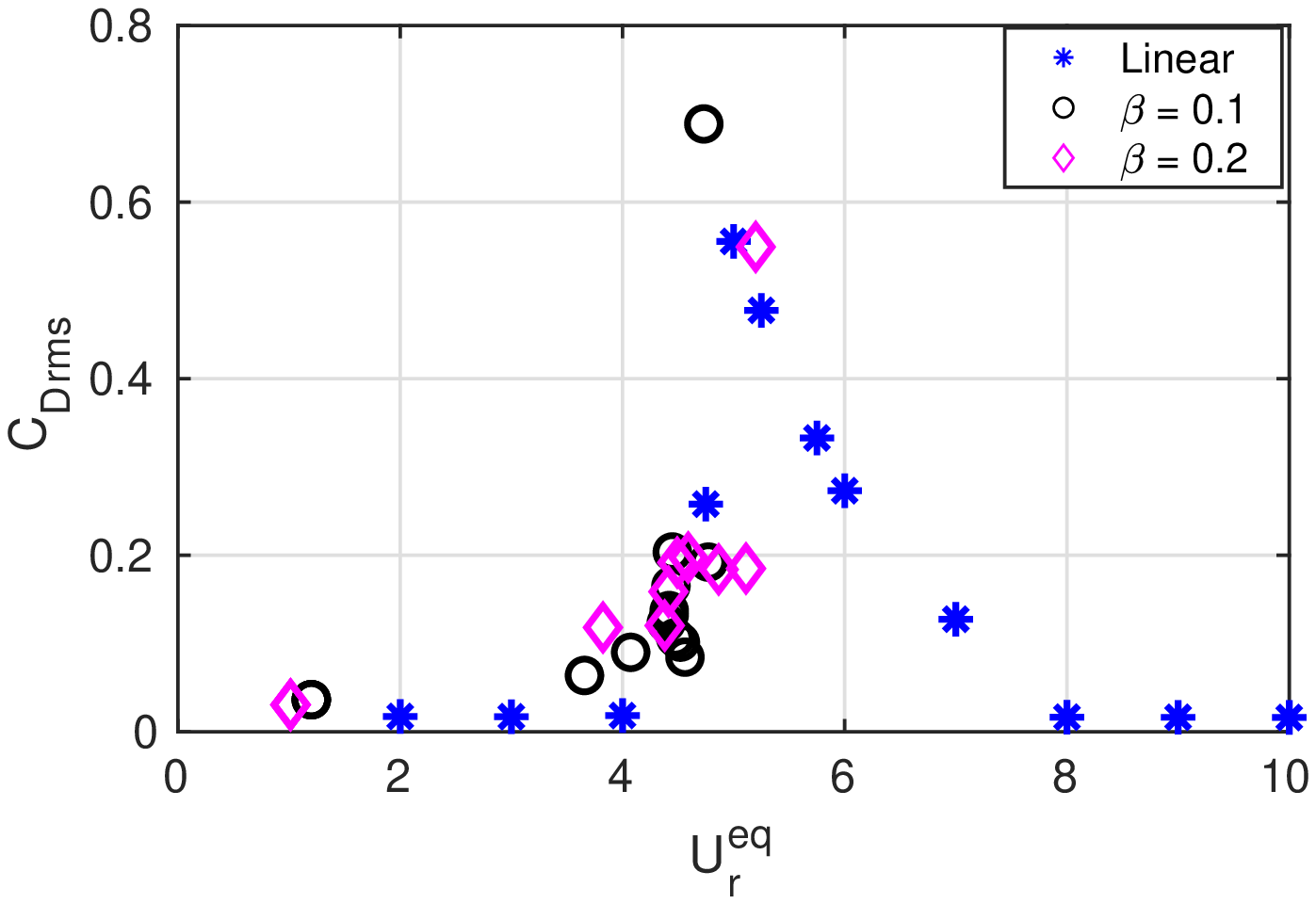}} &
\subfloat[\label{clrmsureq}RMS lift coefficient versus $U_r^{eq}$]{\includegraphics[width=7cm, clip]{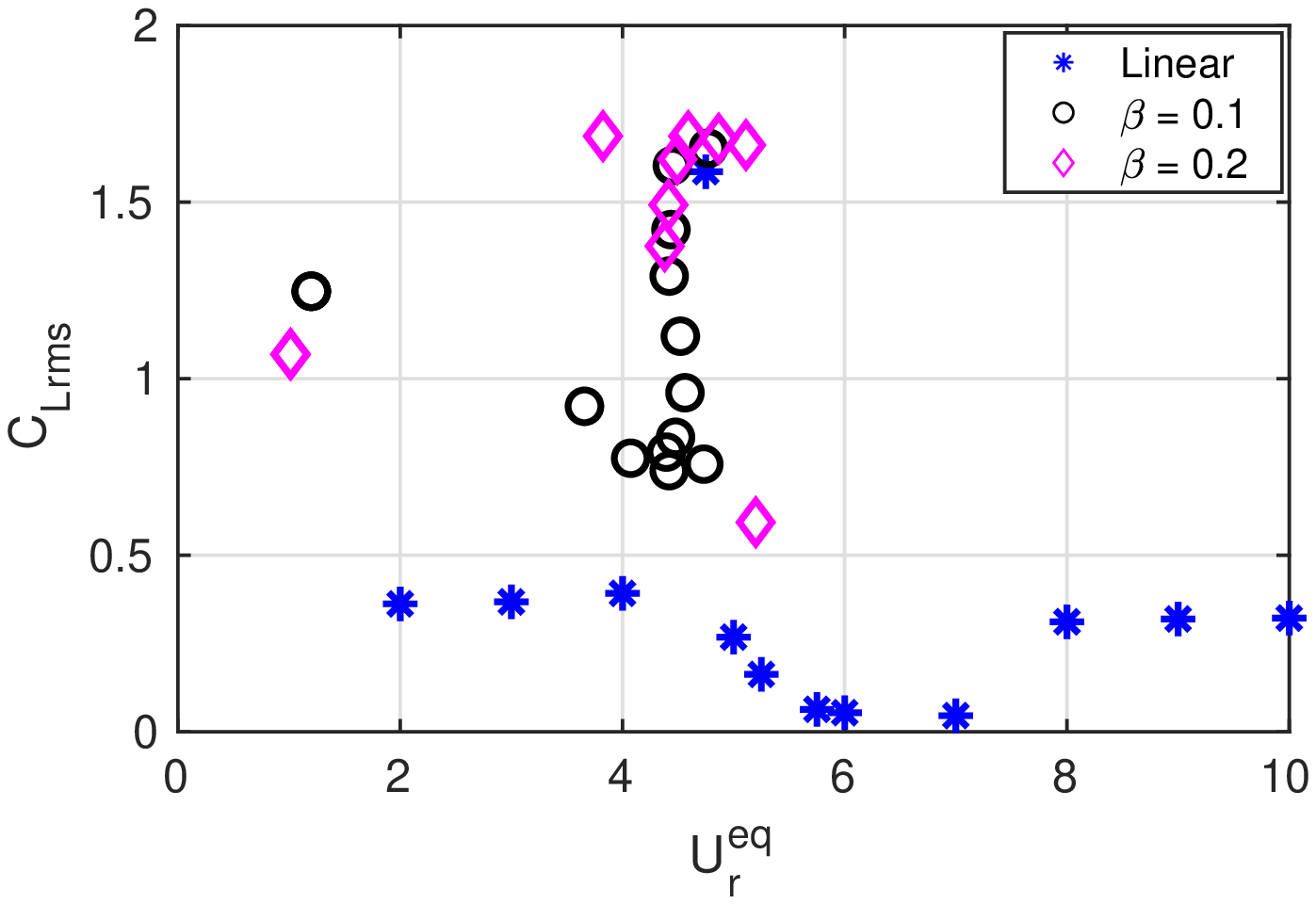}} 
\end{tabular}
\caption{ Plot of displacement amplitude $a_0$, {mean drag coefficient $C_{D_{mean}}$, RMS drag coefficient $C_{D_{RMS}}$ and  RMS lift coefficient $C_{L_{RMS}}$} versus equivalent reduced velocity $U_r^{eq}$ for different types of springs (linear spring, bistable spring with $\beta=0.1$ and bistable spring with $\beta=0.2$). The data point for bistable spring with $\beta=0.2$, $U_r=10$ has not been plotted here due to its large equivalent reduced velocity $U_r^{eq}$. Some of the data points corresponding to desynchronized low amplitude oscillations have also not been plotted here for the same reason.  
%Legends denote Linear spring (\textcolor{blue}{$\ast$}), bistable spring with $\beta=0.1$ ($\circ$), bistable spring with $\beta=0.2$ (\textcolor{magenta}{$\lozenge$}). 
}
\end{figure}

We next report the displacement amplitude $a_0=max(|y|)-|\avg{y}|$, measured with respect to the mean position of the cylinder, for the different spring types and reduced velocities (Fig \ref{ampversusred}). {Here $\avg{y}$ is the mean displacement of the cylinder. Double-well oscillations are indicated by $|\avg{y}|\approx 0$, while single-well oscillations are indicated by $|\avg{y}|\approx\beta$.} 
From Fig \ref{ampur}, it is clear that the maximum value of amplitude, over the whole range of reduced velocities, is almost independent of the type of spring. However, the range of reduced velocity over which significant amplitudes are observed clearly widens for VIV with bistable springs. For VIV with linear spring, $a_0$ remains significant over a small range of $U_r$ ($ 4.75 \leq U_r \leq 7$), and the amplitude falls sharply to zero outside this range. However, in the case of VIV with bistable springs, for $\beta = 0.1$, the range of $U_r$, over which significant amplitudes are observed, varies between  $ 2 \leq U_r \leq 18$. On increasing the value of $\beta$ to $0.2$, the reduced velocity range over which significant amplitude is observed reduces considerably, to  $ 2 \leq U_r \leq 9$. \par

Fig \ref{ureqvsur} shows the equivalent reduced velocity $U_r^{eq}$ (calculated using Eqn \ref{eqredvelrel}), plotted with respect to $U_r$, for different spring types. The presence of spring nonlinearity leads to a significant deviation of the $U_r^{eq}$-vs-$U_r$ curves from linear behaviour. Specifically, $U_r^{eq}$ almost stays constant over a large range for VIV with bistable spring, which therefore also promotes lock-in of the vortex shedding with natural frequency of the structure, and in turn leads to high amplitude oscillations (Fig. \ref{ampur}). The flattening of $U_r^{eq}$ with respect to $U_r$ is similar to results by \cite{Mackowski} for VIV with hardening and softening springs. The anomalous spike in $U_r^{eq}$ at  $U_r=10$ for VIV with bistable spring having inter-well separation $\beta=0.2$ occurs due to lock-in of the fundamental mode of vortex shedding with a harmonic of natural frequency of the structure, as discussed in section \ref{locres}.     

In Fig. \ref{ampurstar}, we plot $a_0$ as a function of the equivalent reduced velocity $U_r^{eq}$, which is based on the natural frequency $\Omega_n$ {(Eqn. \ref{eqredvelrel})}, for all three spring types. Almost all the values of $a_0$ appear to collapse onto the same curve for all types of springs. The major exceptions here are the $U_r=10$ and $U_r=2$ cases for the bistable spring with $\beta=0.2$, and $U_r=2$, $U_r=3$ cases for the bistable springs with $\beta=0.1$. For all these cases, the lack of collapse is probably occuring due to the fact that $a_0$ is quite close to $y_{cr}$, due to which the oscillations will be quite far from being harmonic. There is also a lack of collapse for the cases with highest $a_0$ for each spring type, which may be attributed to change in the vortex shedding pattern from single to double-row of vortices (discussed further in section \ref{vorplot}). The collapse of $a_0$ with respect to $U_r^{eq}$ for the other data points, is in agreement with prior literature on VIV with softening and hardening springs \cite{Mackowski, wang2019effect}. {In Figs. \ref{cdmeanureq}, \ref{cdrmsureq} and Fig. \ref{clrmsureq}, we have also plotted mean drag coefficient $C_{D_{mean}}$, RMS drag coefficient $C_{D_{RMS}}$ and  RMS lift coefficient $C_{L_{RMS}}$ respectively, as a function of equivalent reduced velocity $U_r^{eq}$, for different types of springs. Again, barring a few outliers, we observe a collapse of all the data here points with respect to equivalent reduced velocity $U_r^{eq}$, which is consistent with Fig. \ref{ampurstar}}. \par

\subsection{\label{locres}Lock-in characteristics}
\begin{figure}[] 
\begin{center}
\begin{tabular}{cc}
\subfloat[\label{linearfft} ]{\includegraphics[scale = 0.7]{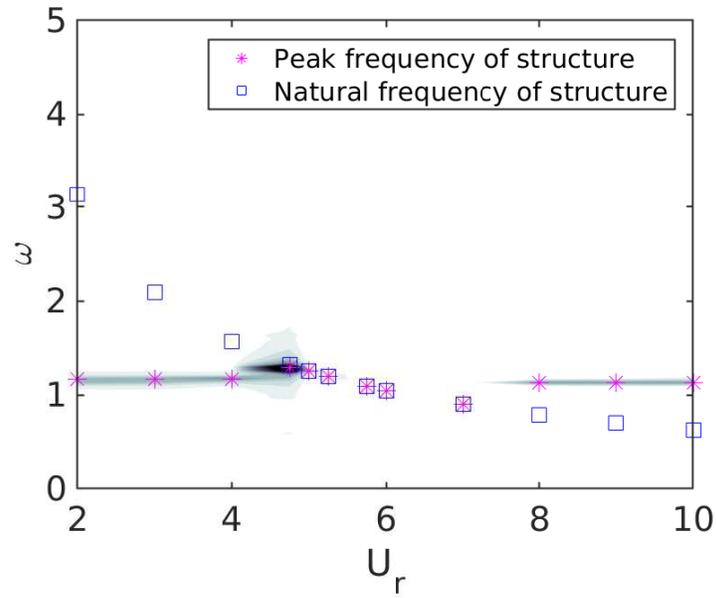}} \\
\subfloat[\label{linearrms} ]{\includegraphics[scale = 0.7]{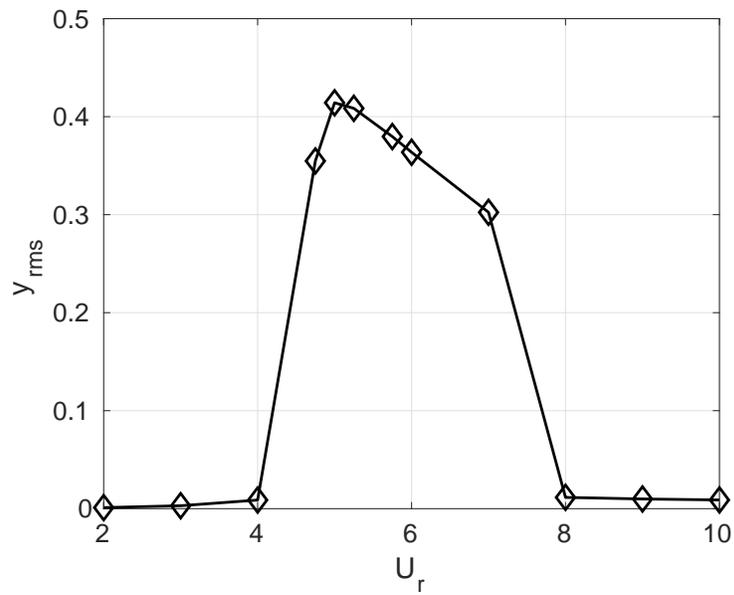}} 

\end{tabular}
\end{center}
\caption{\label{linearlockin} Lock-in characteristics for linear spring. (a) Peak frequency of the PSD of structure, $\omega$ (\textcolor{magenta}{*}), along with natural frequency of structure $\omega_n$ ($\textcolor{blue}{\square}$), superimposed on contour plot of Power Spectral Density of lift force $F$, over a range of $U_r$ (in contour plot, values between data points have been interpolated as a visual guide). (b) Variation of $y_{rms}$ (RMS of $y$), with respect to $U_r$.}
\end{figure}

\begin{figure}[] 
\begin{center}
\begin{tabular}{cc}
 \subfloat[\label{b1fft}  ]{\includegraphics[scale = 0.7]{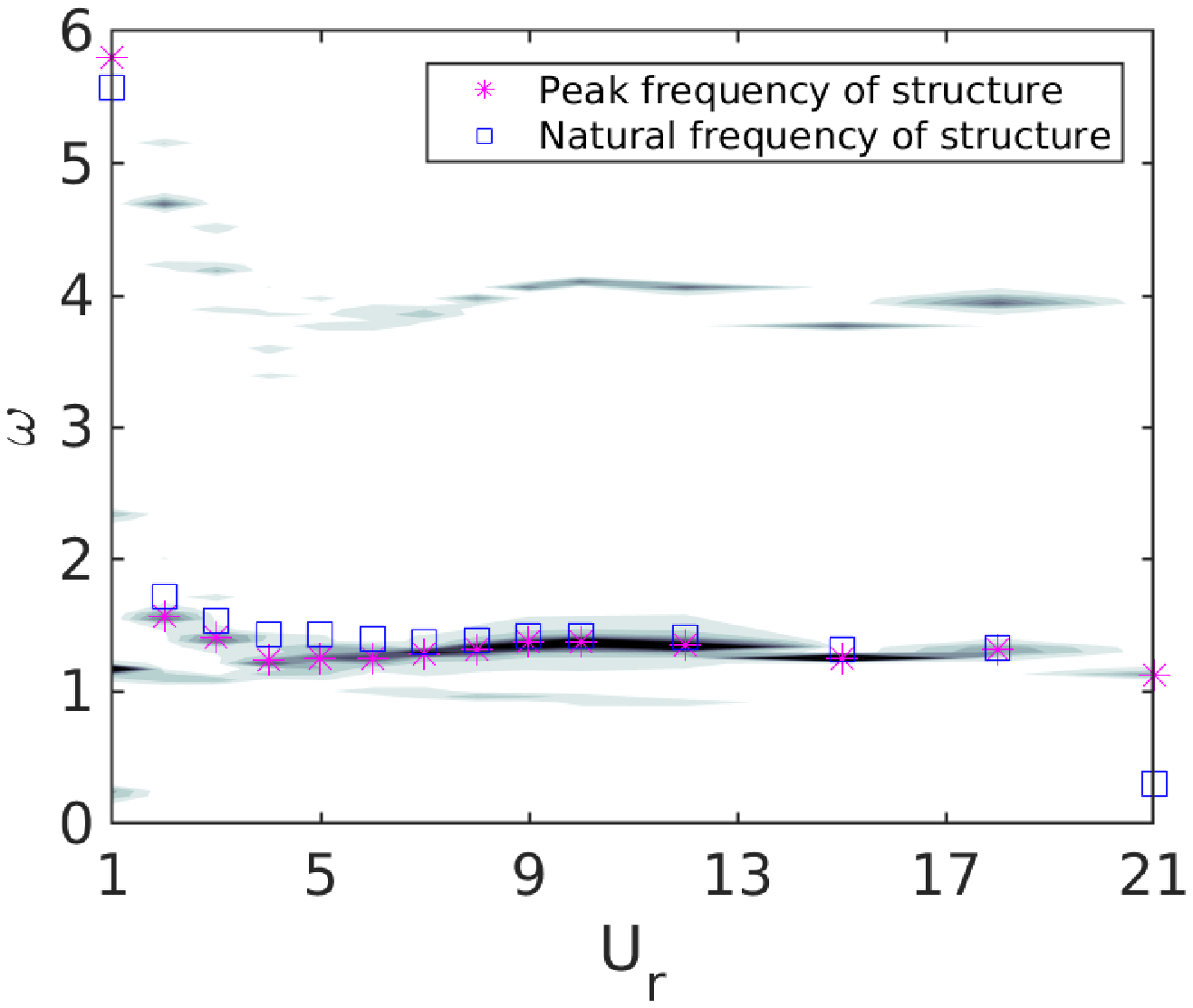}} \\
\subfloat[\label{b1rms}  ]{\includegraphics[scale = 0.7]{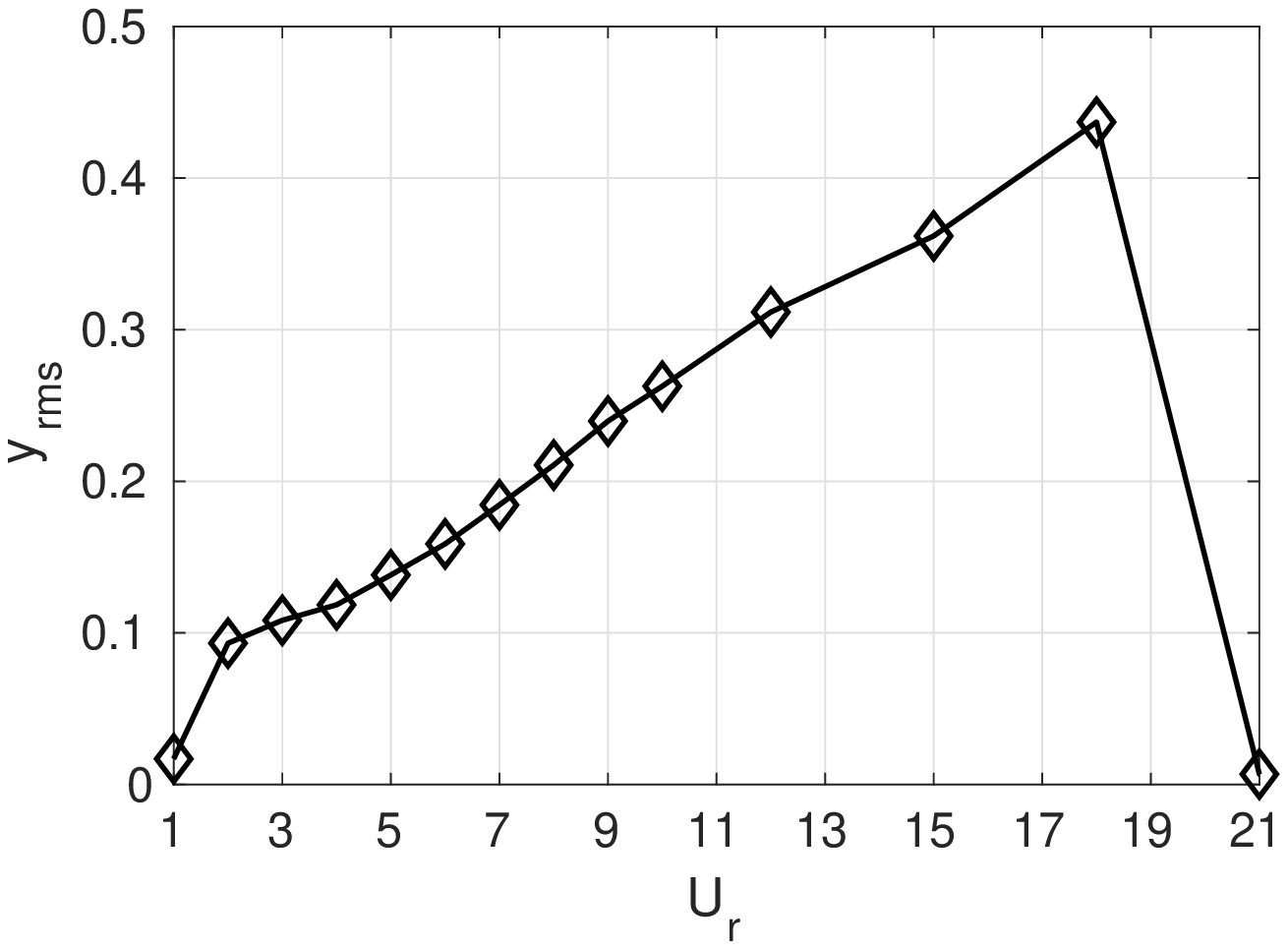}} 
\end{tabular}
\end{center}
\caption{\label{b1lockin} Lock-in characteristics for bistable spring with $\beta=0.1$. Rest of the caption is same as Fig. \ref{linearlockin}}
\end{figure}

\begin{figure}[] 
\begin{center}
\begin{tabular}{cc}
\subfloat[\label{b2fft} ]{\includegraphics[scale = 0.7]{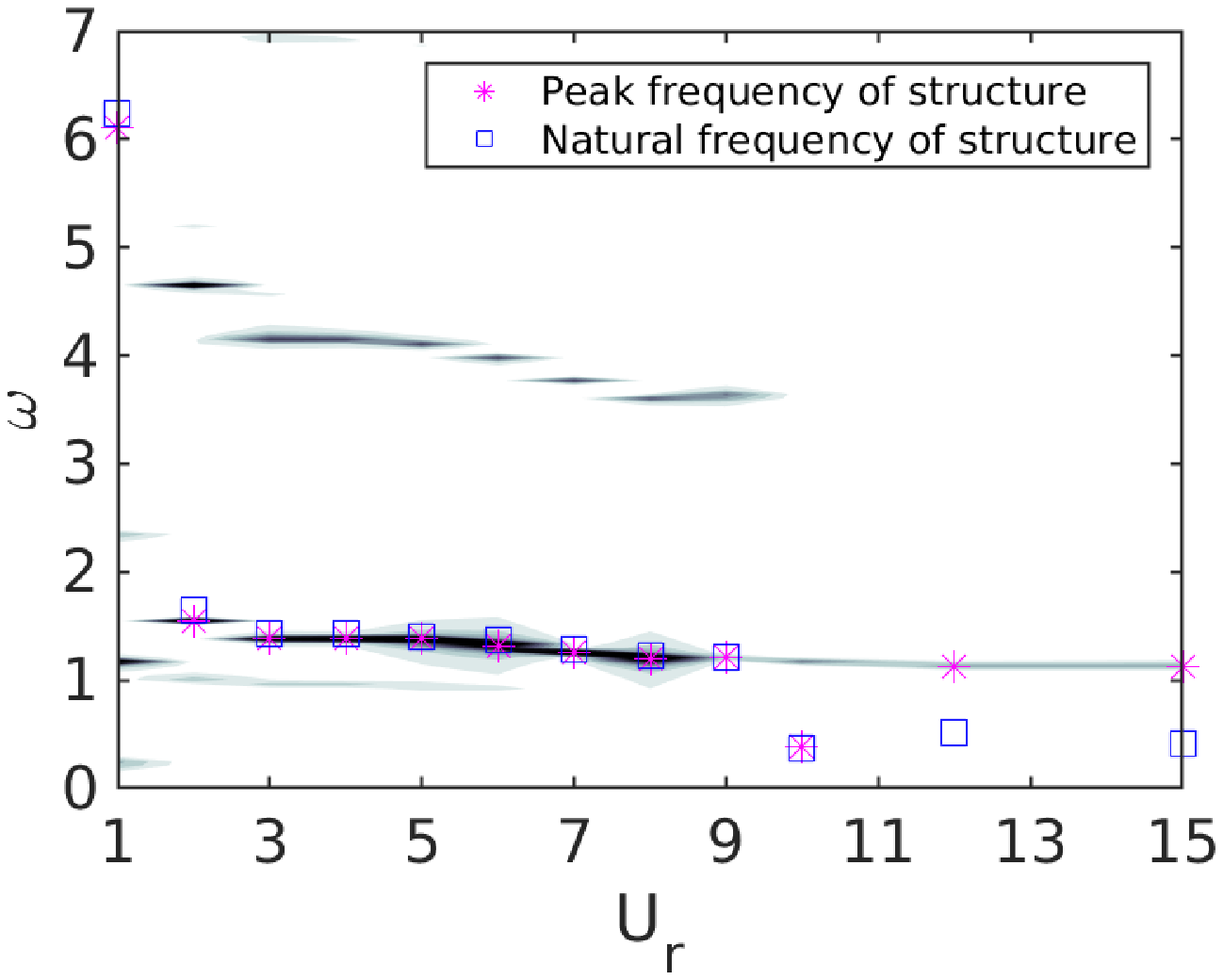}} \\
\subfloat[\label{b2rms} ]{\includegraphics[scale = 0.7]{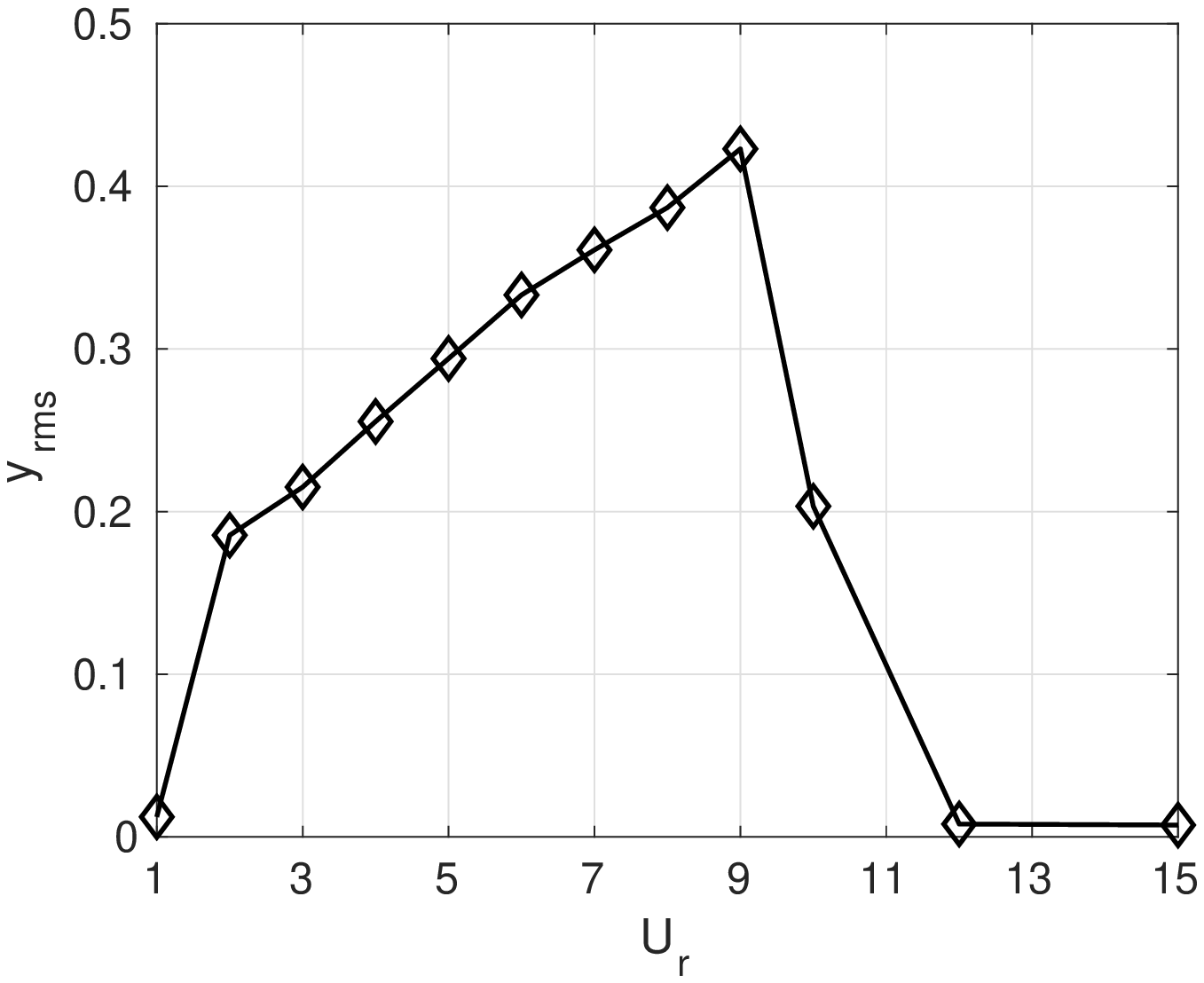}} 
\end{tabular}
\end{center}
\caption{\label{b2lockin} Lock-in characteristics for bistable spring with $\beta=0.2$. Rest of the caption is same as Fig. \ref{linearlockin}}
\end{figure}

Next, in this section, we present the lock-in characteristics for VIV with linear and bistable springs. We are specifically interested in studying the flow regime where the structure locks in with the natural frequency of the spring-mass system. To highlight the synchronization of lift force, structure frequency and natural frequency in the lock-in regime, we plot the isocontour of Power Spectrum Density (PSD) for lift force $F$, over which we superimpose the peak frequency of the PSD of the $\dot{y}$ (structure velocity), along with the natural frequency of the spring-mass system $\omega_n=\Omega_n/\Omega_f$ (Figs. \ref{linearfft},\ref{b1fft},\ref{b2fft}). For bistable springs, we calculate $\omega_n$ based on the maximum $a_0/\beta$, where $a_0$ is obtained from the simulations. We also plot the Root Mean Squared value of $y$ ($y_{rms}$) as a function of $U_r$, which highlight the regions with high amplitude (Figs. \ref{linearrms}, \ref{b1rms}, \ref{b2rms}). 
From Fig \ref{linearfft}, corresponding to VIV with linear spring, we can see that the PSD of lift force has only one strong peak, which locks in with the structure and the natural frequency over $4.75 < U_r < 7$, explaining the large amplitude oscillations in this regime (Fig \ref{linearrms}). Outside the lock-in regime, the structure synchronizes with the lift force, and not the natural frequency of the spring-mass system. \par
For VIV with bistable spring having lower value of inter-well separation ($\beta = 0.1$), we first observe that the PSD of lift force can contain multiple peaks, perhaps due to the non-harmonic nature of the oscillations observed in Figs. \ref{b1pp3}, \ref{b1pp18}. Nevertheless, over the rather wide lock-in regime ($2\leq U_r \leq 18$), only the fundamental frequency of the structure and the lift force appear to synchronize with the natural frequency of the spring mass system. Beyond, $U_r>18$, the fundamental mode of the lift force and the natural frequency of the structure desynchronize, and hence the cylinder undergoes small, single-well oscillations. \par
For VIV with bistable spring having higher value of inter-well separation ($\beta =0.2$), the trends are largely similar (Fig. \ref{b2lockin}) to VIV with bistable springs having smaller inter-well separation $(\beta=0.1$). However, in this case, the lock-in of the structure velocity with the natural frequency occurs over a relatively narrow range. At $U_r = 10$, we observe that the fundamental vortex shedding mode is locking in with a harmonic of the structure frequency (Fig. \ref{b2fft}), leading to lower amplitude compared to the $U_r=9$ case (Fig. \ref{b2rms}). \par
Clearly, for bistable springs, there is a rather large range of $U_r$ over which the fundamental mode of the lift force, very close to the Strouhal frequency of the cylinder $\Omega_f$, is able to lock-in with the natural frequency of the spring-mass system $\Omega_n$. This somewhat anomalous synchronization can be explained by first observing that during the large-amplitude oscillations, the cylinder undergoes double well oscillations. Thus, the increase in range of lock-in for VIV with bistable springs is consistent with the theory presented in section \ref{locinb}, where it was shown that, due to the strong dependence of the natural frequency of bistable springs on the oscillation amplitude during double well oscillations, there is a relatively large range of $U_r$ over which the natural frequency curves intersect the EC curve. The theory also predicted that the lock-in range reduces with increasing $\beta$, which is consistent with the results presented in this section. On the other hand, for VIV with linear springs, since the natural frequency does not depend on oscillation amplitude, therefore the natural frequency curves intersect the EC over a much smaller range in $U_r$.    

\begin{figure}[] 
\begin{center}
\begin{tabular}{cc}
\subfloat[]{\label{lockinall}} {\includegraphics[height=6cm,clip]{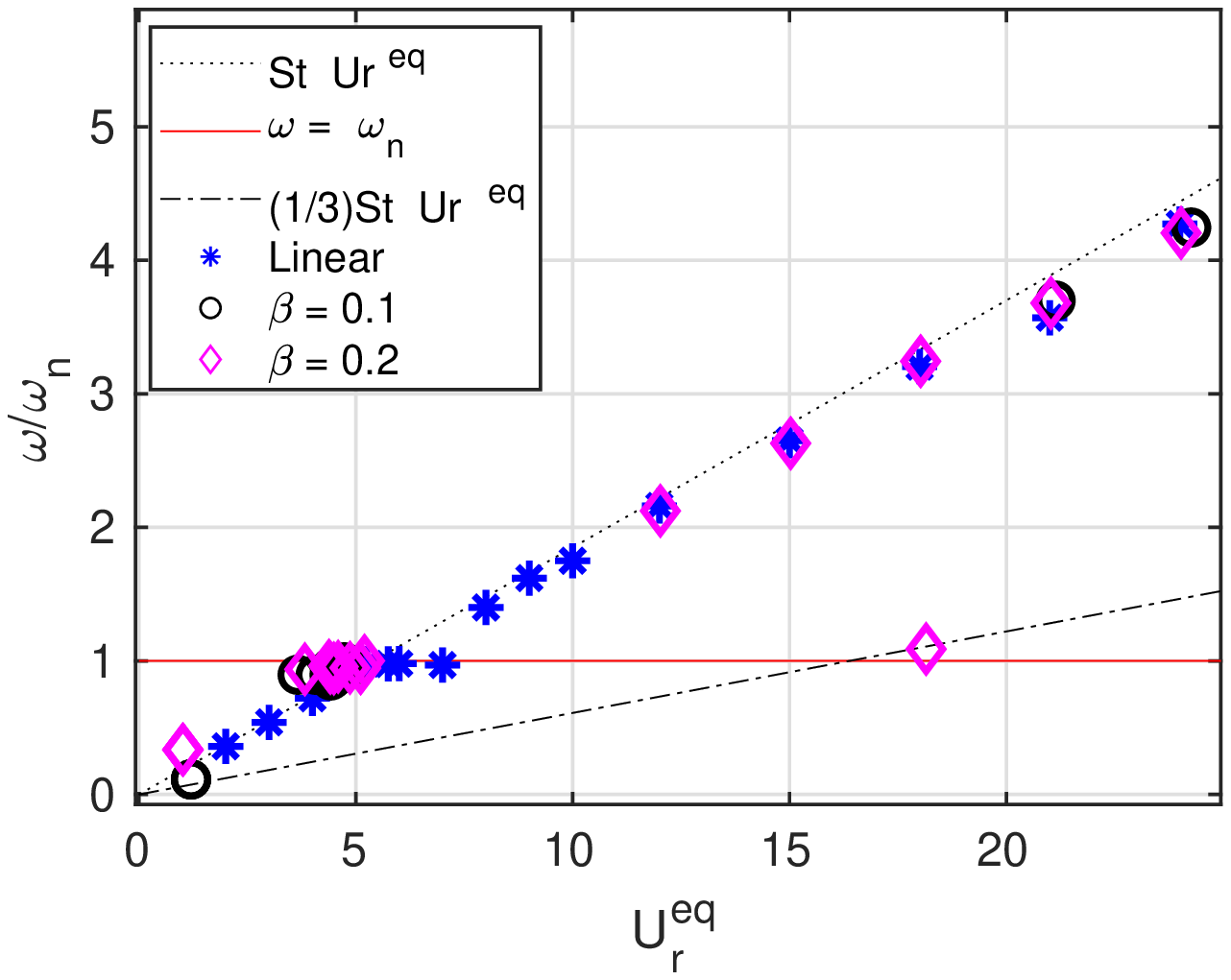}} &  
\subfloat[]{\label{lockinzoom}\includegraphics[height=6cm,clip]{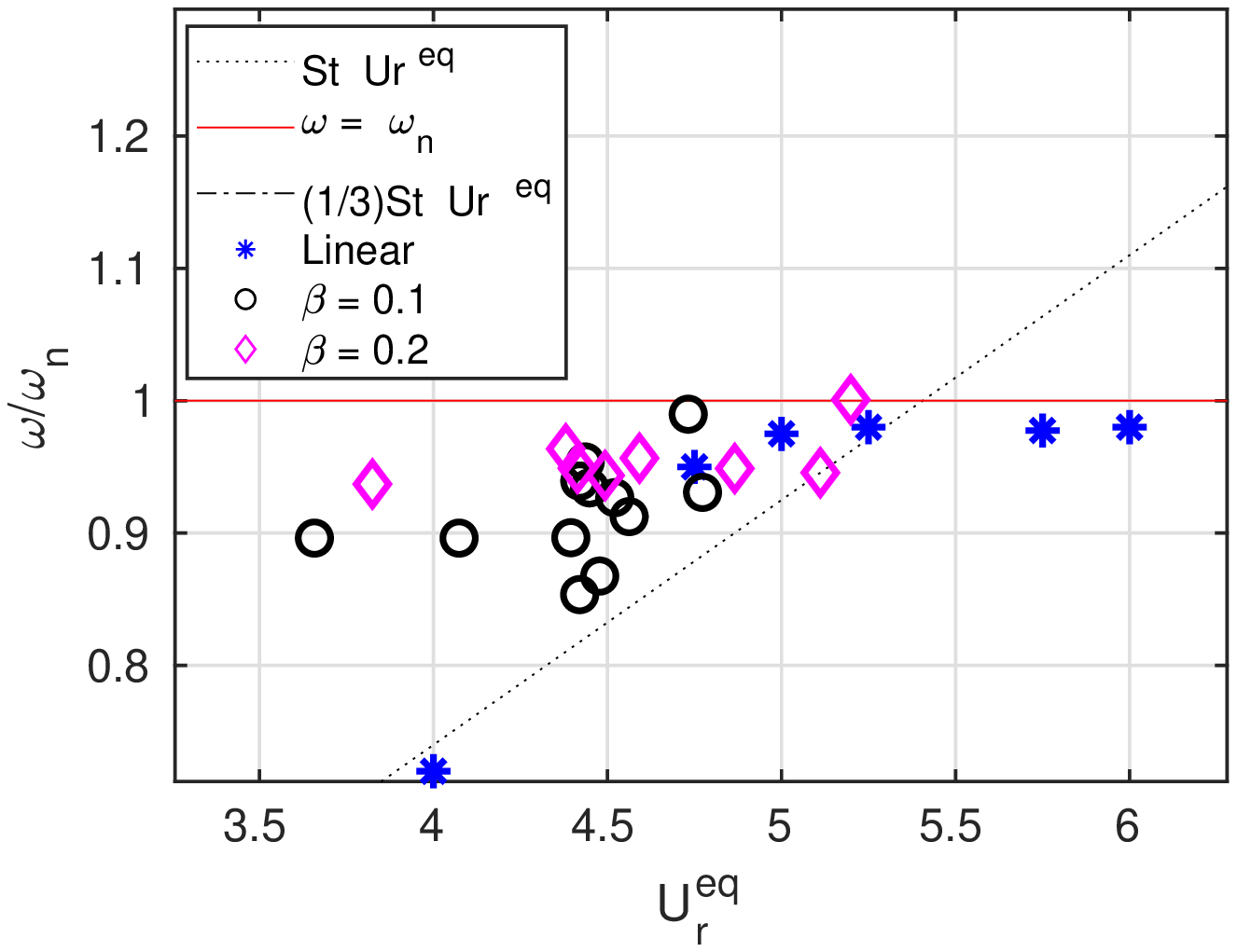}} \\

\end{tabular}
\end{center}
\caption{\label{lockin} Plot showing the lock-in of the structure frequency either with the natural frequency or the vortex shedding frequency. As the points are clustered near the region $U_r^{eq} \approx 5$, Fig. \ref{lockinzoom} zooms the region close to $U_r^{eq} = 5$. The color notation is as, Linear spring (\textcolor{blue}{$\ast$}), $\beta=0.1$ (\textcolor{black}{$\circ$}), $\beta=0.2$ (\textcolor{magenta}{$\diamond$}). }
\end{figure}

We next present a lock-in plot (Fig. \ref{lockin}), consisting of the reduced frequency of the structure $\omega/\omega_n$ plotted against the equivalent reduced velocity $U_r^{eq}$, for the different spring types. Based on the previous discussion, it is not surprising that the data here shows reasonable collapse for the different types of springs. Within the range $3.6 < U_r^{eq} <7 $, the structure locks-in (Fig. \ref{lockinzoom}) with the fundamental natural frequency of the system (i.e. $\omega/\omega_n\approx 1$), whereas outside this range, the structure locks-in with the Strouhal frequency. The sole outlier here is the data point at $U_r=10$ for bistable spring with $\beta=0.2$ (discussed above), which shows a lock-in of the Strouhal frequency with the third harmonic of the natural frequency of the spring-mass system.

\subsection{\label{vorplot}Vortex shedding patterns}

\begin{figure}[] 
%\begin{center}
%\begin{tabular}{ccc}
 \subfloat[ Linear   $U_r = 5$]{\includegraphics[width=0.31\textwidth]{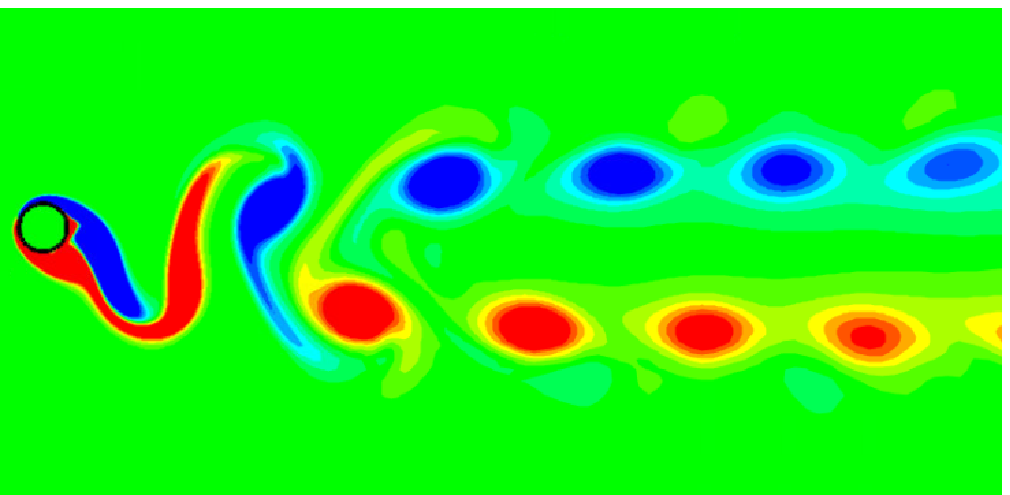}} \hfill
\subfloat[ Linear   $U_r = 5.75$]{\includegraphics[width=0.31\textwidth]{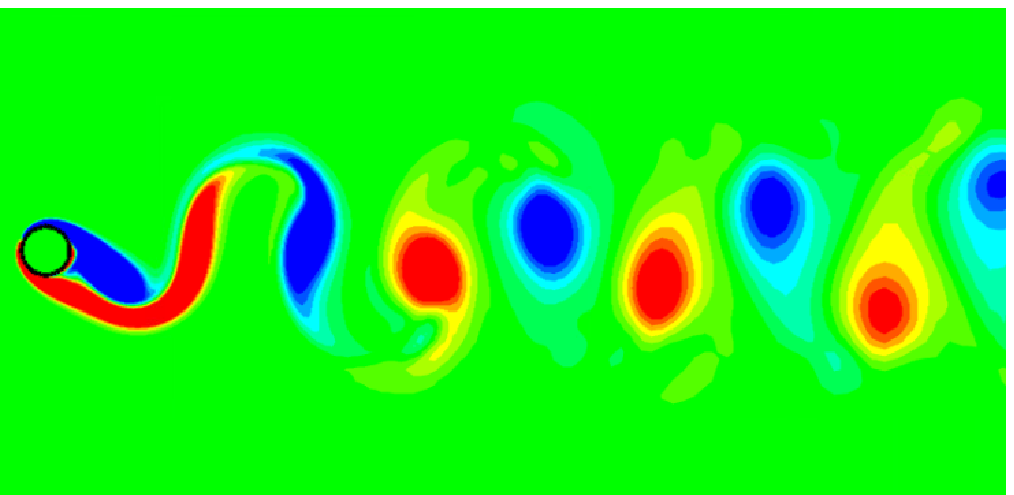}} \hfill 
\subfloat[ Linear   $U_r = 7$]{\includegraphics[width=0.31\textwidth]{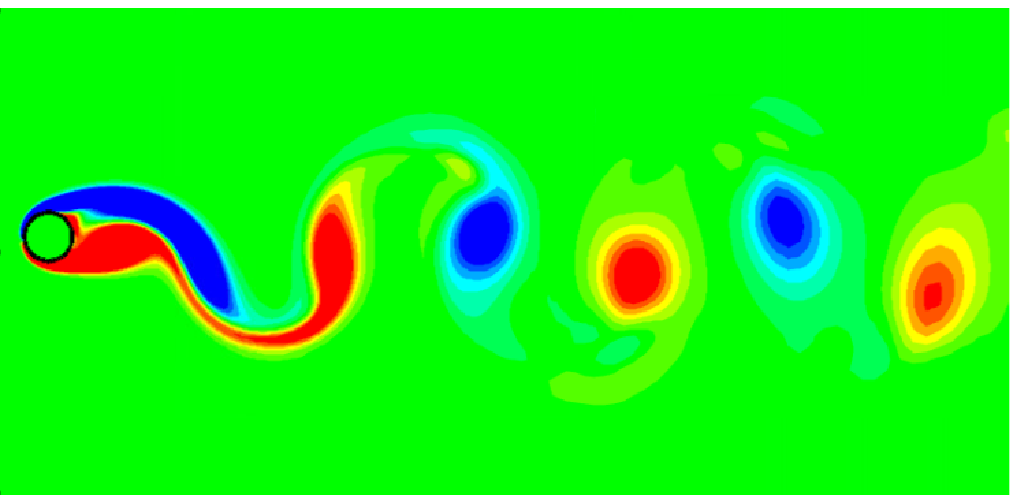}} 
 \par
 \subfloat[ Bistable, $\beta=0.1$, $U_r = 4$]{\includegraphics[width=0.31\textwidth]{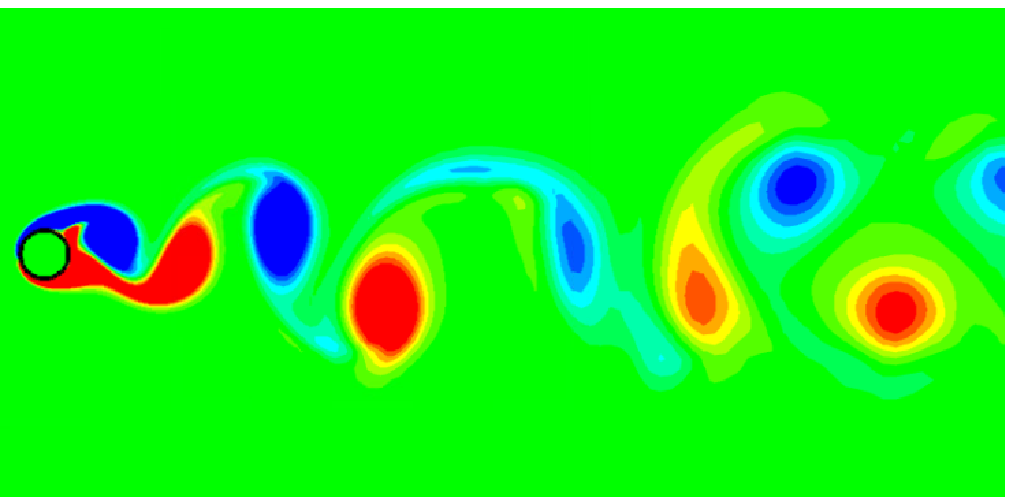}} \hfill
\subfloat[ Bistable, $\beta=0.1$,  $U_r = 9$]{\includegraphics[width=0.31\textwidth]{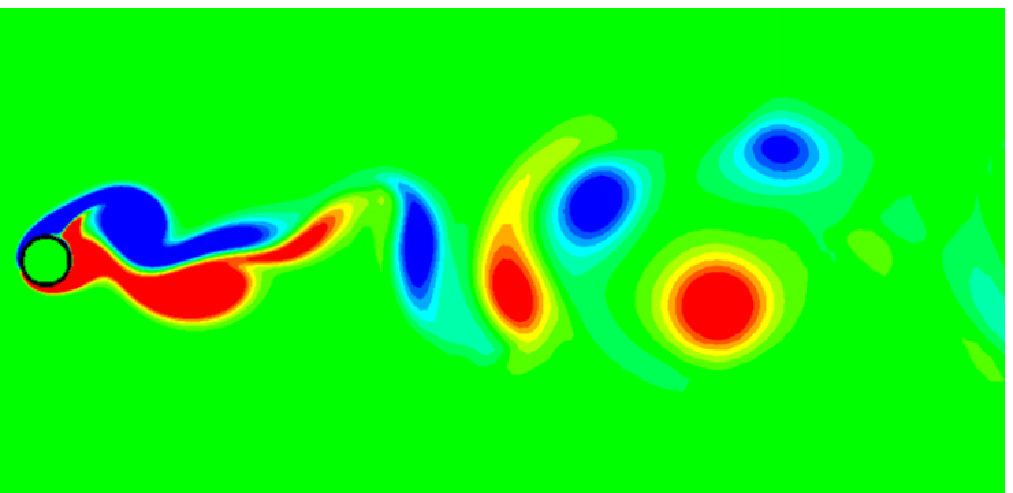}} \hfill
\subfloat[ Bistable, $\beta=0.1$,  $U_r = 18$]{\includegraphics[width=0.31\textwidth]{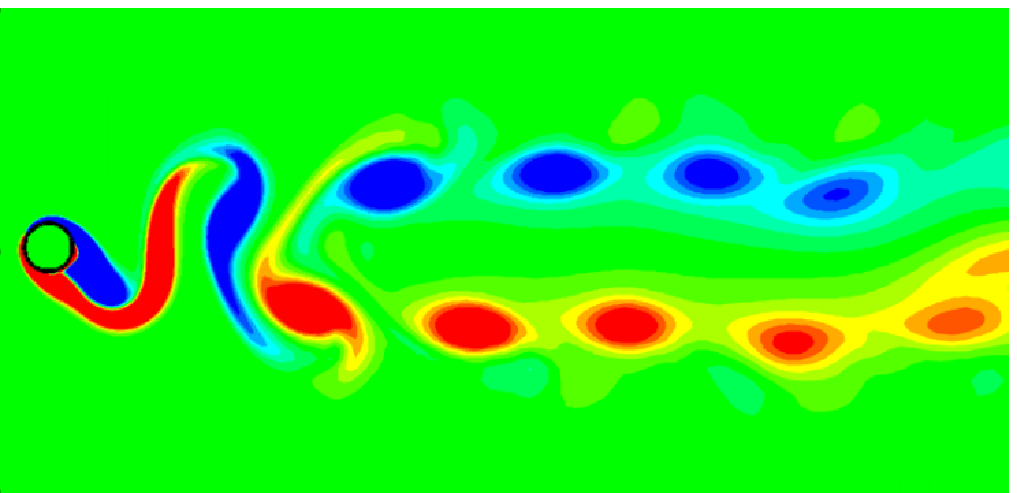}} 
 \par
 \subfloat[ Bistable, $\beta=0.2$, $U_r = 3$]{\includegraphics[width=0.31\textwidth]{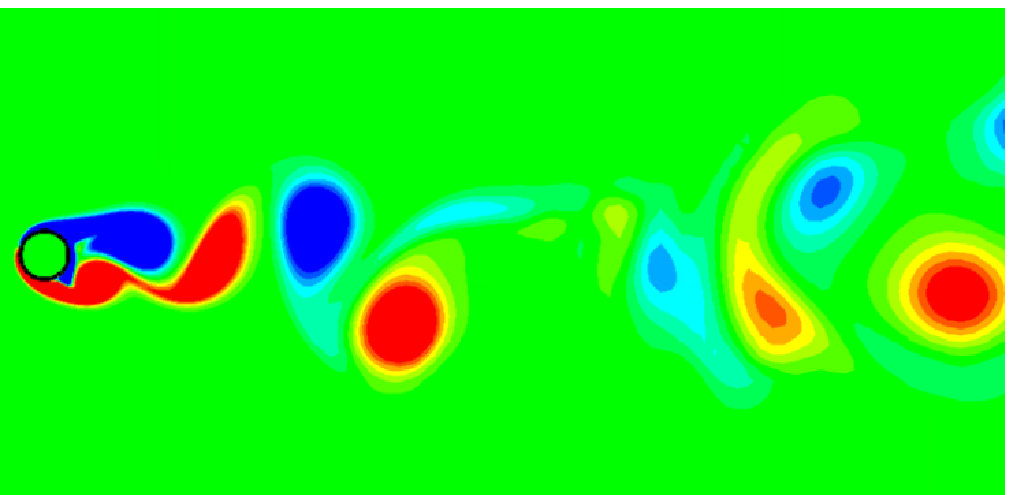}} \hfill
\subfloat[ Bistable, $\beta=0.2$,  $U_r = 6$]{\includegraphics[width=0.31\textwidth]{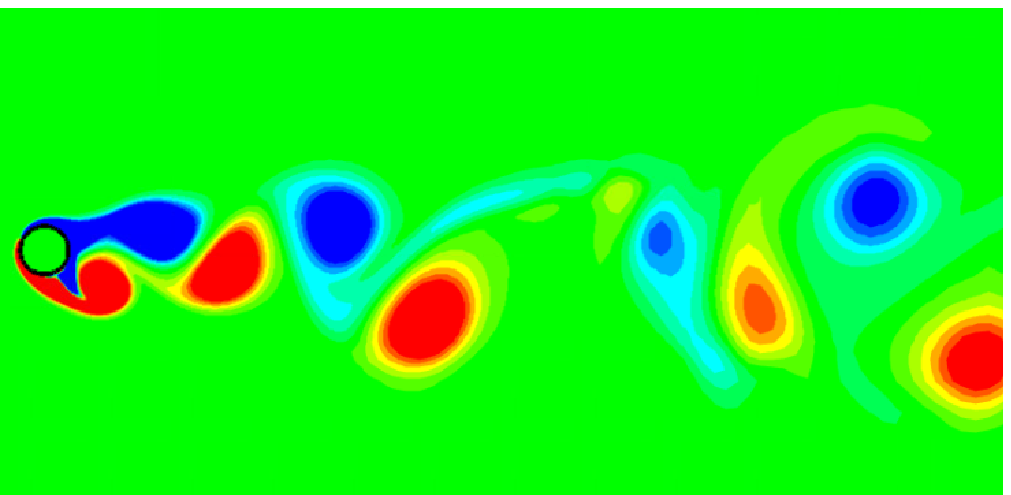}} \hfill
\subfloat[ Bistable, $\beta=0.2$,  $U_r = 9$]{\includegraphics[width=0.31\textwidth]{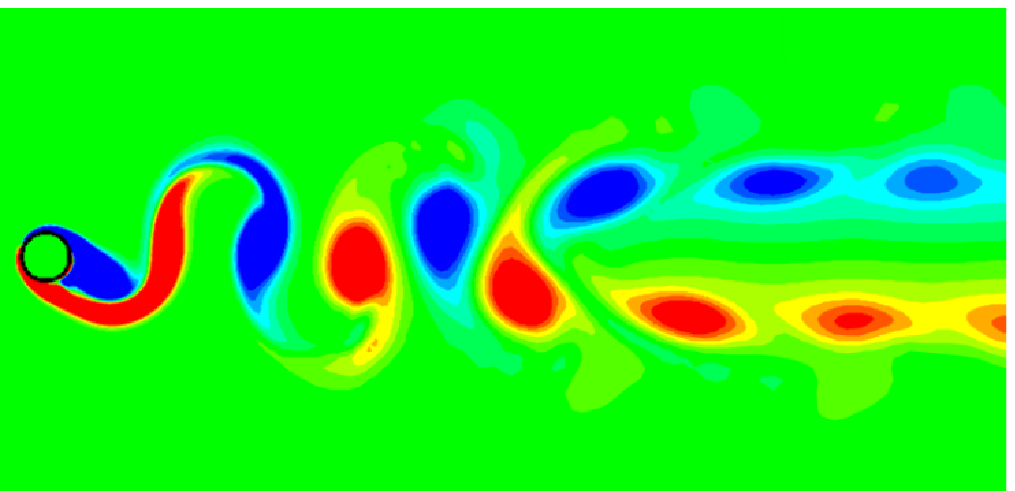}} 
 \par
  \subfloat{\includegraphics[width=1\textwidth]{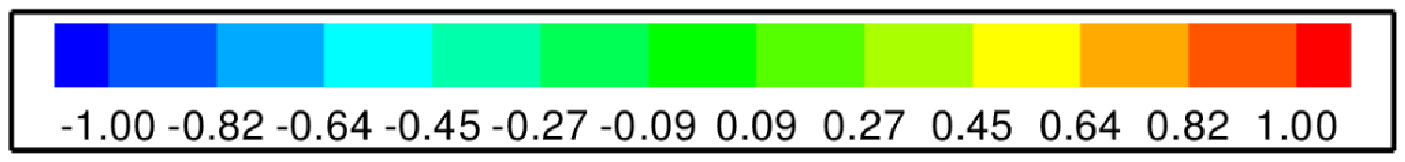}} 

 %\end{tabular}
 
%\end{center}
\caption{\label{vorshedall} Isocontours of vorticity for representative values of reduced velocity $U_r$ for different spring types. The color in the contour plots indicate the value of vorticity, indicated in the color map.}
\end{figure}

In Fig \ref{vorshedall}, we have plotted the vortex shedding pattern for some representative reduced velocities, for all spring types, corresponding to large $a_0$. For all the simulations, we observe different versions of 2S pattern of vortex shedding \cite{williamson2004vortex}, in which 2 vortices of opposite signs are shed at every oscillation cycle. In Fig \ref{ampur}, we can see that for linear spring, $a_0$ is maximum at $U_r=5$, for bistable spring with $\beta=0.1$, $a_0$ is maximum at $U_r=18$, while for bistable spring with $\beta=0.2$, $a_0$ is maximum at $U_r=9$. For these cases, the vortex shedding patterns show that (Fig \ref{vorshedall}(a),(f),(i)), the clock-wise (CW) and counter-clock-wise (CCW) vortices are separated into two rows. For the other cases, with lower $a_0$, the successive CW and CCW vortices are in a single-row configuration. Similar modulation of vortex shedding pattern by the displacement amplitude was observed for VIV with hardening and softening springs by Wang \emph{et al.} \cite{wang2019effect}. The distinct 2-row pattern for high amplitude oscillations also explains the lack of collapse of these data points in the $a_0$-vs-$U_r^{eq}$ graph in Fig. \ref{ampurstar}.   

\begin{figure}[] \centering
%\begin{center}
%\begin{tabular}{cc}
 \subfloat[ Linear  $U_r = U_r^{eq}=4.75$, $a_0=0.51$]{\includegraphics[width = 0.45\textwidth]{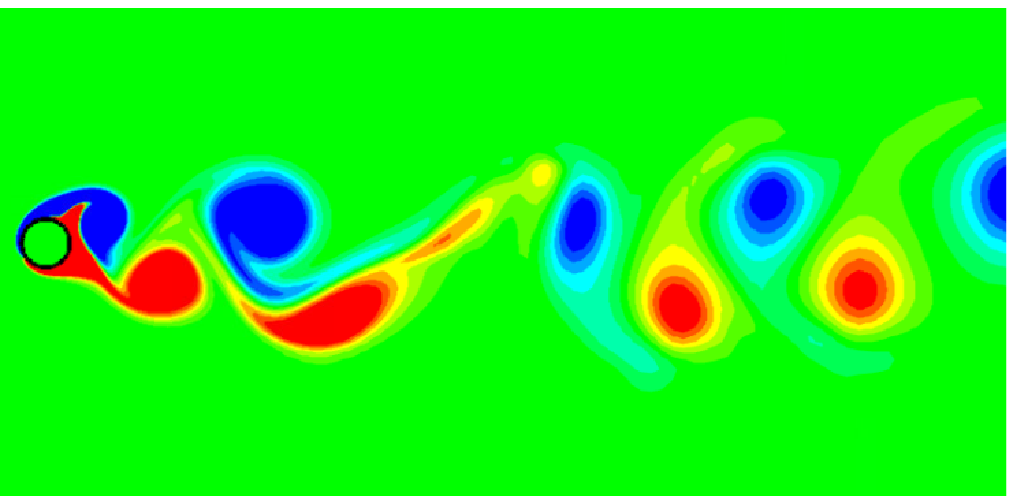}} \hfill
 \subfloat[ Bistable, $\beta=0.1$,   $U_r = 15$, $U_r^{eq}=4.77$, $a_0=0.54$ ]{\includegraphics[width=0.45\textwidth]{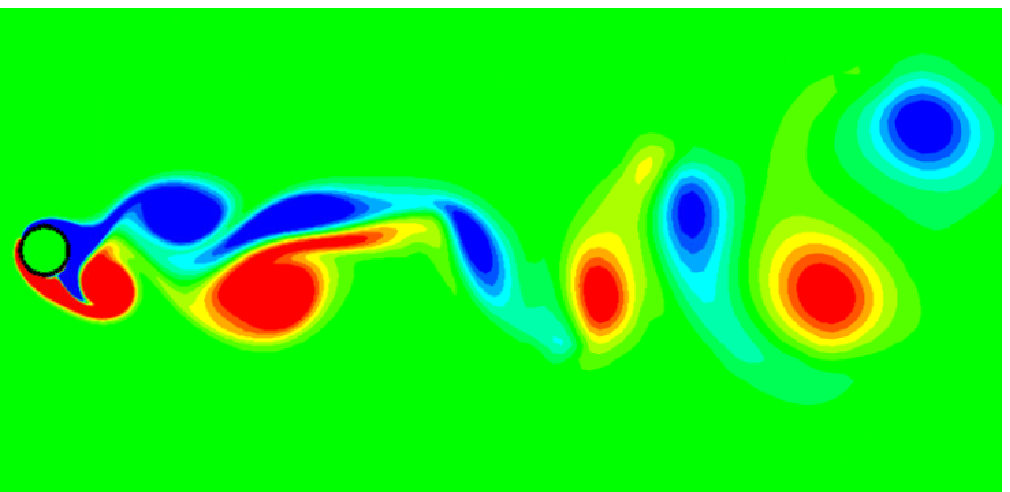}} \par

 \subfloat[ Bistable , $\beta=0.2$,  $U_r =7$, $U_r^{eq}=4.86$, $a_0=0.54$]{\includegraphics[width=0.45\textwidth]{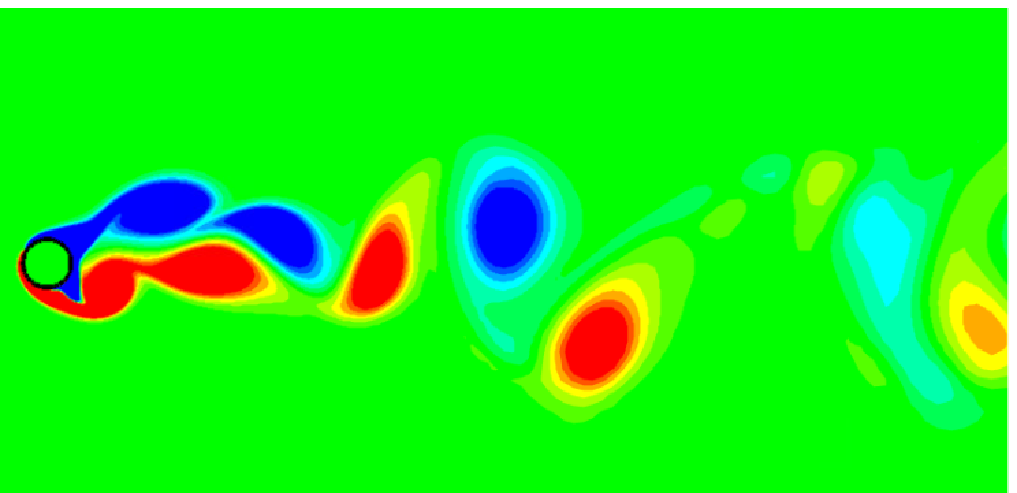}} \hfill
 \subfloat[ Stationary cylinder]{\includegraphics[width=0.45\textwidth]{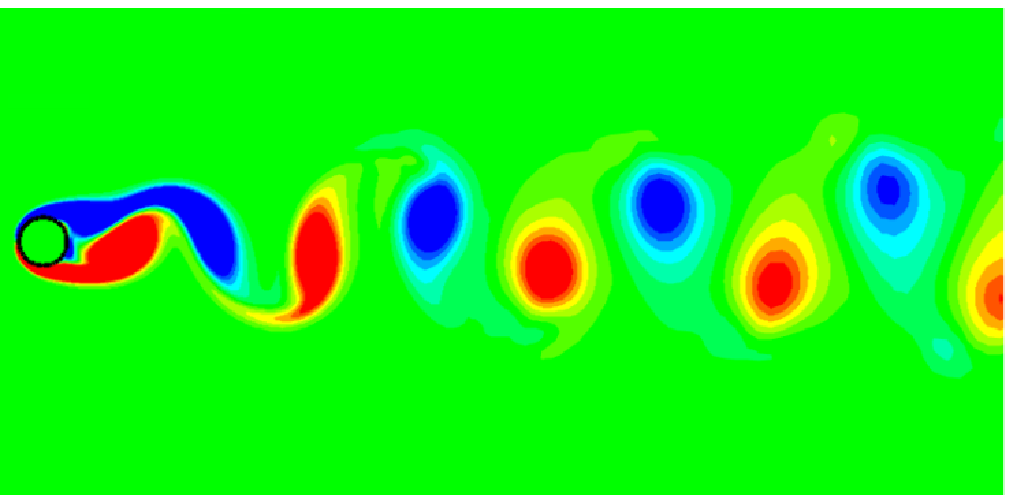}} 
 \par
 \subfloat{\includegraphics[width=0.75\textwidth]{legendnew.eps}} 
 %\end{tabular}
 
%\end{center}
\caption{\label{vorshed} (a)--(c) Vortex shedding patterns for VIV with different springs with $U_r^{eq}\in[4.75,\,4.86]$. (d) Vortex shedding pattern for flow around stationary cylinder. The color in the contour plots indicate the value of vorticity, indicated in the color map. }
\end{figure}

\begin{figure}\centering
 \subfloat[ Bistable, $\beta=0.1$,   $U_r = 10$, $U_r^{eq}=4.4$, $a_0=0.4$]{\includegraphics[width=0.45\textwidth]{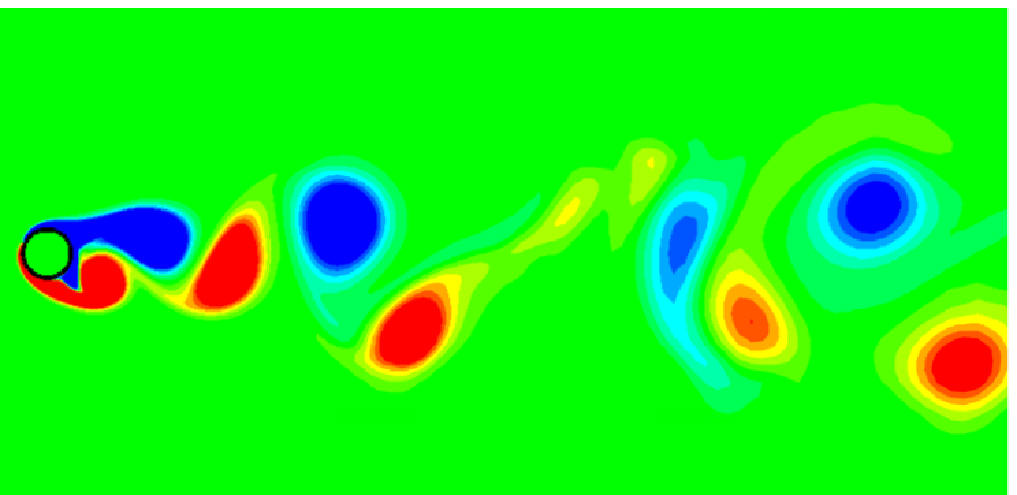}} \hfill
 \subfloat[ Bistable , $\beta=0.2$,  $U_r =4$, $U_r^{eq}=4.41$, $a_0=0.39$]  
 {\includegraphics[width=0.45\textwidth]{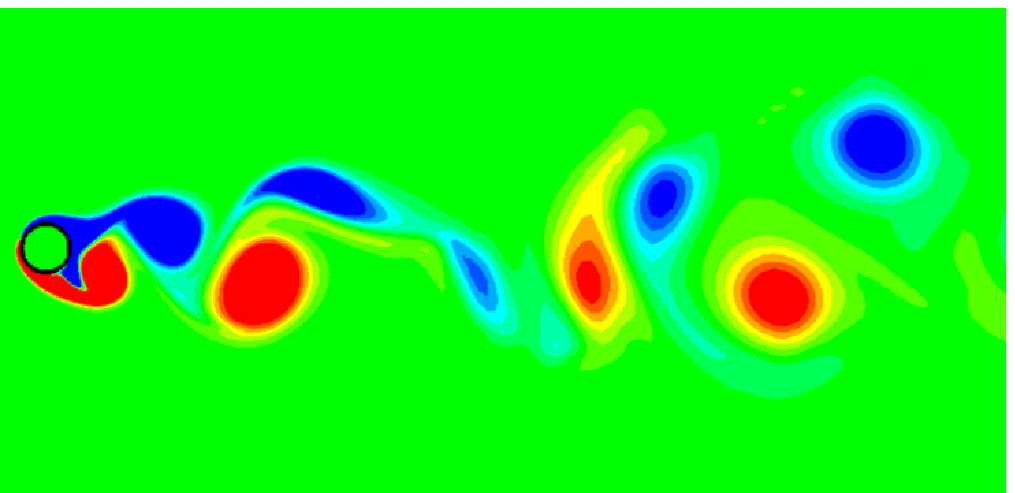}}\par 
\subfloat{\includegraphics[width=.75\linewidth]{legendnew.eps}}
\caption{\label{vorshedbi} Vortex shedding patterns for VIV with bistable spring for $U_r^{eq}\approx 4.4$. The color in the contour plots indicate the value of vorticity, indicated in the color map.}

\label{fig}
\end{figure}

We next compare vortex shedding pattern between the different types of springs for cases with similar values of $U_r^{eq}$ and $a_0$, i.e. which show reasonable collapse in Fig \ref{ampurstar}. Figs \ref{vorshed}(a)--(c), corresponds to cases for which $U_r^{eq}\in [4.75,4.86]$ and $a_0\in[0.51,0.54]$. Clearly, the vortex shedding patterns appear to be quite similar in spite of the different spring types. Compared to the 2S vortex shedding around stationary cylinder (shown in Fig \ref{vorshed}(d) for reference), the vortices appear to be quite distorted, with significant vortex mergers occurring close to the cylinder. In Figs \ref{vorshedbi}(a),(b) we have similarly compared the vortex shedding patterns for VIV with the two different bistable springs, in which $U_r^{eq}\in [4.4,4.41]$ and $a_0\in[0.39,0.4]$ are almost the same. Again, the vortex shedding patterns appear to have several similarities qualitatively. In general we can conclude that the vortex shedding pattern is dictated quite strongly by the amplitude $a_0$ and equivalent reduced velocity $U_r^{eq}$. On the other hand, the patterns appear to be quite insensitive to the rather non-harmonic nature of $y(t)$ during VIV with bistable springs (e.g. Fig \ref{ppfigs}). A similar hypothesis was made by \cite{Mackowski} to explain the collapse of experimental data for VIV with hardening and softening springs.  

\section{\label{ampfreq} Consistency of results with EC based theory}
\begin{figure}[] 
{ \centerline {\includegraphics[width=6in]{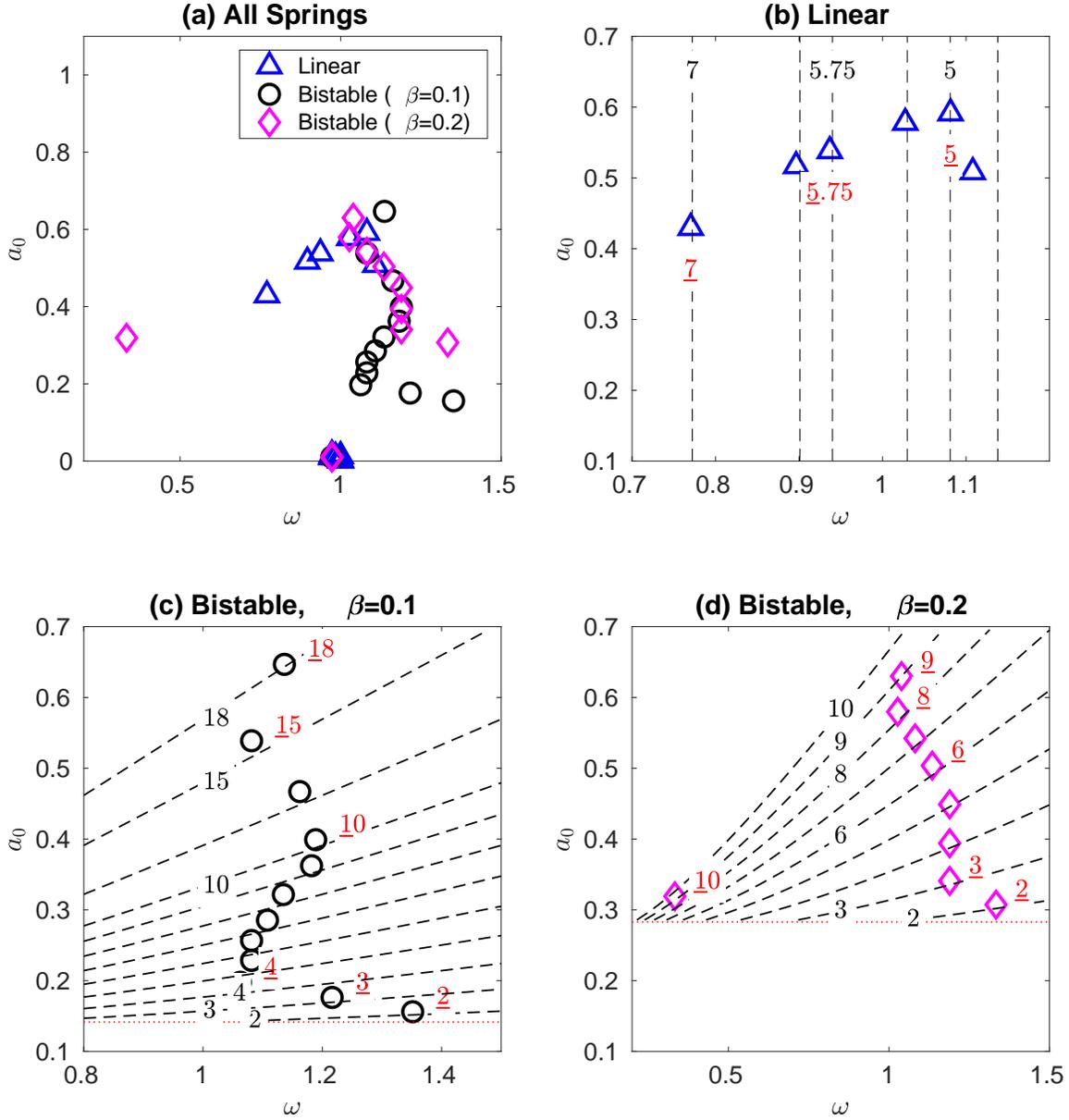}}}
\caption{\label{ampomega} Plot between $a_0$ versus structure frequency $\omega$ for data points corresponding to (a) \emph{all} springs (b) Linear spring only (c) Bistable spring, $\beta=0.1$, (d) Bistable spring, $\beta=0.2$. The dashed black lines in (b),(c),(d) indicate the natural frequency curve $\omega_n=(1/(\St\cdot U_r))F_n^*(a_0/\beta)$, with $F_n^*=1$ for (b) (linear spring) and $F_n^*(x)=F_2(x)$ in Eqn. \ref{fnwell} for (c),(d) (bistable springs, double well oscillations), with $U_r$ varying over values listed in Table \ref{simpar}. In (b),(c),(d), $U_r$ for some of the data points have been indicated with underlined values (red font), $U_r$ for the corresponding natural frequency curve has been indicated with non-underlined values (black font).  }
\end{figure}

We will now discuss whether or not results from our CFD simulations support the theory presented in section \ref{locinb}. In Fig \ref{ampomega}(a), we have plotted the displacement amplitude $a_0$ versus structure frequency $\omega$ for VIV with different types of springs. We observe a collapse of many of the data points onto a curve which qualitatively resembles a large portion of the EC curve predicted by our prior theory (Fig. \ref{ecplot}). The systematic peel-off of some of the data points from the EC can be attributed to the fact that, for some of cases, either $a_0$ is very close to $y_{cr}$, or, the vortex shedding pattern switches from single-row to two-row 2S version at very large amplitudes (discussed in sections \ref{ampchar} and \ref{vorplot}). The maximum value of $a_0$ over all the data points is given by $a_{max}\approx 0.6$, and the EC appears to exist over $0.8\leq \omega \leq 1.2$. The CFD data is, however, not capturing the lower left segment of the EC, perhaps due to the limited number of data points, as well as the large initial displacement used while simulating VIV with bistable springs. Thus the EC probably spans an even higher range over $\omega$. {The EC curve predicted by our theory (Fig. \ref{ecplot}) clearly has a much smaller value of $a_{max}$ compared to the the EC curve represented by the collapsed data points in Fig \ref{ampomega}(a). We therefore do not try to compare the EC curve from theory and CFD simulations on the same graph.} For the EC curve predicted by our theory (Fig. \ref{ecplot}) we observe that the curve is skewed such that the peak in the curve occurs at $\omega<1$. On the other hand, the collapsed data points from the CFD simulations (Fig \ref{ampomega}(a)) show that the peak occurs for $\omega>1$. These differences in the shape of the EC curve imply that the functions $q_0(a_0,\omega)$ and $\psi(a_0,\omega)$ (Eqn \ref{EC}) for our CFD simulations have a different form compared to the same functions derived from the WOM equations, which have in turn been calibrated against data generated at much higher Reynolds numbers \cite{facchinetti2004coupling,pantazopoulos1994vortex,griffin1980vortex}. The existence of 2S vortex shedding pattern during lock-in, along with the insensitivity of the vortex shedding patterns to the spring type, largely explains the collapse of the data points in Fig \ref{ampomega}(a). 

In Figs \ref{ampomega}(b)--(d), we superimpose the natural frequency curves (i.e. $a_0$-vs-$\omega_n$ at different $U_r$) over the $a_0$-vs-$\omega$ data points for each spring type, for the cases with large oscillation amplitudes. {We have used Eq. \ref{omneq} to plot the natural frequency curves for different $U_r$}. We observe that, the $(\omega,a_0)$ data points either intersect or lie in the close vicinity of the respective natural frequency curves (i.e. having same $U_r$ as the data point), which is consistent with the lock-in characteristics discussed earlier (section \ref{locres}). Also, as per the theory discussed in section \ref{locinb}, for VIV with bistable springs, collapsed data points with lower $a_0$ generally correspond to natural frequency curves with lower values of $U_r$ (Figs \ref{ampomega}(c),(d)). Again, the sole exception to this trend is the data point at $U_r=10$ for bistable spring with $\beta=0.2$ (Fig. \ref{ampomega}(d)). Here $a_0$ is relatively small, even though $U_r$ is quite high. Therefore, it is possible for data points to not lie on the EC due to lock-in of vortex shedding mode harmonics of natural frequency of the structure. However, our CFD simulations indicate that such data points exist over a very limited range in $U_r$.
  
\par Finally, we compare the prediction of approximate lock-in range for VIV with bistable springs from Eqn. \ref{approxlockin} with the results from CFD in Table \ref{predictlockin}. The upper limit of the range observed in CFD data is predicted reasonably well by Eqn. \ref{approxlockin} for both springs, but the lower limit is over-predicted by the approximation. The reason for this discrepancy could be that the EC curve in fact has a finite width in $\omega$ whereas Eqn. \ref{approxlockin} assumes the EC to have zero width in $\omega$. The same discrepancy is not seen for the prediction of the upper limit of lock-in range; here the natural frequency curve intersects the tip of the EC, in which case the width of the EC over $\omega$ does not play an important role.      

\begin{table}
\begin{center}
\begin{tabular}{|c|c|c|}
\hline
$\beta$ & CFD & Eqn. \ref{approxlockin}\\
\hline
$0.1$ & $2\leq U_r \leq 18$ & $4.6 \leq U_r \leq 21.2$ \\ 
\hline
$0.2$ & $2\leq U_r \leq 9$ & $4.6 \leq U_r \leq 10.3$ \\
\hline
\end{tabular}
\end{center}
\caption{\label{predictlockin} Prediction of lock-in regime for VIV with bistable springs using Eqn. \ref{approxlockin}, compared to results from CFD. We are using $\St=0.183$, $a_{max}=0.6$ and $C_\infty=0.6$ here.}
\end{table}

\section{\label{conclu}Conclusions}
In this work, we used a solver based on Immersed Boundary Method to simulate free transverse vibrations of a cylinder attached to bistable springs, as well as linear springs, in the presence of uniform fluid flow. The mass ratio was chosen to be relatively high, so that the high-amplitude oscillations typically corresponded to lock-in of the lift force with the natural frequency of the structure. One of the major goals of this work is to examine the validity of our prior theory \cite{badhurshah2019lock}, which predicts the widening of lock-in range for VIV with bistable springs, and which is strictly valid for high mass ratios.
\par The numerical simulations in this paper show that the range of reduced velocities over which the structure oscillates increases significantly for cases with bistable springs, as compared to linear springs. The bistable spring with lower inter-well separation displayed an especially wider lock-in regime. The maximum displacement amplitude ($a_{max}$) appears to be independent of the type of spring. The trends for displacement amplitude $a_0$, as well as the lock-in plot, collapse for different types of springs when $U_r^{eq}$ is used in the abscissa. The vorticity field displays 2S vortex shedding pattern, and the distortions in the pattern, compared to the shedding pattern seen for flow around stationary cylinder, are quite similar for the different spring types. Many of the results here are quite consistent with data from experiments and numerical simulations reported in literature for VIV involving non-linear springs \cite{huynh2017experimental, wang2019effect, Mackowski}. \par Encouragingly, the plots of displacement amplitude versus structure frequency ($a_0$-versus-$\omega$) for different spring types collapse reasonably well, supporting the existence of an Equilibrium Constraint proposed by our prior theory \cite{badhurshah2019lock}. The trends in the data emerging from our simulations, as well as the dependence of lock-in range on the spring non-linearity, may therefore be attributed to the manner in which the natural frequency versus amplitude curves intersect the EC curve. The results from the simulations in this paper, in tandem with our prior theory \cite{badhurshah2019lock}, therefore suggests a way forward towards designing non-linear elastic supports for bluff bodies, with the goal of providing an optimal lock-in range during VIV of the structure.   
\clearpage

\bibliography{FSI_biblio_}

\end{document}